\documentclass[10pt,times,twocolumn]{article}
\usepackage[switch,columnwise]{lineno}

\usepackage{amssymb}
\usepackage{amsthm}
\usepackage{mathtools}
\usepackage[affil-it]{authblk} 
\usepackage[T1]{fontenc} 
\usepackage{amstext}
\usepackage{amsmath} 
\usepackage{multirow}
\usepackage{textcomp} 
\usepackage{tensor} 
\usepackage{indentfirst} 
\usepackage{pdflscape}
\usepackage{array}
\usepackage{shadow}
\usepackage{floatflt}
\usepackage{wrapfig}
\usepackage{picinpar}
\usepackage{framed}
\usepackage[free-standing-units=true]{siunitx}
\DeclareSIUnit{\revolution}{\text{rev}}
\DeclareSIUnit{\lightyear}{\text{ly}}
\DeclareSIUnit{\cycle}{\text{cycle}}
\DeclareSIUnit{\bit}{\text{bit}}
\DeclareSIUnit{\torru}{\text{Torr}} 
\usepackage{longtable,lscape} 
\usepackage{outlines} 
\usepackage{times}
\usepackage{float}
\usepackage{lipsum} 
\usepackage[normalem]{ulem} 
\usepackage{graphicx}    
\usepackage{color}
\usepackage{epsfig}
\usepackage{longtable,lscape} 
\usepackage{rotating} 
\usepackage{hyperref}
\usepackage[left=0.5in, right=0.5in, top=1in, bottom=1in]{geometry}
\usepackage{setspace}
\usepackage{abstract}
\usepackage{tikz} 
\usepackage[siunitx, RPvoltages]{circuitikz}
\usepackage{pgfplots}
\usetikzlibrary{calc,decorations.markings,decorations.text,patterns,arrows.meta,intersections,pgfplots.fillbetween}
\usepackage{xspace} 
\usepackage{xcolor}
\usepackage[showdow, en-US]{datetime2}
\DTMsetstyle{mmddyyyy}
\usepackage{mathrsfs}
\usepackage[T1]{fontenc}

\usepackage{soul} 
\sethlcolor{cyan} 

\makeatletter
\AtBeginDocument{\let\hl\@firstofone}
\makeatother

\usepackage{caption}
\DeclareCaptionLabelSeparator{periodVertLine}{.~\textbar~}
\usepackage{subcaption}

\usepackage{sectsty}
\sectionfont{\normalsize\normalfont\bfseries}
\subsectionfont{\normalsize\normalfont\itshape}
\subsubsectionfont{\normalsize\normalfont\itshape}
\makeatletter
\renewcommand{\@seccntformat}[1]{\csname the#1\endcsname.\quad}
\makeatother

\usepackage{upgreek}
\usepackage{chemmacros}
\usepackage{textgreek} 
\chemsetup{greek=textgreek}
\chemsetup{modules=all}
\chemsetup[reactions]{
	before-tag = R,
	tag-open = ( ,
	tag-close = )
	}
\chemsetup[chemformula]{}

\usepackage{pgf,tikz,ifthen}
\usepackage{pgfplots}
\pgfplotsset{compat=1.18}
\usetikzlibrary{calc,decorations.markings,decorations.text,patterns,arrows.meta,intersections,pgfplots.fillbetween}
 
\newlength \figWidthCol
\newlength \figWidthFull
\newlength \figWidthFullExtra
\newcommand{\sci}[1]{\ensuremath{\times10^{#1}}} 
\newcommand{\app}{{\raise.17ex\hbox{$\scriptstyle\sim$}}} 

\newcommand{\dC}{\textdegree C\xspace} 
\newcommand{\ie}{\textit{i.e.}} 
\newcommand{\eg}{\textit{e.g.}} 
\newcommand{\etc}{\textit{etc.}} 
\newcommand{\etal}{\textit{et~al.}} 
\newcommand{\via}{\textit{via}\xspace} 
\newcommand{\vs}{\textit{vs.\ }} 
\newcommand{\invitro}{\textit{in vitro}\xspace} %

\newcommand{\ji}{\mathrm{i}} 
\newcommand{\overbar}[1]{\mkern 1.5mu\overline{\mkern-1.5mu#1\mkern-1.5mu}\mkern 1.5mu}

\newcommand{\ignore}[1]{}

\usepackage[numbers, square, comma, sort&compress]{natbib}
\usepackage[numbib]{tocbibind}

\usepackage{cuted}


\title{Scalable reflective communication for microscopic electronics}

\author[1]{Matthew F.\ Campbell\thanks{Corresponding author. Email: \href{mailto:cammat@seas.upenn.edu}{cammat@seas.upenn.edu}}}
\author[1]{Kyle Skelil}
\author[2,3]{Chaitanya Karimanasseri}
\author[4,5]{David Gonzalez-Medrano}
\author[6,7]{\authorcr Li Xu}
\author[6]{Jungho Lee}
\author[6]{David T.\ Blaauw}
\author[1]{Igor Bargatin}
\author[4]{Marc Z.\ Miskin\thanks{Corresponding author. Email: \href{mailto:mmiskin@seas.upenn.edu}{mmiskin@seas.upenn.edu}}}
\affil[1]{\footnotesize Department of Mechanical Engineering and Applied Mechanics, University of Pennsylvania, Philadelphia, PA, USA 19104}
\affil[2]{\footnotesize Department of Bioengineering, University of Pennsylvania, Philadelphia, PA, USA 19104}
\affil[3]{\footnotesize Now at: Sidney Kimmel Medical College (SKMC), Thomas Jefferson University, Philadelphia, PA, USA 19107}
\affil[4]{\footnotesize Department of Electrical and Systems Engineering, University of Pennsylvania, Philadelphia, PA, USA 19104}
\affil[5]{\footnotesize Now at: Princeton Innotech, Inc., West Windsor Township, NJ, USA 08550}
\affil[6]{\footnotesize Department of Electrical and Computer Engineering, University of Michigan, Ann Arbor, MI, USA 48109}
\affil[7]{\footnotesize Now at: NVIDIA Corp., Santa Clara, AA, USA 95051}

\date{June 19, 2026}

\begin{document}


\twocolumn[
  \maketitle 
  \begin{onecolabstract}
    \normalsize

    Untethered microscopic electronic circuits hold the potential for extraordinary advances in many fields such as neural transmitting and distributed sensing. 
    However, establishing uplink communications from the microscale back to the macroscopic world remains challenging; existing micro-transmitters are difficult to integrate with semiconductor processing. 
    Here we surmount this obstacle, introducing a strategy for modulating backscattered photons based on the electrochromic polymer PEDOT:PSS (poly(3,4-ethylenedioxythiophene) polystyrene sulfonate) that is scalable to micron-order sizes and manufacturable using standard parallelizable methods. 
    Our devices, which we call SPOTs (submillimeter polymer optical transmitters), actuate at low voltages ($<\pm1$~\si{\volt}), switch in as fast as 10~\si{\micro\second}, can run for millions of cycles, and operate seamlessly in electrolytes.  
    We achieve this design by emphasizing architectural simplicity and mass-manufacturability rather than traditional metrics such as data rates or energy costs. 
    As a demonstration, we develop SPOT-equipped temperature-sensitive photovoltaic-powered foundry-fabricated microchips and use them to wirelessly measure and transmit local temperatures. 
    These results represent an important step toward fully-integrable, micron-scale bidirectional communication.

    \vspace{20pt}
  \end{onecolabstract}
]

\section{Main}\label{S:main}

Recent advances have highlighted the amazing potential of untethered microcircuits for healthcare and biological applications, such as bridging the brain-computer interface~\cite{Marblestone2013-137, Neely2018-64, Barbruni2020-1160, Singer2021-2100664, Moon2021-1430, Lee2024-1110, Lee2025-1259}, sensing and computing at the microscale~\cite{Wu2018-191, Dincer2019-1806739, Cortese2020-9173, Burnett2021-101, Yuan2022-765, Xu2022-1, Shen2026-590}, discovering chemical reactions in massive parallel~\cite{Gorski2025-354}, and controlling robots small enough to move among microorganisms~\cite{Miskin2020-557, Reynolds2022-eabq2296, Hanson2025-e2500526122, Lassiter2025-eadu8009, Khunkitti1998-43, VanHouten2023-937}. These devices hold promise to combine powerful sensing, computation, and locomotive functions into small biofluid-compatible packages, ushering in a new era of investigation and inter-connectivity at the microscale~\cite{Chataut2023-7194, Li2024-127017}. 

Importantly, a critical enabling function for these microcircuits is to communicate with macroscale observers~\cite{Singer2021-2100664}.
Existing microscale communications technologies are wide ranging, including radio frequency (RF) antennas~\cite{Mohan2024-87}, near-field inductive coils~\cite{Barbruni2020-1160}, magnetoelectric (ME) antennas~\cite{Nan2017-296, Luo2024-317}, ultrasound backscattering piezoelectric modules~\cite{Neely2018-64, Liu2025-020702}, micro-light-emitting diodes (\textmu{}LEDs)~\cite{Lin2023-042502}, organic light-emitting diodes (OLEDs)~\cite{Hofmann2011-A1250}, and vertical-cavity surface-emitting lasers (VCSELs)~\cite{Pan2024-229}. Of these, photonic uplink strategies such as \textmu{}LEDs, OLEDs, and VCSELs, by virtue of the quantum nature of light, can be minimized to the smallest lateral on-chip footprints~\cite{Singer2021-2100664}. These optical methods offer strong performance but can be challenging to integrate into CMOS (complementary metal-oxide-semiconductor) circuitry. Their implementation often involves component transfer, wafer bonding, and/or soldering steps; may require encapsulation to prevent leaching of toxic elements; can impose thermal budgets on subsequent processes; can introduce significant on-chip topography; and may require voltage boosting to achieve potentials higher than those native to CMOS electronics~\cite{Perkins2008-13955, Gong2021-842, Chen2022-042005, Espenhahn2022-120200J}\ignore{Pu1999-201}. In contrast to these strategies, we pursue a design that maximizes scalability and minimizes fabrication complexity, at the expense of bandwidth and energy costs. Such a communications platform --- elegant, easy-to-fabricate, fully CMOS-compatible, and microscopic --- would lead to transformative applications in healthcare, chemistry, sensing, and robotics.

\begin{figure*}[p] 
\centering
\includegraphics[width=\figWidthFullExtra]{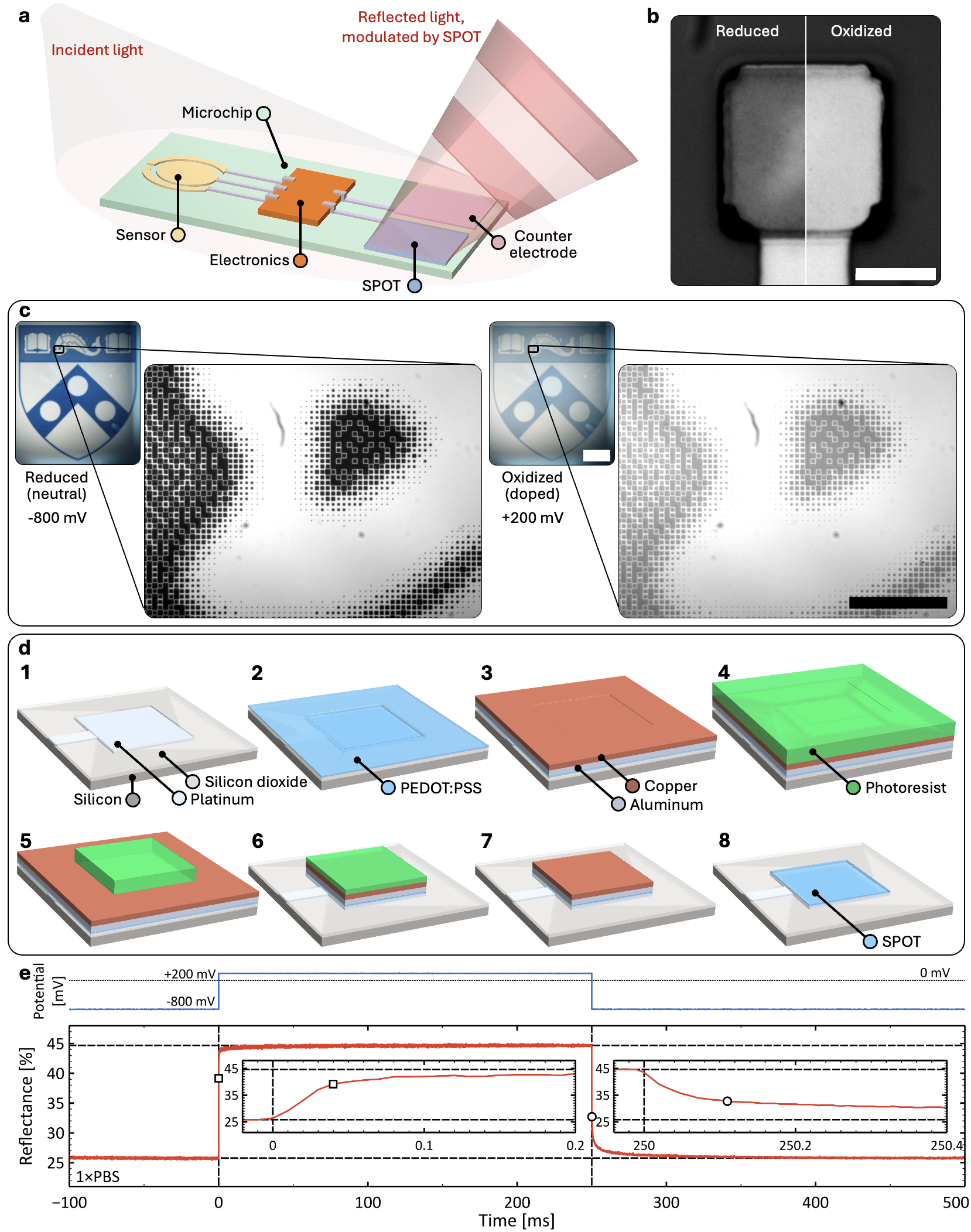}%
\end{figure*}
\begin{figure*}[h!] 
\centering
\caption{\textbf{Operation and fabrication of submillimeter polymer optical transmitters (SPOTs).} 
\textbf{(a)} Concept operational diagram for an arbitrary microchip. A sensor makes a measurement, onboard electronics process and digitize the data, and the resulting voltage is applied to the SPOT (a working electrode coated in PEDOT:PSS) relative to a counter electrode in an electrolyte solution. The voltage alters the polymer's oxidation state, changing its absorptivity and hence modulating the SPOT's reflectivity.
\textbf{(b)} Reduced/de-doped (left) and oxidized/doped (right) states of PEDOT:PSS~($\app45$~\si{\nano\meter})-\ch{Pt} stack in $1\times$PBS, at applied potentials of $-800$~\si{\milli\volt} and $+200$~\si{\milli\volt}, respectively, relative to an \ch{Ag}/\ch{AgCl} counter electrode. The polymer shown is cross-linked by thermal treatment (see Section~\ref{S:fabCrossHeat} in the Supplementary Information)~\cite{Doshi2025-2415827}. 
\textbf{(c)} Demonstration of SPOTs at macroscopic and microscopic scales. Far-left and center-right images show reduced and oxidized states (respectively) of a \si{\centi\meter}-scale chip featuring the University of Pennsylvania coat of arms, consisting of individual $\app170$~\si{\nano\meter}-thick PEDOT:PSS pixels with areas as small as $3\times3$~\si{\micro\meter\squared} (smaller than a red blood cell~\cite{Houchin1958-1185}) on a \ch{Pt} electrode in $1\times$PBS. Center-left and far-right images show enlarged views of the pixelation of the dolphin's eye in the reduced and oxidized states, respectively, in $1\times$PBS. See also Supplementary Videos~1 and~2. 
\textbf{(d)} Schematic diagram showing fabrication protocol (vertical heights exaggerated). 
(1) Begin with bare metal electrode on a \ch{Si} chip; perform \ch{O2} plasma cleaning. (2) Spincoat PEDOT:PSS and bake. (3) Sputter \ch{Al} and \ch{Cu} sequentially. (4) Spincoat photoresist. (5) Pattern photoresist. (6) Ion mill through \ch{Al}, \ch{Cu}, and PEDOT:PSS. (7) \ch{O2}-plasma-ash away photoresist from working electrode. (8) Etch away \ch{Al} and \ch{Cu} from working electrode in dilute \ch{HCl}. The device is now complete. 
\textbf{(e)} Reflective response to $\Delta V = \pm 1$~\si{\volt} electrode potential step changes between -800~\si{\milli\volt} and +200~\si{\milli\volt} (frequency $F=2$~\si{\hertz}). Insets show enlarged views of reflectance rise (oxidation) and fall (reduction), and open symbols show $\frac{1}{e}$ time points. The device tested in (e) consists of a PEDOT:PSS~($\app45$~\si{\nano\meter}, crosslinked with thermal treatment)-\ch{Pt} stack in $1\times$PBS, with lateral dimensions $30\times30$~\si{\micro\meter\squared}. Incident laser light at $\lambda_l=635$~\si{\nano\meter} is focused to a spot of roughly $10\times20$~\si{\micro\meter\squared} on the electrode (see Figure~\ref{F:meausurementSetupDiag} and inset of Figure~\ref{F:electrochemCharExtended}(b) in the Supplementary Information). See also Supplementary Videos~3 and~4. 
Scale bars: (b) 10~\si{\micro\meter}; (c, center) 3~\si{\milli\meter}; (c, right) 300~\si{\micro\meter}. }%
\label{F:setupFabOverview}%
\end{figure*}

Our premise is that photon \textit{backscattering} techniques~\cite{Herle2024-020901, Neely2018-64}\ignore{Lin2021-900} are easier to scale than photon \textit{producing} strategies~\cite{Hofmann2011-A1250, Cortese2020-9173, Lin2023-042502, Pan2024-229}\ignore{DeKoninck2025-63}, as evidenced by technologies such as bar codes~\cite{Ebling2010-4, Gu2011-733}\ignore{Gallo2011-1834}. What is needed is a way to modulate reflected light to convey information, for which we turn to a burgeoning class of electrochromic polymers~\cite{Beaujuge2010-268}. 
Perhaps the most mature of these is PEDOT:PSS (poly(3,4-ethylenedioxythiophene) polystyrene sulfonate), which has found recent uses in technologies such as flexible electronics~\cite{Fan2019-1900813}, organic electrochemical transistors (OECTs)~\cite{Rivnay2018-17086}, optical metasurfaces~\cite{Doshi2025-205} and displays~\cite{Fabiano2017-e1700345, Do2021-106106, Yang2024-2314983}, neural electrodes~\cite{Liu2026-20}, biosensors~\cite{AvilaRamirez2026-e13480}, and more~\cite{Li2024-10575}. 
Among its qualities, PEDOT:PSS is biocompatible~\cite{Li2025-87, Li2025-1192}\ignore{Richardson-Burns2007-1539}, operates organically in electrolytes like those present in biologically-relevant physiological environments~\cite{Alfonso2020-17260, Zhou2022-23505, Zhou2025-6776}, can be readily integrated with antifouling materials~\cite{Wang2017-396, Liu2017-11964, Banas2026-e70492}, exhibits long-term stability~\cite{Schander2016-6174, Duc2018-14, Dijk2020-1900662}, is electrically conductive~\cite{Lin2020-110435, Rebetez2021-2105821, Dingler2022-1600, AlhashmiAlamer2022-35371}, actuates at CMOS-compatible voltages within the electrochemical water window ($V \lesssim \pm1$~\si{\volt})~\cite{Cho2008-900, Stillmaker2017-74, Ehlich2024-53567}, and can be deposited and patterned using standard semiconductor processing techniques~\cite{Ouyang2015-603148, Derek2018-105116, Wang2020-105954, Doshi2024-2271, Park2024-46664, Li2025-4933, Lee2025-e12824} (Figure~\ref{F:setupFabOverview}(b-c). 
These attributes make it a promising candidate to enable simple, reflective uplink communication for microscopic electronics.  

Our proposed strategy is a SPOT (submillimeter polymer optical transmitter). SPOTs comprise a few-nanometer-thick film of PEDOT:PSS atop a mirror-like micron-scale metal electrode immersed in an electrolyte. Their operating principle when implemented on a microcircuit is simple: a stream of information, represented by a CMOS-compatible voltage~\cite{Stillmaker2017-74}, modulates the oxidation state of the polymer, altering its absorptivity and effecting a clearly-observable net reflectivity change. In this way, a microelectronic circuit equipped with a red-blood-cell-sized SPOT~\cite{Houchin1958-1185} (Figure~\ref{F:setupFabOverview}(c)) can optically communicate with a macroscopic observer (Figure~\ref{F:setupFabOverview}(a)). 

SPOTs' simple design makes them straightforward to fabricate (Figure~\ref{F:setupFabOverview}(d); see Section~\ref{S:fabSPOTs} in the Supplementary Information). Beginning with a chip featuring exposed metal electrodes and encapsulated wiring, we clean the surface with \ch{O2} plasma,  spincoat a mixture of 5\% ethylene glycol in an aqueous dispersion of PEDOT:PSS, and subsequently bake it (125~\dC) to dry the film.  We then sputter protective \ch{Al} and \ch{Cu} coatings sequentially over the polymer layer, and conduct photolithography to coat just the working electrode areas with photoresist. We next ion mill through the metal and PEDOT:PSS layers to remove the polymer everywhere but on the working electrodes, remove the residual photoresist using \ch{O2} plasma ashing, and etch away the remaining \ch{Al-Cu} coating in dilute \ch{HCl}. This leaves the chip in its original state with PEDOT:PSS patterned only onto the desired electrode surfaces. Notice that all of the fabrication steps occur at low temperatures ($T<200$~\dC); this means SPOTs can be implemented on circuitry that has strict thermal budgets after all other processing is complete. Also, since SPOTs are agnostic to their method of fabrication, many of the other demonstrated PEDOT:PSS patterning strategies in the literature could be used instead of the route presented here~\cite{Ouyang2015-603148, Derek2018-105116, Wang2020-105954, Doshi2024-2271, Park2024-46664, Li2025-4933, Lee2025-e12824}.  

\section{Results}\label{S:results}

Having fabricated SPOTs, we explore their time responses by measuring the temporal signal of reflected $635$~\si{\nano\meter} laser light in response to a potential step change (here and elsewhere, voltages are referenced to an \ch{Ag}/\ch{AgCl} counter electrode; see also Figure~\ref{F:meausurementSetupDiag} in the Supplementary Information). Results for a 30~\si{\micro\meter}-square SPOT with a $\app45$~\si{\nano\meter} polymer film stepping between $V = +200$ and $-800$~\si{\milli\volt} in $1\times$PBS are shown in Figure~\ref{F:setupFabOverview}(e) (see also Supplementary Videos~5 and~6). At time $t=0$~\si{\milli\second}, we step the voltage from $V=-800$~\si{\milli\volt} to $V=+200$~\si{\milli\volt}, which causes the PEDOT:PSS to oxidize and become more transparent, thereby increasing the reflectance signal (at $635$~\si{\nano\meter}, the reflectance changes monotonically with voltage~\cite{Lin2020-110435, Rebetez2021-2105821, Dingler2022-1600, Keene2023-eadi3536}; see Figures~S11 and~S12 in the Supplementary Information). Conversely, at $t=250$~\si{\milli\second}, we step the voltage back down, causing the polymer to reduce and become more absorbing, decreasing the reflectance. The PEDOT:PSS reacts quickly to the $\Delta V = \pm 1$~\si{\volt} steps, with $\frac{1}{e}$ reflectance rise (oxidation) and fall (reduction) times of $\tau_{e,o}=40$~\si{\micro\second} and $\tau_{e,r}=110$~\si{\micro\second}, respectively. That the oxidation time is faster than the reduction time has been observed by others~\cite{Paulsen2020-2003404, Rebetez2021-2105821, Keene2023-eadi3536} (see also Section~\ref{S:additionalVstepData} in the Supplementary Information).

The time scales shown in Figure~\ref{F:setupFabOverview}(e) suggest that operation at a frequencies approaching $F=10$~\si{\kilo\hertz} may be possible. To examine this, we characterize a 30~\si{\micro\meter}-square SPOT with a $\app35$~\si{\nano\meter} polymer film by now driving it with a 1~\si{\volt} peak-to-peak sine wave at frequency $F$ and measuring the required current amplitude $A_I$ and resulting relative reflectance amplitude $A_R$ (see Section~\ref{S:ExptSetupFdep} in the Supplementary Information). The experimental results are shown in Figure~\ref{F:electrochemChar} and in Figure~\ref{F:electrochemCharExtended} in the Supplementary Information. 

\begin{figure}
\centering
\includegraphics[width=\figWidthCol]{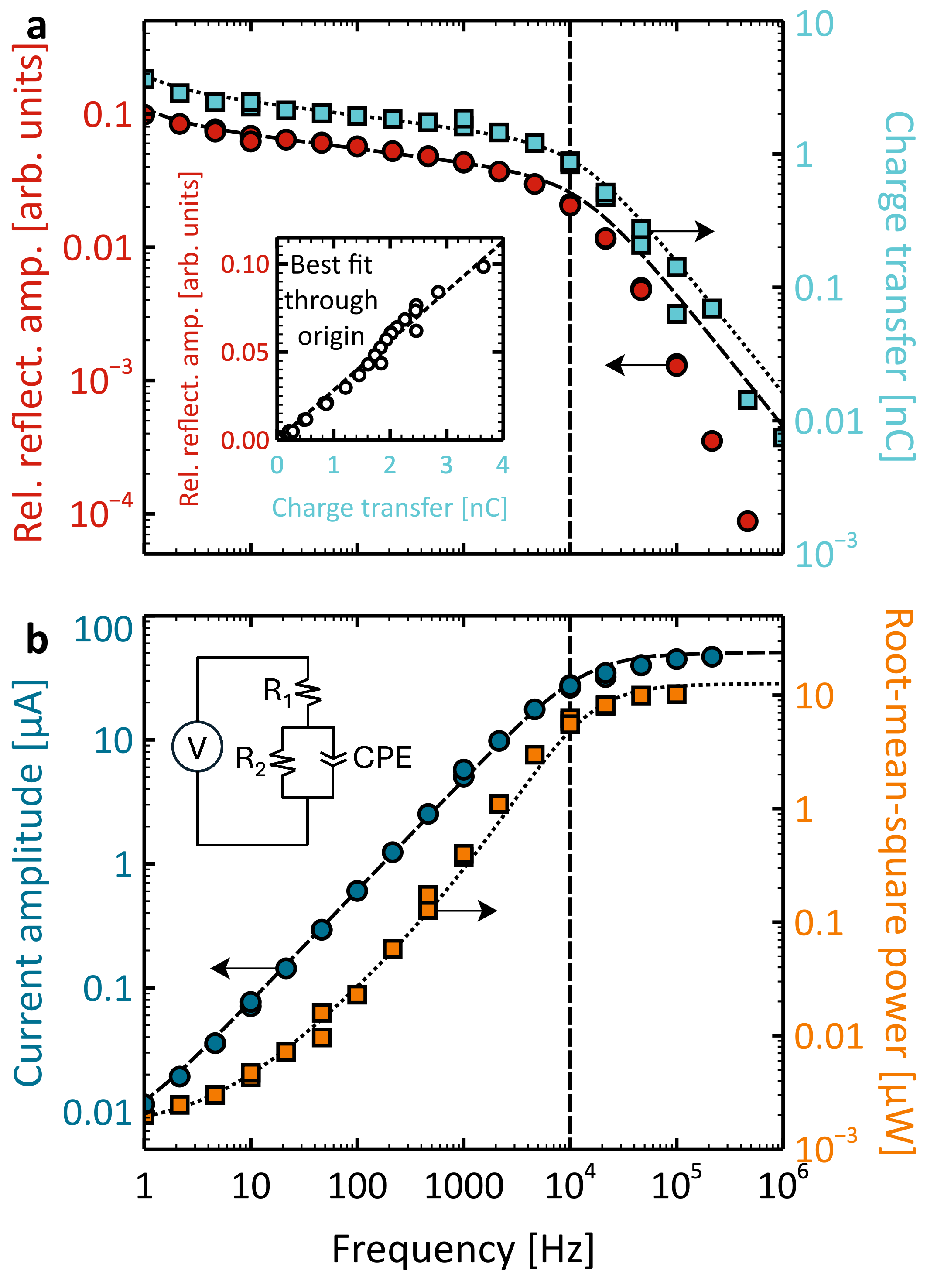}%
\caption{\textbf{Frequency-dependent characterization a SPOT.} 
Device consists of a PEDOT:PSS~($\app35$~\si{\nano\meter})-\ch{Pt} stack in $1\times$PBS, with lateral dimensions $30\times30$~\si{\micro\meter\squared}. Incident laser light at $\lambda_l=635$~\si{\nano\meter} is focused to a spot of roughly $10\times20$~\si{\micro\meter\squared} on the electrode. The input potential is a sinusoidal voltage between maximum and minimum values of  $+200$~\si{\milli\volt} and $-800$~\si{\milli\volt}, respectively. 
\textbf{(a)} Relative reflectance amplitude, defined as half of the peak-to-peak signal divided by the average value, and charge transferred per half cycle. Inset in (a) shows linear relation between these quantities. Vertical dashed line shows critical frequency $F_c$.
\textbf{(b)} Required current amplitude and root-mean-square power. Inset in (b) shows circuit model used to produce dashed and dotted line fits, consisting of a resistor $R_1=9.9$~\si{\kilo\ohm} in series with the parallel combination of a resistor $R_2=80$~\si{\mega\ohm} and a constant phase element (CPE) with parameters $Y_0 = 3.7$~\si{\nano\siemens\second\tothe{0.9}} and $n=0.9$. 
See also Figure~\ref{F:electrochemCharExtended} and Section~\ref{S:modelFreqElectroOpticalData} in the Supplementary Information.
}%
\label{F:electrochemChar}%
\end{figure}

Our measurements highlight two important frequency regimes of operation, separated by a critical frequency $F_c \approx 10$~\si{\kilo\hertz}.  At $F<F_c$, the current amplitude $A_I$ increases with $F$ and the relative reflectance amplitude $A_R$ stays roughly constant. When operating in this regime, an increase in speed comes at the cost of greater required power but not a significant decrease in the optical signal-to-noise ratio (SNR). The required RMS (root-mean-square) power at $F=10$~\si{\kilo\hertz} is about 10~\si{\nano\watt\per\micro\meter\squared}; for context, 1~\si{\nano\watt} is roughly the amount of solar power incident upon 1~\si{\micro\meter\squared}. As such, the necessary power at 10~\si{\kilo\hertz} could comfortably be supplied by a 10\% efficient photovoltaic (PV) array with an area 30~times that of the SPOT under a modest microscope illumination of 2~suns ($\sim 2000$~\si{\watt\per\meter\squared}) --- an area ratio exemplified in the singulated microcircuits depicted in Figure~\ref{F:demonstrationFigure}(a). 

The performance is markedly different for $F>F_c$, where $A_R$ can be seen to decrease rapidly even though $A_I$ remains constant (Figure~\ref{F:electrochemChar}(a)). We find $F_c$ to vary inversely with device side length $L$ and capacitance $C$, and within the limited range of device sizes we tested it spanned from about 1~\si{\kilo\hertz} to 34~\si{\kilo\hertz} (see Section~\ref{S:FreqScaling}, Figure~\ref{F:linearRelationships}, and Table~\ref{T:dataFitParams} in the Supplementary Information). Although the achievable modulation bandwidth is modest compared to emitter-based optical systems~\cite{Lin2023-042502, Hofmann2011-A1250, Pan2024-229}, frequencies up to and including $F\sim10$~\si{\kilo\hertz} are nevertheless useful for many biomedical applications. For example, 10~\si{\kilo\hertz} would allow for transmission of spatially-resolved measurements in \invitro environments such as organoids, including ion and enzyme concentrations (1-100~\si{\hertz})~\cite{Takebe2019-956, Hofer2021-402}; would be sufficient for optical reporting of the computer memory state in microrobots (10-1000~\si{\hertz})~\cite{Lassiter2025-eadu8009};  and even approaches the timescale required for neural recording ($\gtrsim 10$~\si{\kilo\hertz})~\cite{Moon2021-1430, Marblestone2013-137, Kim2025-4032, Li2025-87}. 

The right axis of Figure~\ref{F:electrochemChar}(a) shows the charge transferred per half-cycle $Q_h$, integrated from the measured current (see Section~2.2 in the Supplementary Information). The trends in the charge transfer $Q_h$ are similar to the relative reflectance amplitude $A_R$, a relationship that is highlighted in the inset of this figure. That $A_R$ increases linearly with $Q_h$ was first suggested by Keene, Rao, and Malliaras~\cite{Keene2023-eadi3536}, who pointed out that the absorption coefficient of PEDOT:PSS at $\lambda_l=635$~\si{\nano\meter} increases monotonically with voltage~\cite{Rebetez2021-2105821} (see also Figures~S11 and~S12 in the Supplementary Information) and that, in the voltage range $\langle -800, +200 \rangle$~\si{\milli\volt}, PEDOT:PSS behaves capacitively, with charge and potential linearly related~\cite{Keene2023-eadi3536}. The $A_r$---$Q_h$ link is in principle not obvious, because the polymer can assume three charge states (PEDOT\textsuperscript{0} (neutral), PEDOT\textsuperscript{+1} (polaron), and PEDOT\textsuperscript{+2} (bipolaron)) but the absorption coefficient of PEDOT\textsuperscript{0} at $635$~\si{\nano\meter} is several times that of PEDOT\textsuperscript{+1} and PEDOT\textsuperscript{+2}~\cite{Rebetez2021-2105821}. 
In our case, the linear relationship is fortuitous because it allows a simple optical measurement to provide direct insight into the electronic state of the polymer, a fact that has been exploited by others as well~\cite{Paulsen2020-2003404, Rebetez2021-2105821, Keene2023-eadi3536, Li2025-eadt5186}. For example, the decrease in the relative reflectance amplitude $A_R$ for $F>F_c$ reflects the fact that the charge available for changing the polymer's redox state $Q_h$ also decreases as $F$ increases.

We find that the frequency-dependent performance of SPOTs can be approximated by a simple Randles lumped circuit model consisting of a resistor $R_1$ in series with the parallel combination of a second resistor $R_2$ and a constant phase element with parameters $Y_0$ and $n$ (approximating the polymer's capacitance), as shown in the inset of Figure~\ref{F:electrochemChar}(b) and discussed in Section~\ref{S:modelFreqElectroOpticalData} in the Supplementary Information. The model's results are shown in dashed and dotted lines in Figure~\ref{F:electrochemChar} and Figure~\ref{F:electrochemCharExtended} in the Supplementary Information. Similar simple circuit models have been used by others to capture the behavior of PEDOT:PSS~\cite{Koutsouras2017-2321, Wang2021-5226, Smolka2023-107528}. In the Supplementary Information, we show that this model can be used to explain the impact of experimenting in $1\times$PBS \vs $10\times$PBS (Figure~\ref{F:pbsConcentration}), the effect of the counter electrode configuration (Figure~\ref{F:refElectrode}), and the influence of the SPOT lateral size (side length) on the current amplitude (Figure~\ref{F:areaScaling}). Additionally, using the model, we can define the critical frequency $F_c$ as that where the real and imaginary components of the impedance  have the same magnitude, near where the roll-off in the reflectance amplitude occurs. This leads to scalings of $F_c \sim \frac{1}{L}$ and $F_c \sim \frac{1}{C}$ with device side length $L$ and capacitance $C$, respectively, consistent with our observations (see Section~\ref{S:FreqScaling} and Figure~\ref{F:areaScaling} in the Supplementary Information). Lastly, the model leads to a value of the volumetric capacitance of PEDOT:PSS of $C^\ast \sim 38.5$~\si{\farad\per\centi\meter\cubed} for the experiment of Figure~\ref{F:electrochemChar} (see Section~\ref{S:addlFig2} in the Supplementary Information), similar to other values in the literature~\cite{Kurra2014-17058, Rivnay2015-e1400251, Proctor2016-1433, Volkov2017-1700329, Tybrandt2017-eaao3659, Bianchi2020-11252}. For completeness, we note that detailed models of PEDOT:PSS have also been proposed~\cite{Stavrinidou2013-244501, Volkov2017-1700329, Tybrandt2017-eaao3659, Rebetez2021-2105821, Cucchi2022-4514, Keene2023-eadi3536, Unigarro2025-12329}, but we prefer the simplicity of the Randles model for the engineering purposes of this work.

The circuit model is not only useful in the electrical domain; it can provide important optical insights as well. Using the linear relationship between the relative reflectance amplitude $A_R$ and the charge transferred per half-cycle $Q_h$ (inset of Figure~\ref{F:electrochemChar}(a)), one can predict the scaling of $A_R$ with $F$. The maximum data rate and energy cost per bit transmitted can then be assessed, given an estimate of the noise floor and reflected intensity for a particular optical configuration (Figure~\ref{F:electrochemCharExtended} in the Supplementary Information). As an example, if the SPOT used in Figure~\ref{F:electrochemChar} had dimensions of $10\times20$~\si{\micro\meter\squared} (\ie, the size of the laser spot) instead of $30\times30$~\si{\micro\meter\squared}, the minimum energy cost would decrease from $E_b\sim90$~\si{\pico\joule\per\bit} at $F\sim150$~\si{\hertz} to $E_b\sim20$~\si{\pico\joule\per\bit} at $F\sim300$~\si{\hertz}, allowing for an increase in transmission rate for a lower energy penalty (see Section~\ref{S:addlFig2} in the Supplementary Information). For context, these energy costs are about two orders of magnitude higher than those required by VCSELs~\cite{Larisch2020-18931}.

The results shown in Figures~\ref{F:electrochemChar} and~S4 (Supplementary Information) depend somewhat on the specific optical arrangement, which essentially comprises an illumination source, the SPOT to modulate the reflected intensity, an optical train to collect and focus the reflected photons, and a photodiode or camera to measure the resulting intensity. We focused primarily on laser illumination and photodiode detection for this work, but similar signal levels could likely be attained using full-frame light-emitting diode illumination and video camera monitoring, allowing for parallel readout from a large assembly of untethered microchips~\cite{Cortese2020-9173, Hanson2025-e2500526122}. 

\begin{figure*}
\centering
\includegraphics[width=\figWidthFull]{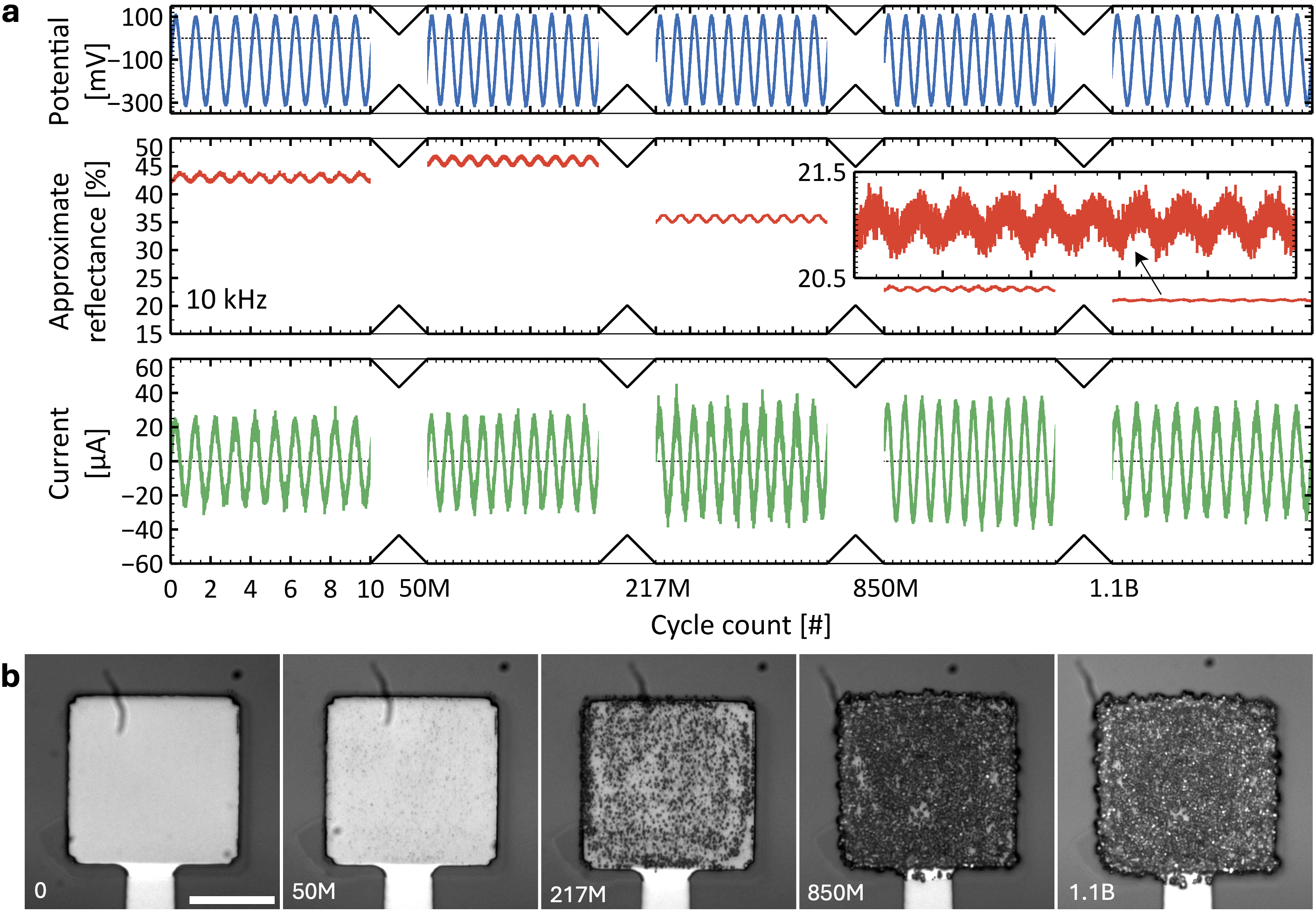}%
\caption{\textbf{Results of durability study.} 
Device consists of a PEDOT:PSS~($\app84$~\si{\nano\meter})-\ch{Pt} stack in $1\times$PBS, with lateral dimensions $40\times40$~\si{\micro\meter\squared}. The polymer is cross-linked using GOPS (see Section~\ref{S:fabCrossGOPS} in the Supplementary Information)~\cite{Tseng2024-13384}, and the driving signal applied to the working electrode is a 10~\si{\kilo\hertz} sine wave between maximum and minimum potentials of +100~\si{\volt} and -300~\si{\volt}, respectively.
\textbf{(a)} Driving voltage (top), approximate reflectance at $\lambda_l=635$~\si{\nano\meter} (middle), and current (bottom) at select intervals.
\textbf{(b)} Montage of micrographs showing the SPOT at intervals from panel~(a). 
See also Figure~\ref{F:additionalDurationFigs} in the Supplementary Information.
Scale bar (b): 20~\si{\micro\meter}.
}%
\label{F:duration}%
\end{figure*}

\begin{figure*}
\centering
\includegraphics[width=\figWidthFullExtra]{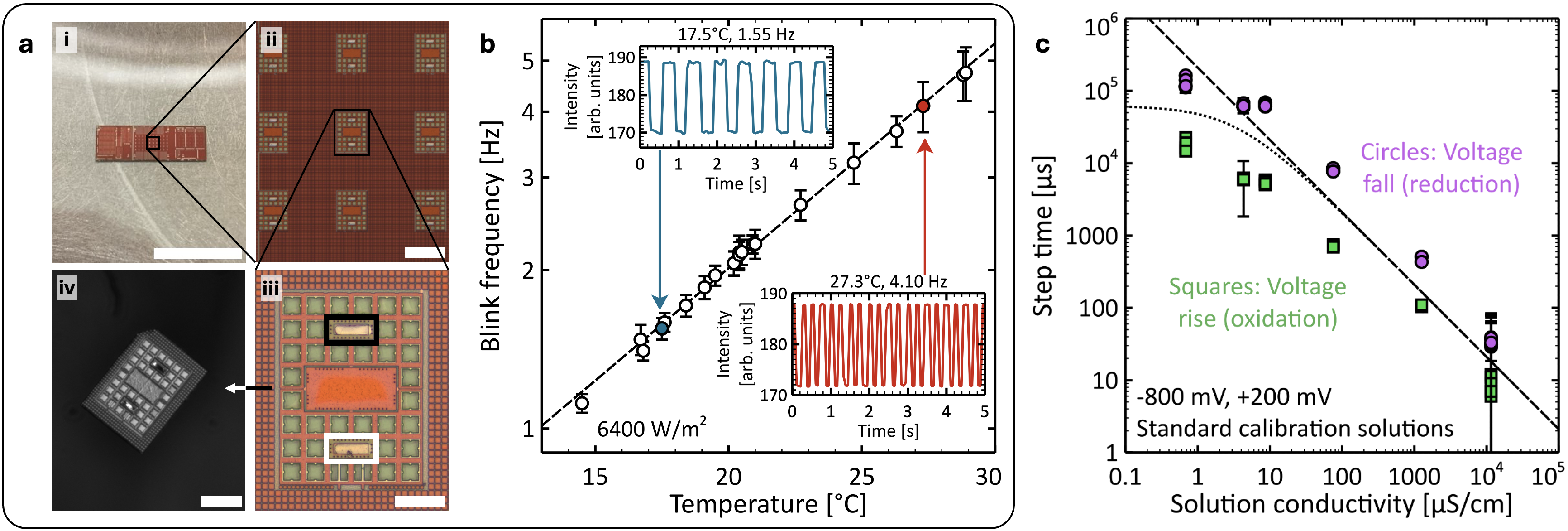}%
\caption{\textbf{Demonstrations of SPOTs.}
\textbf{(a)} Images of back-end fabrication steps to add SPOTs to CMOS temperature-sensitive microchips. (i) Chip containing microcircuits prior to PEDOT:PSS processing; (ii) micrograph of individual circuits on chip; (iii) micrograph of a single microcircuit with a 50~\si{\nano\meter} PEDOT:PSS film atop the working electrode (the black rectangle denotes the working electrode, the white rectangle denotes the counter electrode, the green squares are photovoltaic panels (PVs), and the ratio of the PV area to the working electrode area is about 28); (iv) micrograph of singulated microcircuit released in $1\times$PBS solution.  
\textbf{(b)} Relationship between blink frequency and temperature for one microcircuit in $1\times$PBS at a light intensity of 6400~\si{\watt\per\meter\squared}, prior to chip dicing. The error bars represent standard deviations for the measurements. Insets show sample intensity data obtained from two experiments at 17.5\dC (blue) and 27.3\dC (red). 
\textbf{(c)} Time $\left(\frac{1}{e}\right)$ required for reflectance rise (oxidation) and fall (reduction) for $\Delta V = \pm 1$~\si{\volt} electrode potential step changes between -800~\si{\milli\volt} and +200~\si{\milli\volt}, as a function of the solution conductivity $\sigma_s$ for several off-the-shelf standard conductivity calibration solutions. The error bars represent standard deviations for the measurements. 
The device for (c) consists of a PEDOT:PSS~($\app 123$~\si{\nano\meter})-\ch{Pt} stack with lateral dimensions $10\times10$~\si{\micro\meter\squared}. Incident laser light at $\lambda_l=635$~\si{\nano\meter} is focused to a spot of roughly $10\times20$~\si{\micro\meter\squared} on the electrode; note that the laser focal point size is slightly larger than the electrode size.
The dashed line shows the inverse scaling of the response time with solution conductivity, and the dotted line shows the harmonic mean of the solution conductivity-dependent dashed line with a longer constant time that may represent a secondary charging route at the lowest conductivity values (see Section~\ref{S:stepLowSolnCond} in the Supplementary Information). 
Scale bars: (a): (i) 10~\si{\milli\meter}, (ii) 200~\si{\micro\meter}, (iii) 50~\si{\micro\meter}, (iv) 100~\si{\micro\meter}.
}%
\label{F:demonstrationFigure}%
\end{figure*}

We test the longevity of SPOTs by driving them with a sinusoidal voltage between maximum and minimum potentials of $+100$~\si{\milli\volt} and $-300$~\si{\milli\volt}, respectively, at $F=10$~\si{\kilo\hertz}. Figure~\ref{F:duration} shows results for a SPOT that ran continuously for more than 1.1~billion cycles over the course of over 31~\si{\hour} (see also Figure~\ref{F:additionalDurationFigs} in the Supplementary Information, and Supplementary Videos~7 and~8). To our knowledge, this duration is orders of magnitude longer than other PEDOT:PSS optical switching devices demonstrated to date, summarized in a recent review by Wang and Liu~\cite{Wang2023-100036}. Although the average reflectance $\overbar{\varrho}$ and relative reflectance amplitude $A_R$ drop by about 50\% and 90\%, respectively, throughout this time, the modulation is still discernible, as seen in the inset to Figure~\ref{F:duration}(a). We note that the lifetime we observe is partially due to some aspects of our test protocol: (1) our frequency ($F=10$~\si{\kilo\hertz}) is about five orders of magnitude higher than some others ($F \sim 25$~\si{\milli\hertz}.), allowing us to accrue more cycles in much less time; (2) our voltage range ($\langle -300, +100 \rangle$~\si{\milli\volt}) is smaller than those of others ($\langle -1, +1 \rangle$~\si{\volt}), keeping the polymer well within a reversible operation window; and (3) we allow the SPOT to operate essentially until its end of life, whereas other studies tested their devices as long as they retained a large fraction of their maximum and minimum absorptivity values (in reduced and oxidized states, respectively). 

We observe that, throughout the experiment, the working electrode develops dark spots, and that the density of those spots increases with time. Low-depth-of-field microscopy at different working distances, using the focal plane to pinpoint the out-of-chip dimension, suggests that these dark areas are roughened PEDOT:PSS clusters that scatter away the incoming light. The degradation is isolated to the working electrode being driven; neighboring SPOTs on the same chip that are unexcited but still submerged in the electrolyte retain their original appearance (see Figure~\ref{F:additionalDurationFigs} in the Supplementary Information). We speculate that the degradation is associated with mechanical fatigue as the PEDOT:PSS swells and shrinks when hydrated cations from the electrolyte move in and out (respectively) during the cyclic reduction and oxidation (respectively) processes~\cite{Proctor2016-1433, Volkov2017-1700329, Fabiano2017-e1700345, Savva2018-12023, Biessmann2018-9865, Modarresi2020-6267, Rebetez2021-2105821, Dingler2022-1600, Lyu2023-746, Sedghamiz2023-5512, Doshi2025-205}. The change in appearance may also be coupled to the so-called ``self healing'' properties of PEDOT:PSS~\cite{Li2020-2002853}. 

These ideas point to several pathways for extending SPOTs' durability, including improving adhesion of the polymer to the metal electrode using electrografting of salts~\cite{Smolka2023-107528}, enhancing stability of the polymer through chemical treatments~\cite{Adilbekova2024-2405094}, or pursuing other crosslinking methods~\cite{Tang2021-1436}; we recommend future research on SPOT longevity. For the time being, we note that even cycle counts of 200~million are sufficient for many long-term tasks. For instance, this would provide an uninterrupted 1~\si{\kilo\hertz} data uplink from a microrobot for more than two days~\cite{Lassiter2025-eadu8009}. Furthermore, SPOTs could be parallelized across many microcircuits, mitigating the impact of individual device degradation in large-scale systems. 

We provide two practical demonstrations of SPOTs, beginning with wirelessly communicating local temperatures measured by CMOS microcircuits. Such an application is important for health monitoring, point-of-care diagnostics, and biology~\cite{Cortese2020-9173, Pradhan2021-21, Chen2024-e30649}. Our CMOS-fabricated microcircuits, each with footprints of $150\times200$~\si{\micro\meter\squared}, feature photovoltaics for power, an onboard internal clock, and two electrodes that pulse alternately between voltages of roughly $+1$~\si{\volt} and $-1$~\si{\volt} at a frequency that is proportional to temperature. We obtain $4\times12$~\si{\milli\meter\squared} chips from a commercial foundry, each containing 50~microcircuits~\cite{Xu2022-1}, perform a \ch{SiO2} encapsulation process to protect the circuitry, and then use our standard process (see Section~\ref{S:fabSPOTs} in the Supplementary Information) to apply PEDOT:PSS to one of the two electrodes (hereafter referred to as the working electrode) on each of the microcircuits in parallel (Figure~\ref{F:demonstrationFigure}(a)). We then immerse the chip in $1\times$PBS and measure the blink frequency of the working electrode of one of the microcircuits as a function of the incident light intensity and temperature (see Supplementary Video~9 and Figure~\ref{F:demonstrationFigure}(b)). Thereafter, we coat the chip with \ch{Al}, \ch{Cu}, and photoresist, dice it to release the individual microcircuits, gently dissolve away the coatings to obtain individual circuits, and observe them operating again in $1\times$PBS. This example highlights the ability of SPOTs to provide a minimal-overhead, back-end compatible uplink communications pathway for standard CMOS microcircuits. 

SPOTs can be useful not only to communicate scientific information, but gather it, too. 
Figure~\ref{F:demonstrationFigure}(c) revisits the experiment of Figure~\ref{F:setupFabOverview}(e), showing measurements of the time $\left(\frac{1}{e}\right)$ required for reflectance rise (oxidation) and fall (reduction) for $\Delta V = \pm 1$~\si{\volt} electrode potential step changes between -800~\si{\milli\volt} and +200~\si{\milli\volt}, as a function of the solution conductivity $\sigma_s$ for several off-the-shelf standard conductivity calibration solutions. 
We observe that, in the limit of high electrolyte concentrations (the right side of the plot), the step times scale inversely with solution conductivity. The relationship between the reflectance and the transferred charge (Figure~\ref{F:electrochemChar}(a)) allows us to interpret this in terms of the electrical properties of the SPOT; specifically, for the purposes of this discussion, we consider that the optical rise time and the electrical rise time are equivalent. Then, the inverse relationship between rise times and solution conductivity follows from the scaling of the solution resistance, which for square micro-electrodes with side length $L$ is $R_s \sim \frac{\sqrt{\pi}}{4 \sigma_s L}$~\cite{Newman1966-501} (see Section~\ref{S:FreqScaling} in the Supplementary Information). This relationship also leads to a volumetric capacitance of $C^\ast \sim 39$~\si{\farad\per\centi\meter\cubed}, again similar to other values in the literature and to our value from the experiment of Figure~\ref{F:electrochemChar}~\cite{Kurra2014-17058, Rivnay2015-e1400251, Proctor2016-1433, Volkov2017-1700329, Tybrandt2017-eaao3659, Bianchi2020-11252} (see Section~\ref{S:stepLowSolnCond} in the Supplementary Information). At lower concentrations (below $\sigma_s \sim 10$~\si{\micro\siemens\per\centi\meter}), the step times begin increasing more slowly, which may reveal an alternate charging pathway, as discussed in Section~\ref{S:stepLowSolnCond} in the Supplementary Information. 

Figure~\ref{F:demonstrationFigure}(c) illustrates that, given an experimental or theoretical calibration curve, a measurement of the rise or fall time of a SPOT can be used to determine solution conductivity - a problem of practical importance in microrobot~\cite{Hanson2025-e2500526122} and microbiology~\cite{Record1998-143} experiments. Importantly, whereas conventional conductivity meters rely on a measurement of current between immersed electrodes~\cite{Jones1933-1780, Brinkmann2003-346, Orru2013-035903, Zhang2014-055005}, a signal that drops as conductivity decreases, SPOTs make use of an optical reflectance measurement in the time domain, wherein the time increases with decreasing conductivity. For instance, a $\sigma_s=1$~\si{\micro\siemens\per\centi\meter} solution measured with the experimental setup used in Figure~\ref{F:demonstrationFigure}(c) would lead to a reduction step time of $\sim 100$~\si{\milli\second} --- easily resolvable with an off-the-shelf photodiode. However, an electrical conductivity meter using an electrode geometry consistent with the size of the SPOT tested here (\eg, two $ 10\times 10$~\si{\micro\meter\squared} electrodes positioned 10~\si{\micro\meter} apart, yielding a cell constant of $K=1000$~\si{\per\centi\meter}) employing a $A_V=0.5$~\si{\volt} amplitude square wave potential would measure a current amplitude of just 500~\si{\pico\ampere}, a level requiring noise-suppressing equipment~\cite{Kim2013-52}.
The range over which our SPOT-based solution conductivity meter exhibits linearity is $\sigma_s \gtrsim 10$~\si{\micro\siemens\per\centi\meter}, which is relevant for applications in biological solutions or implanted devices (compare to the conductivity of $1\times$PBS, which is roughly $15$~\si{\milli\siemens\per\centi\meter}~\cite{HassanpourTamrin2025-61568}).
%
%
%
We have observed that the relative reflectance amplitude $A_R$ tends to decrease with decreasing conductivity, but this can likely be overcome by repeated averaging. 
Lastly, we note that SPOTs hold the potential to produce time-and-spatially-resolved conductivity data, because an array of devices can be probed simultaneously using a single microscope video camera. 



\section{Discussion}\label{S:discussion}

SPOTs as implemented today have lower data rates and require greater power than \textmu{}LEDs~\cite{Lin2023-042502}, OLEDs~\cite{Hofmann2011-A1250}, and VCSELs~\cite{Pan2024-229}, but they can be fabricated in parallel using back-end CMOS-compatible processes and introduce minimal on-chip topography ($\lesssim 200$~\si{\nano\meter} for SPOTs compared to $\sim 900$~\si{\nano\meter} for \textmu{}LEDs~\cite{Cortese2020-9173} or $\sim 7$~\si{\micro\meter} for VCSELs~\cite{Pu1999-201}). Also, SPOTs involve more fabrication steps than radio frequency (RF) antennas~\cite{Mohan2024-87} and near-field inductive coils~\cite{Barbruni2020-1160}, and require a power supply (\eg, photovoltaic panels), unlike some magnetoelectric (ME) antennas~\cite{Nan2017-296, Luo2024-317} and ultrasound backscattering piezoelectric modules~\cite{Neely2018-64, Liu2025-020702}; however, SPOTs remain useful at smaller lateral sizes ($\lesssim 10$~\si{\micro\meter} for SPOTs compared to $\gtrsim 100$~\si{\micro\meter} for RF/ME antennas, inductive coils, and piezoelectric modules~\cite{Singer2021-2100664}). Future iterations can harness recent materials and fabrication advances to operate in air using hydrogel layers~\cite{Lu2019-1043, Takalloo2019-60, Zhang2020-1904752, Lin2024-14740}; exhibit higher signal-to-noise ratios using retro-reflectors~\cite{Bender1973-229, Chalasani2011-1158, Wheaton2021-8504, Han2022-eabn0602}; achieve greater bandwidths by changing the polymer mixture and processing~\cite{Rivnay2016-11287, Ko2019-1900710, Gu2022-14679, Wang2023-100036}; and combine optical communication and sensing functions into the same electrode~\cite{Liang2021-2100061, Wang2023-100036, Burtscher2024-7, Kim2025-4032, Li2025-87}. In addition, we estimate that, using commercial silicon foundries, SPOTs can be built for much less than one cent ($\ll$US\$0.01) per device (see Section~\ref{S:costEst} in the Supplementary Information), making them cost-competitive with other micro-communications platforms such as \textmu{}LEDs or VCSELs~\cite{Cortese2020-9173}. The fabrication protocols and device characteristics we introduce here can immediately be leveraged to integrate SPOTs into a wide array of devices, enabling simple yet transformative bidirectional micro-macro connectivity in medicine~\cite{Marblestone2013-137, Neely2018-64, Barbruni2020-1160, Singer2021-2100664, Moon2021-1430, Lee2025-1259}, chemistry~\cite{Gorski2025-354}, sensing~\cite{Cortese2020-9173, Chataut2023-7194, Li2024-127017}\ignore{Wu2018-191, Dincer2019-1806739}, and robotics~\cite{Miskin2020-557, Hanson2025-e2500526122, Lassiter2025-eadu8009}\ignore{Reynolds2022-eabq2296}.


\section{Methods}\label{S:methods}

We produce the SPOTs, test chips, and temperature sensing microcircuits using standard CMOS processes. Full details are provided in the Supplementary Information. To test our devices, we reflect $\lambda_l=635$~\si{\nano\meter} laser light off of a SPOT and collect it with a photodiode. We drive the SPOT with a potential applied to it relative to a \ch{Ag}/\ch{AgCl} counter electrode in the electrolyte (often $1\times$PBS) and simultaneously measure the current delivered using a transimpedance amplifier. Information on the benchtop setup and our signal processing is provided in the Supplementary Information. 

\section{Data availability}\label{S:dataAvail}

All raw and processed data are available in the Supplementary Information and/or from the corresponding authors upon reasonable request.

\section{Acknowledgments}\label{S:acknowledgments}

This work was carried out in part at the Singh Center for Nanotechnology at the University of Pennsylvania, which is supported by the NSF National Nanotechnology Coordinated Infrastructure Program under grant NNCI-2025608.
It was also carried out in part in the Nanofabrication Facility at the University of Delaware and the Micro/Nanofabrication Center at Princeton University.
M.F.C.\ is partially supported by the National Institutes of Health under grant number K25-AI-166040-01. 
The authors wish to thank the scientific staff at the Singh Center for Nanotechnology at the University of Pennsylvania and at the Nanofabrication Facility at the University of Delaware for fabrication advice and useful discussions. 
In addition, the authors thank 
Dr.\ Lucas C.\ Hanson for assistance in printing photomasks, 
Dr.\ Abhilasha Kamboj for assistance in photographing the University of Pennsylvania coat of arms chip, 
Dr.\ Siyoung Lee for assistance in preliminary ion milling characterization experiments and in fabricating the PDMS liquid barriers for the SPOT characterization experiments, 
Dr.\ William H.\ Reinhardt for assistance in preliminary electrical measurements of the temperature-sensing chips, and
Dr.\ Michael F.\ Reynolds for developing the ion mill-based PEDOT:PSS fabrication routine used in this work.  

\section{Author contributions}\label{S:authCont}
M.F.C.\ and M.Z.M.\ conceptualized the study and conceived the methodology. 
M.F.C.\ and K.S.\ performed the fabrication.
M.F.C., K.S., C.K., and D.G.-M.\ performed the characterization and optical measurements.
M.F.C.\ analyzed the data. 
L.X., J.L., and D.T.B.\ designed the temperature-sensing microchips. 
M.F.C.\ made the figures. 
M.F.C., D.T.B., I.B., and M.Z.M.\ wrote the manuscript. 

\section{Competing interests}\label{S:compInt}

M.F.C.\ and M.Z.M.\ are co-inventors of a patent application (\#63/908,097) about SPOTs. The other authors declare no competing interests.

\footnotesize
\bibliography{pedotpssReflectorSources}
\normalsize


\pagebreak
\begin{strip} 
	\begin{center}
	\Large \textit{Supplementary information:}\\Scalable reflective communication for microscopic electronics
	\end{center}
	\vspace{2cm}
\end{strip}

\setcounter{section}{0}
\setcounter{equation}{0}
\setcounter{figure}{0}
\setcounter{table}{0}
\makeatletter
\renewcommand{\thesection}{S\arabic{section}}
\renewcommand{\theequation}{S\arabic{equation}}
\renewcommand{\thefigure}{S\arabic{figure}}
\renewcommand{\thetable}{S\arabic{table}}

\bigskip

\section{Fabrication}\label{S:fabrication}

\subsection{Fabrication protocol for SPOTs}\label{S:fabSPOTs}

Our fabrication protocol for SPOTs involves depositing the PEDOT:PSS, covering the polymer with protective films, patterning the polymer, and removing the protective films.  

We deposit the PEDOT:PSS as follows, starting with a \ch{Si} chip featuring exposed \ch{Pt} electrodes and wiring encapsulated with \ch{SiO2}. 
We begin by assembling a mixture of 5\% v/v ethylene glycol (EG, \#324558, MilliporeSigma) in an aqueous dispersion of PEDOT:PSS (PH 1000, Ossila Ltd.) and sonicating it for 5~\si{\minute} to ensure adequate mixing. 
Meanwhile, we expose the chip surface to \ch{O2} plasma in a Jupiter II RIE plasma etcher (150~\si{\watt}, 90~\si{\second}), which has a cleaning effect, increases wettability, and improves adhesion by the PEDOT:PSS.
Then, we spincoat the PEDOT:PSS mixture (static deposition through a 0.2~\si{\micro\meter} filter (\#431219, Corning Inc.), 500~\si{\revolution\per\second} for 10~\si{\second} then 3000~\si{\revolution\per\second} for 60~\si{\second}) and subsequently bake it on a hot plate (125~\dC, 15~\si{\minute}) to dry the film. 
We typically deposit polymer films with order 30-60~\si{\nano\meter} thickness because these thin values result in faster performance (see Section~\ref{S:FreqScaling}). 
To achieve thicker polymer films, we repeat this spin-bake sequence as needed. 
We use a razor blade to remove a small amount of the PEDOT:PSS film from a non-electrode area and use a stylus profilometer (KLA Tencor P7 2D profiler) to measure its thickness. 

We then sputter protective 50~\si{\nano\meter} \ch{Al} and 50~\si{\nano\meter} \ch{Cu} coatings sequentially over the PEDOT:PSS film in a Denton Explorer 14 magnetron sputterer (\ch{Al}: 200~\si{\watt} DC, 3~\si{\milli\torru}, 3~\si{\angstrom\per\second}; \ch{Cu}: 400~\si{\watt} DC, 3~\si{\milli\torru}, 6~\si{\angstrom\per\second}; ultimate pressure 5\sci{-6}~\si{\torru}). 
We verify the thickness of these and other sputtered metal coatings using stylus profilometry. 
These metal films prevent harm to the polymer film during the photolithography development steps~\cite{Williams1996-256, Williams2003-761, Ouyang2014-1822, Ouyang2015-603148}. 
We next spincoat the chip with photoresist (MicroPOSIT S1813, Kayaku Advanced Materials, static deposition, 500~\si{\revolution\per\second} for 5~\si{\second} then 3000~\si{\revolution\per\second} for 45~\si{\second}) and perform a soft bake on a hot plate (115~\dC, 60~\si{\second}) to drive out the casing solvent (we do not use an adhesion promoter for this lithography step). 

We pattern the polymer using photolithography and ion milling.  
We expose the chip through a photomask (SUSS MicroTec MA6 Gen3 Mask Aligner, 155~\si{\milli\joule\per\centi\meter\squared}, hard contact) and develop it (AZ 300 MIF, MicroChemicals GmbH, 60~\si{\second} with agitation, followed by deionized water (DI \ch{H2O}) rinse and \ch{N2} blow-dry), leaving photoresist only on top of the working electrode areas. Note that the \ch{Cu} protects the \ch{Al} from the AZ 300 MIF developer. 
Next, we ion mill through the \ch{Al}, \ch{Cu}, and PEDOT:PSS layers to remove the polymer everywhere but on the working electrodes (IntlVAC Nanoquest II IBE, 400~\si{\volt} potential, 400~\si{\watt} RF power, 510~\si{\milli\ampere} beam current, \ch{Ar} bombardment, 45\textdegree\xspace tilt, 10~\si{\revolution\per\minute} platen spin rate, platen chilled to 20~\dC, $\sim32$~\si{\nano\meter\per\minute} etch rate), being careful not to etch too far into the \ch{SiO2} encapsulation or other non-PEDOT:PSS-coated exposed \ch{Pt} electrode areas. 
To verify complete removal, we use a Keithley 2450 source measure unit (SMU) to test the resistivity of the chip on an area coated by \ch{SiO2}; the resistance is of the \si{\giga\ohm} scale if all of the PEDOT:PSS has been successfully milled away. 

Lastly, we remove the remaining protective films. 
To remove the photoresist, we use \ch{O2} plasma ashing (Jupiter II RIE plasma etcher, 150~\si{\watt}, 6-10~\si{\minute}). 
We remove the \ch{Al} and \ch{Cu} from the working electrode areas by wet etching (10\% v/v 12N \ch{HCl} (Transene Co., Inc.) in DI \ch{H2O}, 30~\si{\minute})~\cite{Williams1996-256, Williams2003-761}. 
The acid may also have the added impact of increasing the conductivity of the PEDOT:PSS film, but we have not characterized this here~\cite{Biessmann2019-1800654}. 
The chips are thus left in their original state with PEDOT:PSS patterned on the desired electrode surfaces. 

We are aware that other PEDOT:PSS deposition~\cite{Buga2022-045004, Niederhoffer2023-011002, Li2023-604, Mousavi2023-2201282, Shin2024-116418, Slaughter2024-14944, Hik2023-143136} and patterning~\cite{Ouyang2015-603148, Derek2018-105116, Wang2020-105954, Doshi2024-2271, Park2024-46664, Li2025-4933, Lee2025-e12824} strategies have been demonstrated, most of which would be suitable alternatives. 
Our use of the particular methods and tools listed above reflects our familiarity with, and the availability of, them in our facilities. 

\subsection{Fabrication protocol for metalized chips}\label{S:fabMetalChips}

Our fabrication procedure to create the metalized chips for characterizing SPOTs involves depositing a base \ch{SiO2} insulation layer on the \ch{Si} substrate, patterning metal electrodes and wiring, and encapsulating the wires with patterned \ch{SiO2}. 

We begin with a 10~\si{\centi\meter} prime single-side polished \textit{p}(\ch{B})-doped 525~\si{\micro\meter}-thick \ch{Si} wafer (Wafer World, Inc., resistivity 1-20~\si{\ohm\centi\meter}). 
First, we deposit about 600~\si{\nano\meter} \ch{SiO2} on the surface using plasma enhanced chemical vapor deposition (PECVD; Oxford Instruments PlasmaLab System 100, 350~\dC, 300~\si{\nano\meter\per\minute}) to provide insulation from the \ch{Si} substrate. We verify the thickness of this and other \ch{SiO2} films using reflectometry (KLA Instruments Filmetrics F40). We then cleave the wafer into 2~\si{\centi\meter} square chips and clean them with \ch{O2} plasma (Jupiter II RIE plasma etcher, 150~\si{\watt}, 90~\si{\second}). 

Next, we fabricate the metal electrodes and wires.  To do this, on each chip, we deposit an adhesion promoter in a vapor-phase oven (Genesis HMDS Vapor Prime Oven, hexamethyldisilazane (HMDS), 150~\dC), spincoat negative-tone photoresist (APOL-LO 3202, KemLab, Inc., static deposition, 500~\si{\revolution\per\second} for 5~\si{\second} then 4000~\si{\revolution\per\second} for 45~\si{\second}), and perform a soft bake on a hot plate (115~\dC, 2~\si{\minute}). 
We then expose the chips through a photomask (SUSS MicroTec MA6 Gen3 Mask Aligner, 140~\si{\milli\joule\per\centi\meter\squared}, hard contact), perform a post-exposure bake on a hot plate (115~\dC, 60~\si{\second}) to promote crosslinking of the photoresist in the exposed areas, and develop them (AZ 917 MIF, MicroChemicals GmbH, 75~\si{\second} with agitation, followed by DI \ch{H2O} rinse and \ch{N2} blow-dry). 
After an \ch{O2} plasma descumming step (Jupiter II RIE plasma etcher, 150~\si{\watt}, 90~\si{\second}), we sputter 10~\si{\nano\meter} \ch{Ti} (adhesion layer) and 100~\si{\nano\meter} \ch{Pt} in a Denton Explorer 14 magnetron sputterer (\ch{Ti}: 350~\si{\watt} DC, 3~\si{\milli\torru}, 2~\si{\angstrom\per\second}; \ch{Pt}: 140~\si{\watt} DC, 3~\si{\milli\torru}, 2.6~\si{\angstrom\per\second}; ultimate pressure 5\sci{-6}~\si{\torru}) and perform liftoff by sonication in Remover PG (Kayaku Advanced Materials, Inc., 80~\dC, 60~\si{\minute}). 
When removing the chips from the Remover PG bath, we briefly plunge them into a second Remover PG bath to remove any scum film that may adhere to the chips from the first bath, and then we rinse them with DI \ch{H2O} and dry them with \ch{N2}. 
We follow the liftoff process with another \ch{O2} plasma descumming step (Jupiter II RIE plasma etcher, 150~\si{\watt}, 90~\si{\second}).  

We insulate the wires using \ch{SiO2}, in a process that involves first depositing 200~\si{\nano\meter} using PECVD (Oxford Instruments PlasmaLab System 100, 350~\dC, 300~\si{\nano\meter\per\minute}), and second, depositing about 20~\si{\nano\meter} using atomic layer deposition (ALD; Cambridge Nanotech S200, 200~\dC, 0.7~\si{\angstrom\per\cycle}, Bis(diethylamino)silane (BDEAS) and \ch{O3} precursors) to backfill any pinholes in the coating. 
We pattern the insulation by depositing HMDS (Genesis HMDS Vapor Prime Oven, vapor phase, 300~\dC), spin-coating photoresist (MicroPOSIT S1813, Kayaku Advanced Materials, static deposition, 500~\si{\revolution\per\second} for 5~\si{\second} then 4000~\si{\revolution\per\second} for 45~\si{\second}), baking the chips on a hot plate (115~\dC, 60~\si{\second}), aligning and exposing a pattern through a photomask (SUSS MicroTec MA6 Gen3 Mask Aligner, 100~\si{\milli\joule\per\centi\meter\squared}, vacuum contact), developing the photoresist (AZ 300 MIF, MicroChemicals GmbH, 60~\si{\second} with agitation, followed by DI \ch{H2O} rinse and \ch{N2} blow-dry), descumming the chips (Jupiter II RIE plasma etcher, 150~\si{\watt}, 60~\si{\second}), and performing a post-development bake on a hot plate (115~\dC, 90~\si{\second}) to reflow the photoresist. 
This series of steps coats the entire top surface of the chips with photoresist, except for the electrode areas in the centers of the chips and probe pads at their perimeter.  
Next, we wet-etch the chips in 6:1 buffered oxide enchant (BOE, Transene Company, Inc., 2.5~\si{\minute}, 2~\si{\nano\meter\per\second} etch rate), and use reflectometry (KLA Instruments Filmetrics F40) and electrical resistance (Keithley 2450 SMU) to ensure complete removal of the \ch{SiO2} on the electrodes and probe pads.  

The finished chips each have eight pairs of encapsulated wires in a circular pattern, extending radially from open electrodes in their centers to open probe pads on their perimeters. 
Each electrode pair consists of a \ch{Pt} $300\times300$~\si{\micro\meter\squared} counter electrode (normally not used, but see Section~\ref{S:impactCounterElec}) adjacent to a second square \ch{Pt} working electrode. 
The eight working electrodes have side lengths 10, 20, 30, 40, 50, 100, 200, and 300~\si{\micro\meter}. 

\subsection{Crosslinking with GOPS}\label{S:fabCrossGOPS}

We crosslink the PEDOT:PSS on some SPOTs using (3-Glycidyloxypropyl)Trimethoxysilane (GOPS, Thermo Scientific, \#216541000)~\cite{Tseng2024-13384}. 
The procedure follows that of the non-cross-linked polymers, except the mixture we deposit consists of 1\% v/v dimethyl sulfoxide (DMSO, Fisher Chemical, D139-1) and 5\% v/v GOPS in an aqueous dispersion of PEDOT:PSS (PH 1000, Ossila Ltd.).  
Note that this mixture does not include ethylene glycol (EG). 
After spin-coating the chip with this mixture, we bake it on a hotplate at 130~\dC for 10~\si{\hour}. 

\subsection{Crosslinking \via heat treatment}\label{S:fabCrossHeat}

We crosslink other SPOTs using thermal treatment~\cite{Doshi2025-2415827}. 
In this case, we spincoat the standard 5\% v/v EG/PEDOT:PSS mixture onto the chip, briefly bake it on a hotplate at an elevated temperature (180~\dC, 2~\si{\minute}), and then wash it. 
The washing protocol consists of soaking the chip in DI \ch{H2O} for 11~\si{\minute} and then rinsing in DI \ch{H2O} for a few seconds; this washing sequence is performed three times. 
These washing steps remove excess PSS from the polymer. 

\subsection{Fabrication protocol for temperature-sensitive microchips}\label{S:fabClockbot}

We obtain custom-fabricated 55~\si{\nano\meter} 
CMOS chips (triple-well deeply depleted channel (DDC) process, United Semiconductor Japan Co., Ltd.) for our temperature-sensing measurements. As received, these chips are 50~\si{\micro\meter} thick,  have lateral dimensions of $4\times12$~\si{\milli\meter\squared}, and each contain 50 temperature-sensitive $150\times200$~\si{\micro\meter\squared} microcircuits (the microcircuits we use in this experiment occupy only a fraction of the total area, the rest of which is used for other circuits)~\cite{Xu2022-1}. Our back-end process steps for each chip include mounting the chip to a handle wafer, encapsulating the circuitry, depositing and etching the PEDOT:PSS to form SPOTs, and dicing to obtain individual microscopic sensors.  

We begin by cleaning the as-received chips with \ch{O2} plasma (Jupiter II RIE plasma etcher, 150~\si{\watt}, 3~\si{\minute}). 
We then encapsulate the chips by depositing \ch{SiO2} using PECVD (200~\si{\nano\meter}; Oxford Instruments PlasmaLab System 100, 200~\dC, 300~\si{\nano\meter\per\minute}) and ALD (20~\si{\nano\meter}; Cambridge Nanotech S200, 200~\dC, 0.7~\si{\angstrom\per\cycle}, Bis(diethylamino)silane (BDEAS) and \ch{O3} precursors). 
This protects the circuitry but leaves the electrodes covered as well; note the lower temperature of the PECVD step compared to the macroscale chips for SPOT testing introduced in Section~\ref{S:fabMetalChips}. 
We next adhere the chips to 50~\si{\milli\meter} sapphire carrier wafers. 
To do this, we clean the carrier wafers using \ch{O2} plasma (Jupiter II RIE plasma etcher, 150~\si{\watt}, 3~\si{\minute}), spincoat negative-tone photoresist (NR7-3000P, Futurrex, Inc., static deposition, 500~\si{\revolution\per\second} for 5~\si{\second} then 3000~\si{\revolution\per\second} for 20~\si{\second}), place the microchips in the center of each wafer (one chip per wafer), tap the chips with tweezers to promote adhesion, bake the wafers on a hot plate (90-125~\dC, 2~\si{\minute}), develop the wafers until all the photoresist except for that which is underneath each chip is removed (RD6, Futurrex, Inc.), rinse the wafers with DI \ch{H2O} and dry them with \ch{N2}, and finally bake the wafers on a hot plate (90-125~\dC, 3~\si{\minute}).

To make electrical contact, we must remove the \ch{SiO2} encapsulation above the electrodes and subsequently sputter metal into the resulting holes, creating sealed, electrically-conductive junctions. 
We define the etch areas over the electrodes using photolithography, which involves depositing an adhesion promoter in a vapor-phase oven (Genesis HMDS Vapor Prime Oven, hexamethyldisilazane (HMDS), 150~\dC), spin-coating photoresist (MicroPOSIT S1827, Kayaku Advanced Materials, static deposition, 500~\si{\revolution\per\second} for 5~\si{\second} then 3000~\si{\revolution\per\second} for 45~\si{\second}), soft-baking the chips on a hot plate (115~\dC, 90~\si{\second}), aligning and exposing a pattern (SUSS MicroTec MA6 Gen3 Mask Aligner, 180~\si{\milli\joule\per\centi\meter\squared}, soft contact), developing the chips (AZ 300 MIF, MicroChemicals GmbH, 60~\si{\second} with agitation, followed by soaking for 60~\si{\second} each in two sequential DI \ch{H2O} baths with light agitation, rinsing with DI \ch{H2O} (20~\si{\second}), and drying with \ch{N2}), descumming the chips (Jupiter II RIE plasma etcher, 150~\si{\watt}, 3~\si{\minute}), and performing a reflow bake on a hot plate (115~\dC, 3~\si{\minute}). 
Thereafter, we use reactive ion etching (PlasmaPro 80 RIE, Oxford Instruments plc, \ch{CF4}, 6~\si{\minute}, $\sim45$~\si{\nano\meter\per\minute} etch rate) to cut completely through the \ch{SiO2} in the areas not covered with photoresist, and, with the same photoresist coating in place, sputter 20~\si{\nano\meter} \ch{Ti} and 150-200~\si{\nano\meter} \ch{Pt} to fill the holes (Denton Explorer 14 magnetron sputterer, \ch{Ti}: 350~\si{\watt} DC, 3~\si{\milli\torru}, 2~\si{\angstrom\per\second}; \ch{Pt}: 140~\si{\watt} DC, 3~\si{\milli\torru}, 2.6~\si{\angstrom\per\second}; ultimate pressure 5\sci{-6}~\si{\torru}). 
Following this, we sonicate the chips (still adhered to the sapphire handle wafers) in Remover PG (Kayaku Advanced Materials, Inc., 60~\dC, 30~\si{\minute}) to lift off the excess metal. 
This sonication process also removes the chips from the handle wafer by dissolving the NR7-3000P photoresist. 
When removing the chips from the Remover PG bath, we briefly plunge them into a second Remover PG bath to remove any scum film that may adhere to the chips from the first bath, and then we rinse them with DI \ch{H2O} and dry them with \ch{N2}. Finally, we adhere the chips back onto the carrier sapphire wafers (see above) and descum them using \ch{O2} plasma (Jupiter II RIE plasma etcher, 150~\si{\watt}, 90~\si{\second}).

Our subsequent PEDOT:PSS deposition and SPOT patterning procedure mirrors that of the test chips introduced in Section~\ref{S:fabSPOTs}, beginning with the \ch{O2} plasma cleaning step and concluding with the dilute \ch{HCl} etch of the \ch{Al} and \ch{Cu} metal coatings. 
After these steps, the microcircuits on the chips, which are still attached to the handle wafer, can be tested under a microscope in $1\times$PBS (phosphate buffered saline). 

Lastly, we dice the chips to release individual microsensor motes. 
With the chips still attached to the handle wafers, we first sputter protective 50~\si{\nano\meter} \ch{Al} and 50~\si{\nano\meter} \ch{Cu} coatings sequentially over the PEDOT:PSS film in a Denton Explorer 14 magnetron sputterer (\ch{Al}: 200~\si{\watt} DC, 3~\si{\milli\torru}, 3~\si{\angstrom\per\second}; \ch{Cu}: 400~\si{\watt} DC, 3~\si{\milli\torru}, 6~\si{\angstrom\per\second}; ultimate pressure 5\sci{-6}~\si{\torru}). 
We then carefully use a razor blade to trace the perimeter of each chip, removing the metal on all four sides.  
This allows us to gently remove the chips from the handle wafer by soaking them (no sonication) in Remover PG (Kayaku Advanced Materials, Inc., 25~\dC, 4~\si{\hour}), which dissolves the underlying photoresist. Next, we affix the chips face up using acetone-soluble adhesive tape (BGF7080, AI Technology, Inc.) to 10~\si{\centi\meter} \ch{Si} handle wafers, spin on a protective layer of photoresist (AZ 3300, Integrated Micro Materials, static deposition, 500~\si{\revolution\per\second} for 5~\si{\second} then 3000~\si{\revolution\per\second} for 40~\si{\second}), and bake the handle wafers on a hot plate  (65~\dC, 1~\si{\minute}) to harden the photoresist. After this, we remove the acetone-soluble adhesive tape and chips together from the \ch{Si} wafers and place these stacks onto UV-curable dicing tape (\#1020R-11.5, Ultron Systems, Inc.), press the air bubbles out, bake the chip-tape-tape stacks (65~\dC, 1~\si{\minute}), and allow the stacks to cool. We then dice the chips to cut out the individual microcircuit motes (Advanced Dicing Technologies 7100, \ch{Ni} hub blade, 5~\si{\milli\meter\per\second} feed rate, 28,000~\si{\revolution\per\minute}). 
Following this step, we remove the microchip motes from the soluble adhesive tape by heating them on a hotplate at 130~\dC for about 10~\si{\minute}; when the tape becomes wet/gummy, we soak the chip-tape stacks in acetone (Fisher Chemical, A949-4) in a shallow PYREX (Corning, Inc.) Petri dish to fully remove the diced motes from the tape.  
Our next steps are a series of in-place solution transfers that dilute and replace one chemical with another while leaving the diced microcircuit motes submerged in the same dish. 
For each solution change, we first pipette the majority of the initial solution out of the dish, and subsequently dilute the remainder of the dish with the next liquid. 
We perform this removal-backfilling series in triplicate for each solution transfer. 
The order of solutions is acetone (Fisher Chemical, A949-4, to remove the chips from the dicing tape), DI \ch{H2O}, isopropyl alcohol (Millipore Sigma , 8.18766), DI \ch{H2O}, Remover PG (Kayaku Advanced Materials, Inc., to remove the photoresist), DI \ch{H2O}, and finally dilute (1N) \ch{HCl} (Fisher Chemical, SA48-1, to etch away the \ch{Al} and \ch{Cu} films). 
We allow the chips to soak in the \ch{HCl} overnight to fully etch away the metal coatings, and then use the in-place solution transfer technique to transfer from \ch{HCl} to DI \ch{H2O} and finally to 1$\times$PBS (\#SH30256.02, Cytiva Life Sciences) for testing. 

\subsection{Fabrication protocol for logo demonstration}\label{S:fabLogo}

Our fabrication protocol to create macroscopic images consisting of micron-sized SPOTs, such as the University of Pennsylvania coat of arms shown in Figure~\ref{F:setupFabOverview}(c) in the main article, is similar to that for individual devices introduced in Section~\ref{S:fabSPOTs}. After spincoating PEDOT:PSS onto a \ch{Pt}-coated \ch{Si} chip (as outlined above), we sputter a protective \ch{Cu} layer (in later fabrication iterations we experimented with using \ch{Cu} only (no \ch{Al}) as a protective coating) and coat the chip with S1813 photoresist. We then use a Durham Magneto Optics MicroWriter ML3 Pro laser writer (1~\si{\micro\meter} resolution, 405~\si{\nano\meter} wavelength, 50~\si{\milli\joule\per\centi\meter\squared} resist sensitivity, 0~\si{\micro\meter} focus correction) to directly expose a pixel pattern on the chip that we determine from an image (in this example, the coat of arms logo) using Floyd–Steinberg dithering~\cite{Floyd1976-75} in Layout Editor software. After developing the photoresist (AZ 300 MIF, MicroChemicals GmbH, 75~\si{\second} with agitation, followed by DI \ch{H2O} soak (30~\si{\second}), DI \ch{H2O} rinse, and \ch{N2} blow-dry), we descum the pattern using \ch{O2} plasma ashing (Jupiter II RIE plasma etcher, 150~\si{\watt}, 2~\si{\minute}) and ion mill the chip to remove the \ch{Cu} and PEDOT:PSS everywhere not coated in photoresist (IntlVAC Nanoquest I IBE, 400~\si{\volt} potential, 200~\si{\watt} RF power, 150~\si{\milli\ampere} beam current, \ch{Ar} bombardment, 45\textdegree\xspace tilt, 10~\si{\revolution\per\minute} platen spin rate, platen chilled to 20~\dC, $\sim56$~\si{\nano\meter\per\minute} etch rate), being careful not to etch too far into the \ch{Pt} electrode. We complete the chips using the process outlined earlier, including ashing the remaining photoresist away and etching the \ch{Cu} using dilute \ch{HCl}. 

\section{Frequency-dependent measurements}\label{S:infoOnFdependent}

\subsection{Experimental setup for frequency-dependent measurements}\label{S:ExptSetupFdep}


\begin{figure}
\centering
\includegraphics[width=\figWidthCol]{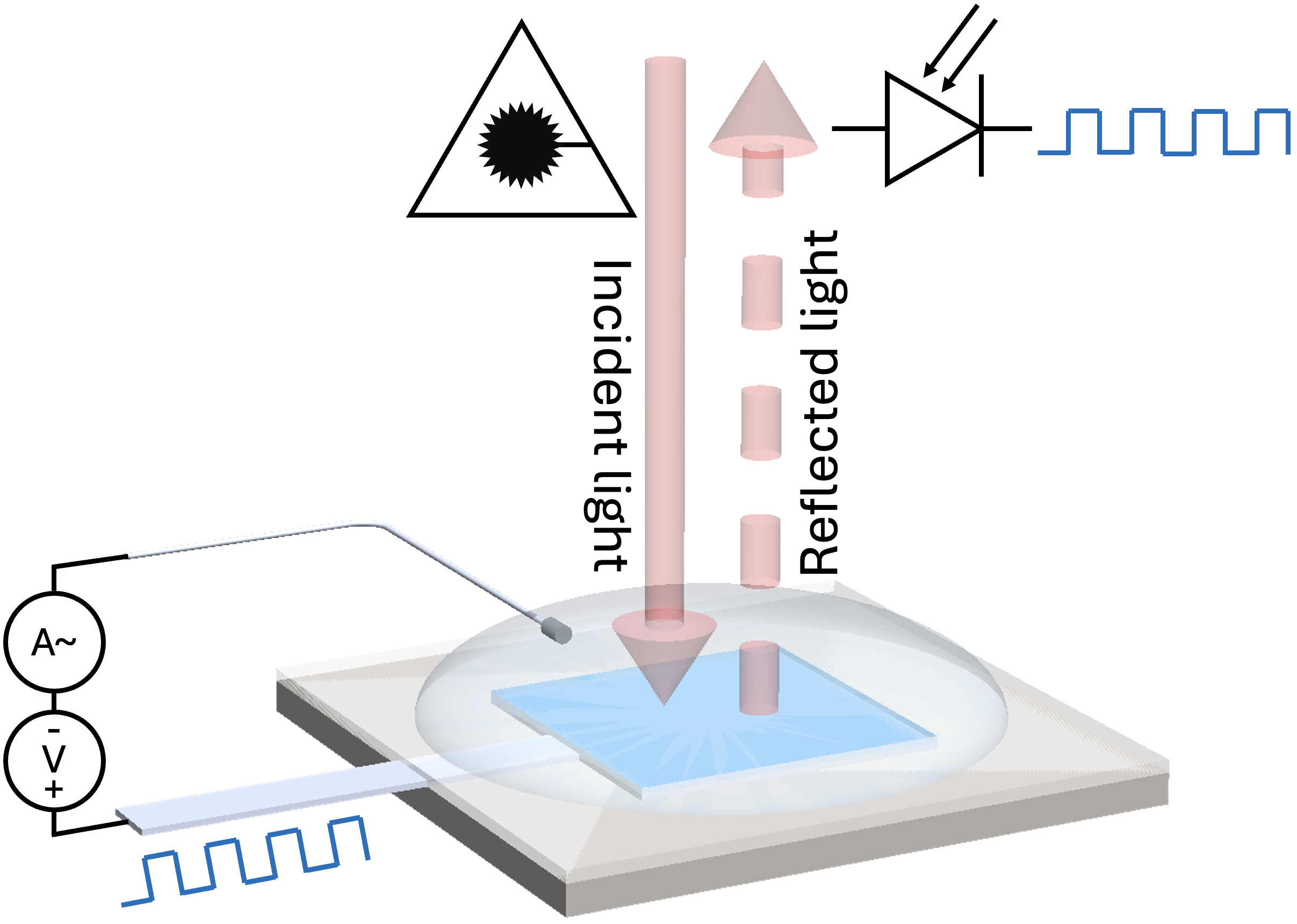}%
\caption{\textbf{Schematic diagram of essential experimental setup.} 
Laser light at $\lambda_l=635$~\si{\nano\meter} is directed to a SPOT, and the reflectance is measured using a photodiode. A potential is applied to the SPOT relative to a \ch{Ag}/\ch{AgCl} counter electrode in the electrolyte, and a transimpedance amplifier placed in series monitors the current.
}%
\label{F:meausurementSetupDiag}%
\end{figure}

We characterize SPOTs on chips in a custom opto-electrical experimental setup, depicted in Figure~\ref{F:meausurementSetupDiag}. 
Our optical arrangement involves a light-emitting diode (LED) light source (Thorlabs, Inc., SOLIS-3C with DC2200 driver), an immersion lens (Olympus Co., LUMPLFLN40XW \#34-557), and a digital camera (Thorlabs, Inc., Zelux CS165MU1). 
We use beam splitters to add a second coaxial optical train to the setup, consisting of a diode laser (Thorlabs, Inc., CPS635S, $\lambda_l=635$~\si{\nano\meter}) and photodetector (Thorlabs, Inc., PDA100A2); the arrangement allows us to direct the laser beam through the immersion lens and to independently position it within the viewing frame, so that we can align it onto a SPOT. 
The $\lambda_l=635$~\si{\nano\meter} wavelength of this laser is ideal because it is near the peak of the absorbance spectrum for the neutrally-charged polymer (PEDOT\textsuperscript{0})~\cite{Rebetez2021-2105821} and near the wavelength at which the ratio of the reflectance signal at an applied potential of $+200$~\si{\milli\volt} to that at $-800$~\si{\milli\volt} is maximized (see Figure~\ref{F:plotSpectra}). 
The laser spot through the immersion lens has dimensions of roughly $10\times20$~\si{\micro\meter\squared}, and its intensity when focused on the SPOT is roughly 1500~\si{\watt\per\meter\squared}, as measured using a Thorlabs S120C meter through a 10~\si{\micro\meter}-diameter aperture (Thorlabs Inc, PD10D) at $\lambda=635$~\si{\nano\meter}. 
We set the gain on the photodetector at a range that provides sufficient response times for our experiments ($800$~\si{\kilo\hertz} bandwidth). 
We calibrate the reflectance of our laser intensity measurements using a 525~\si{\micro\meter}-thick \ch{Si} chip with a $\sim68$~\si{\nano\meter} \ch{Au} layer (5~\si{\nano\meter} \ch{Cr} adhesion layer), deposited \via electron-beam evaporation (Kurt J.\ Lesker Co., PVD75; \ch{Cr}: 2~\si{\angstrom\per\second}; \ch{Au}: 5~\si{\angstrom\per\second}; ultimate pressure 3\sci{-6}~\si{\torru}) and verified by reflectometry (KLA Instruments Filmetrics F40). 
This provides the corrected reflectance values $\varrho_{corr}$ according to
\begin{equation}
    \varrho_{corr} = \varrho_{\ch{Au},theo} \frac{V_{p,chip}-V_{p,chip,dark}}{V_{p,\ch{Au}}-V_{p,\ch{Au},dark}}, 
    \label{E:correctReflect}
\end{equation}
where $V_{p,chip}$, $V_{p,chip,dark}$, $V_{p,\ch{Au}}$, and $V_{p,\ch{Au},dark}$ are the photodetector output voltage values (corresponding to measured light intensity values) when viewing the SPOT with the laser on, when viewing the SPOT with the laser off, when viewing the \ch{Au} chip with the laser on, and when viewing the \ch{Au} chip with the laser off, respectively.  Also,  $\varrho_{\ch{Au},theo}$ is the theoretical reflectivity of the \ch{Au} chip at the laser wavelength ($\lambda_l=635$~\si{\nano\meter}) calculated according to the transfer-matrix method~\cite{Macleod2017-book, Campbell2022-90} using optical constants from the literature~\cite{Segelstein1981-thesis, Olmon2012-235147, Sytchkova2021-111530, Schinke2015-067168}. 
We calculate the relative reflectance amplitude $A_R$ using
\begin{equation}
    A_R= \frac{V_{p,chip,max}-V_{p,chip,min}}{2V_{p,chip,average}},
    \label{E:relReflectChange}
\end{equation}
where $V_{p,chip,max}$, $V_{p,chip,min}$, and $V_{p,chip,average}$ are the highest, lowest, and average photodetector output voltage values at a given frequency, respectively. The error introduced by not strictly converting the photodetector voltages to reflectance values is only about 1\%. 
When conducting optical tests with the laser on, we turn off the LED to increase the signal to noise ratio (SNR) on the photodetector.  

Our electrical characterization setup typically consists of a function generator (BK Precision 4052), a transimpedance amplifier (Stanford Research Systems SR570), a 1~\si{\milli\meter} \ch{Ag}/\ch{AgCl} pellet (Warner Instruments E-205) as a counter electrode, and a digital oscilloscope (Picoscope 5442D) operated with DC (direct current) coupling. 
We split the voltage output of the function generator, directing half to the oscilloscope as a reference and half to the chip \via a probe needle (Signatone, Co., SE-T) oriented with a three-axis manipulator (Signatone, Co., SP-100). 
We use a poly(dimethylsiloxane) (PDMS) barrier to confine the solution (unless otherwise specified, 1$\times$PBS (\#SH30256.02, Cytiva Life Sciences), volume about 500-700~\si{\micro\liter} measured using a micropipette (VWR International, Co.)) to a small bubble directly over the chip.  
In this way, the probe needle, which contacts a pad on the perimeter of the chip, which in turn is connected to the working electrode (\ie, the SPOT) \via an encapsulated wire, does not need to be immersed in the solution.
In addition, this confines the fluid to an area very close to the SPOT, minimizing parasitic capacitive coupling between the electrical lead on the chip and the liquid through the \ch{SiO2} encapsulation layer. 
We measure the function generator voltage, the transimpedance amplifier signal, and the photodetector signal simultaneously on the oscilloscope. 
We typically apply potentials in the range $\langle -800, +200\rangle$~\si{\milli\volt} to the SPOTs because these bounds provide good optical contrast within the reversible operation window of PEDOT:PSS (see Figure~\ref{F:rampRef}). 
We conducted our measurements using a two-electrode setup, with the \ch{Ag}/\ch{AgCl} electrode serving both reference and counter electrode functions. This allowed our experiments to better mirror the configuration of SPOTs on microcircuits, which would likewise have just two electrodes. As suggested by Figure~\ref{F:electrochemChar}(b) in the main article, the voltage drop across the solution (with resistance $R_s \sim 10$~\si{\kilo\ohm}) becomes significant (roughly 10\% of the $A_V=0.5$~\si{\volt} potential amplitude) at current amplitudes of $A_I \sim 5$~\si{\micro\ampere} and larger. Additional information on two-electrode electrochemistry is available elsewhere~\cite{BardFaulkner2001-text, Harris2023-722}.

\subsection{Processing frequency-dependent data}\label{S:processFreqElectroOpticalData}

\begin{figure}
\centering
\includegraphics[width=\figWidthCol]{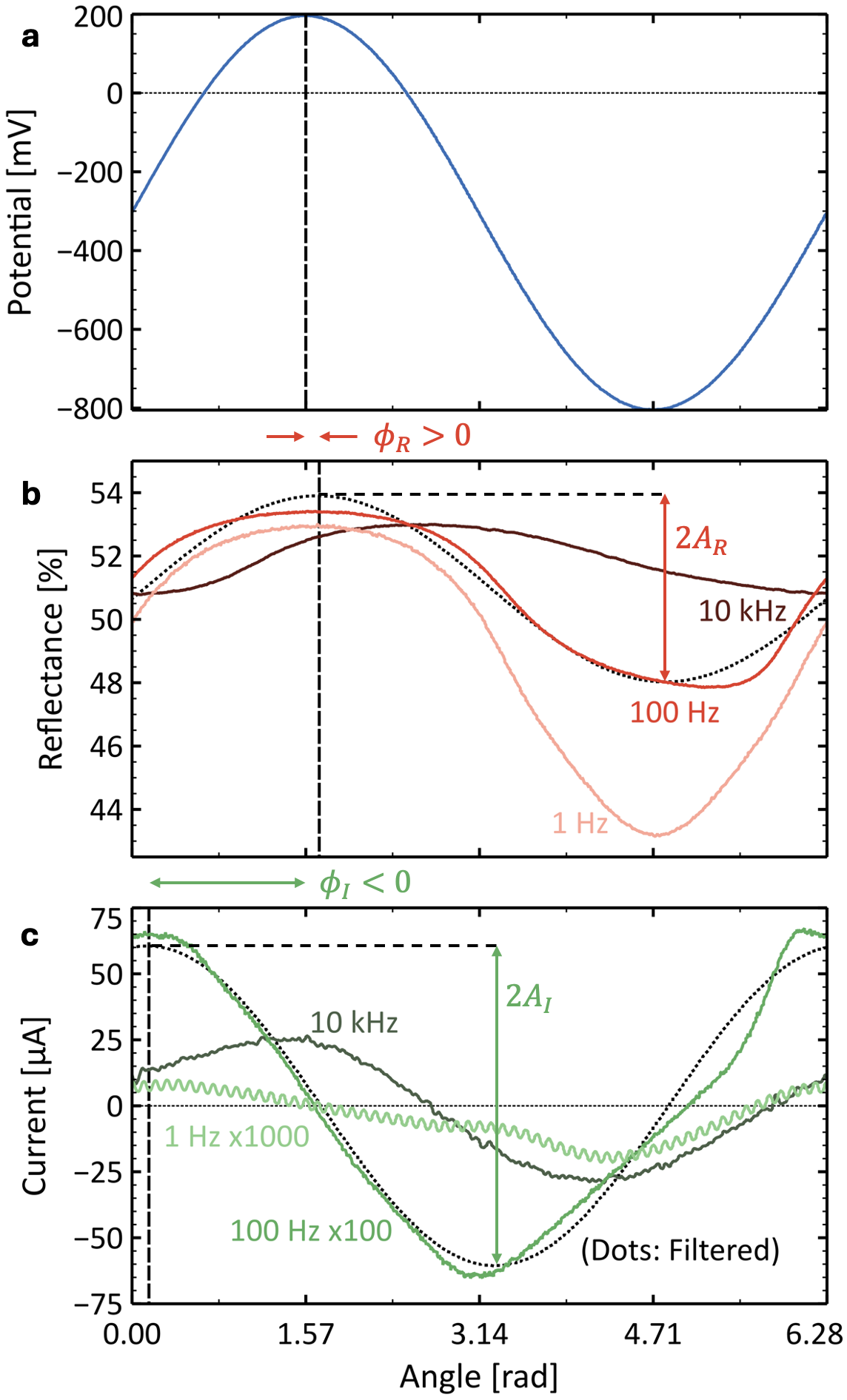}%
\caption{\textbf{Example frequency-dependent data.} Device consists of a PEDOT:PSS~($\app35$~\si{\nano\meter})-\ch{Pt} stack in $1\times$PBS, with lateral dimensions $30\times30$~\si{\micro\meter\squared}. 
The abscissa is depicted in cycle angle (radians) to make it time-invariant, so that three frequencies (1~\si{\hertz}, 100~\si{\hertz}, and 10~\si{\kilo\hertz}) can be shown simultaneously. Records shown with solid lines in this figure are time-synchronous averages of 100 cycles to reduce noise. 
\textbf{(a)} The input potential is a sinusoidal voltage between maximum and minimum values of  $+200$~\si{\milli\volt} and $-800$~\si{\milli\volt}, respectively. 
\textbf{(b)} Measured reflectance signal from incident laser light at $\lambda_l=635$~\si{\nano\meter}. The dotted record shows the filtered version of the 100~\si{\hertz} wave (baseline-shifted to the average value of the raw data), and on it the reflectance amplitude $A_R$ (Equation~\ref{E:sineModelFit}) and phase shift $\phi_R$ (Equation~\ref{E:phaseShiftR}) are shown. 
\textbf{(c)} Measured current signal; note the scale factors listed on the graph for the 1~\si{\hertz} ($I\times1000$) and 100~\si{\hertz} ($I\times100$) records. The noise in the 1~\si{\hertz} record has a characteristic frequency of 60~\si{\hertz} and is presumably from other electronics in the laboratory. The dotted record shows the filtered version of the 100~\si{\hertz} wave, and on it the current amplitude $A_I$ (Equation~\ref{E:sineModelFit}) and phase shift $\phi_I$ (Equation~\ref{E:phaseShiftI}) are shown. 
}%
\label{F:avgWaveform}%
\end{figure}

We process the electrochemical and optical data in MATLAB R2024b software. At each drive frequency $F$ we gather at least 100 cycles so that we can perform averaging, and we collect data by increasing from low to high frequency, optimizing the gain factor of the transimpedance amplifier at each step.  Our calculations involve filtering and averaging the data, fitting a sinusoidal model, and deriving relevant quantities. We filter the data using a narrow bandpass filter at the drive frequency, clipping ten cycles at the beginning and end to avoid ringing effects. We then use time-synchronous averaging to produce a single representative period of the filtered wave at each frequency. In addition, we also time-synchronous average the unfiltered data, which allows us to check our filtered results and to see chemistry-related signatures in the electrical and optical domains. Example voltage, reflectance, and current data are provided in Figure~\ref{F:avgWaveform}. 

After filtering and averaging the data, we fit a sine model of the form 
\begin{equation}
    \Upsilon_w = A_w \sin \left(2 \pi F t - \phi_{a,w} \right),
    \label{E:sineModelFit}
\end{equation}
where $\Upsilon_w$ is the measured quantity, $t$ is time, $A_w$ is the amplitude (defined as half the difference between the maximum and minimum values), $\phi_{a,w}$ is the phase angle of the wave, and subscript $w$ denotes either the drive voltage ($V$), the reflectance ($R$), or the current ($I$). (Note that for clarity we will use the terminology $\Upsilon_I \rightarrow I$ for current, $\Upsilon_R \rightarrow \varrho$ for reflectance, and $\Upsilon_V \rightarrow V$ for voltage.) Equation~\ref{E:sineModelFit} assumes that we have subtracted any DC offset such that the signals are shifted to a zero-baseline (this is accomplished through the filtering). 

From here, we derive several important quantities. The phase shifts of the reflectance and current relative to the drive voltage are 
\begin{align}
    \phi_R &= \phi_{a,R} - \phi_{a,V} \label{E:phaseShiftR} \\
    \phi_I &= \phi_{a,I} - \phi_{a,V}. \label{E:phaseShiftI}
\end{align}
Note that in this convention, $\phi>0$ corresponds to a phase lag in time. The impedance quantities are 
\begin{align}
    Z_{mag} &= \frac{A_V}{A_I} = |Z| \label{E:Zmag} \\
    Z &= Z_{mag} \exp \left( \ji \phi_I \right) \label{E:Z} \\
    Z_{re} &= \Re\left(Z\right) \label{E:Zre} \\
    Z_{im} &= \Im\left(Z\right) \label{E:Zim} .
\end{align}
where the vertical bars denote the magnitude of a complex number, $\ji$ is the imaginary number, and $\Re$ and $\Im$ denote the real and imaginary components. The charge transferred as a function of time is 
\begin{equation}
    Q = \int_0^t I dt
    \label{E:chargeVec}
\end{equation}
and the charge transferred per half cycle is 
\begin{equation}
    Q_h = \max(Q) - \min(Q) .
    \label{E:chargeHalfA}
\end{equation}
Note that, for a sine wave, this is equivalent to 
\begin{equation}
    Q_{h,sine} = \frac{A_I}{\pi F} . 
    \label{E:chargeHalfB}
\end{equation}

To find the signal-to-noise ratio (SNR) of the reflectance data, we need an estimate of the noise power and the signal power.  We estimate the noise power $\Phi_n$ by filtering out the first eight harmonics (\ie, $f$, $2f$, $3f$, \ldots, $8f$) of the raw photodetector voltage signal using two-level Butterworth filters, clipping the first ten cycles to avoid transient effects, and finding the root-mean-square (RMS) value of the result shifted to a zero-average baseline. 
\begin{equation}
    \Phi_n = \sqrt{\frac{1}{N_c}\sum_{i=1}^{N_c} \left(V_{p,i,butter,clip}-\frac{1}{N_c}\sum_{i=1}^{N_c} V_{p,i,butter,clip}\right)^2}
    \label{E:noiseRMSpower}
\end{equation}
Here, $N_c$ is the number of points in the Butterworth-filtered clipped wave and $V_{p,i,butter,clip}$ is the photodetector voltage value of the $i$\textsuperscript{th} point in the Butterworth-filtered clipped wave.  We estimate the power of the signal $\Phi_s$ by finding the root-mean-square (RMS) value of the bandpass-filtered signal, also shifted to a zero-average baseline. 
\begin{equation}
    \Phi_s = \sqrt{\frac{1}{N}\sum_{i=1}^N \left(V_{p,i,BP}-\frac{1}{N}\sum_{i=1}^N V_{p,i,BP}\right)^2}
    \label{E:sigRMSpower}
\end{equation}
Here, $N$ is the number of points in the bandpass-filtered wave and $V_{p,i,BP}$ is the photodetector voltage value of the $i$\textsuperscript{th} point in the bandpass-filtered wave. The SNR, denoted $R_{SN}$, is the ratio of these quantities:
\begin{equation}
    R_{SN} = \frac{\Phi_s}{\Phi_n}.
    \label{E:snr}
\end{equation}

We estimate the required RMS power, denoted $ P_{rms}$, by
\begin{equation}
    P_{rms} = \left(\frac{A_V}{\sqrt{2}} \right) \! \left(\frac{A_I}{\sqrt{2}} \right) \cos\left( \phi_I \right)
    \label{E:rmsPwr}
\end{equation}
According to the Shannon-Hartley theorem~\cite{Rioul2014-4892}, the maximum bit rate, denoted $F_{b,m}$, can be estimated by
\begin{equation}
    F_{b,m} = F \log_2 \! \left(1+R_{SN}\right).
    \label{E:Fbm}
\end{equation}
Finally, we estimate the energy cost per bit transmitted according to 
\begin{equation}
    E_b = \frac{P_{rms}}{F_{b,m}}.
    \label{E:Eb}
\end{equation}

\subsection{Circuit model for frequency-dependent data}\label{S:modelFreqElectroOpticalData}

\begin{figure}
\centering
\begin{tikzpicture}[]
\draw (0,0) to[R=$R_1$] ++(2,0) coordinate(n1) -- ++(0,0.5) coordinate(in);
\draw (in) -- ++(1,0) coordinate(cpe1) ++(0.2,0) coordinate(cpe2) 
-- ++(0.8,0) coordinate(out)
(cpe1) -- ++(-0.2,0.4) (cpe1) -- ++(-0.2,-0.4)
(cpe2) -- ++(-0.2,0.4) (cpe2) -- ++(-0.2,-0.4)
(cpe1)  ++(0,0.4) node[above]{CPE ($Y_0$, $n$)};
\draw (n1) -- ++(0,-0.5) to[R, l_=$R_2$] ++(2,0) -- ++(0,0.5) coordinate(n2) -- (out) (n2) --++(1,0);
\end{tikzpicture}
\caption{\textbf{Simple circuit model.} Randles circuit with resistor $R_1$ in series with the parallel combination of  resistor $R_2$ and constant phase element (CPE) with parameters $Y_0$ and $n$.}%
\label{F:simpleCircuit}%
\end{figure}
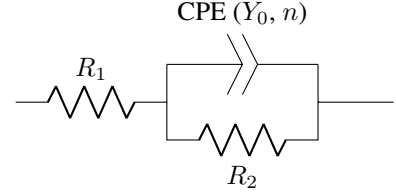

To fit the frequency-dependent data, we use a model consisting of a resistor $R_1$ in series with the parallel combination of a resistor $R_2$ and a constant phase element (CPE) with parameters $Y_0$ and $n$, whose impedance is obtained by
\begin{equation}
Z_{cpe}=\frac{1}{\left( 2 \pi F \ji \right)^n Y_0}
\label{E:Zcpe}
\end{equation}
(see Figure~\ref{F:simpleCircuit} and the inset of Figure~\ref{F:electrochemChar}(b) in the main text). The equation for the overall circuit impedance therefore has the form
\begin{equation}
    Z_{sim} = R_1 + \frac{1}{\frac{1}{R_2} + \left(2 \pi F \ji \right)^n Y_0} .
    \label{E:modelImpedance}
\end{equation}
We fit this model to our complex impedance data by brute force. The resulting $Z_{sim}$ data are comparable to the $Z$ data found in Equation~\ref{E:Z}, and the real and complex components can be obtained by using  $Z_{sim}$ for $Z$ in Equations~\ref{E:Zre} and~\ref{E:Zim}. The impedance magnitude is available through Equation~\ref{E:Zmag}, and thereafter, given the amplitude of the applied voltage ($A_V=0.5$~\si{\volt} for the frequency-dependent experiments, \ie, 1~\si{\volt} peak-to-peak), we calculate the current amplitude $A_I$. We obtain the simulated phase angle of the current using
\begin{equation}
    \phi_I = \arctan \! \left(\frac{Z_{im}}{Z_{re}}\right). 
    \label{E:phaseCurrentModel}
\end{equation}
From there, we obtain the other simulated quantities using Equations~\ref{E:chargeHalfB}-\ref{E:Eb}. 

\subsection{Scaling of important quantities}\label{S:FreqScaling}

We define the critical frequency $F_c$ as that where $Z_{re} = - Z_{im}$ (in the frequency range where the roll-off in the relative reflectance amplitude occurs), \ie, where $\phi_I=\frac{-\pi}{4}$. Applying this to Equation~\ref{E:phaseCurrentModel} using the model of Equation~\ref{E:modelImpedance} and simplifying (in the approximation that $Z_{cpe} \ll R_2$ at high $F$), we find
\begin{equation}
    F_c \simeq \frac{1}{2\pi}\left(\frac{\sin\left(\frac{n\pi}{2}\right)-\cos\left(\frac{n\pi}{2}\right)}{R_1 Y_0}\right)^{\!\frac{1}{n}}.
    \label{E:criticalFrequency}
\end{equation}
We use Equation~\ref{E:criticalFrequency} to calculate $F_c$ from our experimental data. From here, approximating the constant phase element as a simple capacitor ($n\rightarrow1$, $Y_0\rightarrow C$), we have 
\begin{equation}
    F_c \approx \frac{1}{2\pi R_1 C}, 
    \label{E:criticalFrequencyRC}
\end{equation}
which shows the inverse scaling of the critical frequency $F_c$ with SPOT capacitance $C$ (Figure~\ref{F:linearRelationships}(a)). In the Randles circuit model, $R_1$ captures the solution resistance $R_s$, which for a square microelectrode with side length $L$ is given by
\begin{equation}
    R_s \sim \frac{\sqrt{\pi}}{4 \sigma_s L} ,
    \label{E:RsScaling}
\end{equation}
where $\sigma_s$ is the solution resistance~\cite{Newman1966-501}. Note the inverse scaling between $R_s$ and $L$ (Figure~\ref{F:linearRelationships}(c)). Also, the capacitance $C$ for a square SPOT with thickness $h$ and area $L^2$ is related to the volumetric capacitance of PEDOT:PSS $(C^\ast)$ by
\begin{equation}
    C = h L^2 C^\ast 
    \label{E:volCap}
\end{equation}
(see Figure~\ref{F:linearRelationships}(d)). Combining Equations~\ref{E:criticalFrequencyRC}-\ref{E:volCap}, we have 
\begin{equation}
    F_c \sim \frac{2 \sigma_s}{\pi^{3/2} L h C^\ast} , 
    \label{E:scaleFc}
\end{equation}
which shows the inverse scaling of the critical frequency $F_c$ with the product of the SPOT thickness $h$ and side length $L$ (Figure~\ref{F:linearRelationships}(b)). Noting that the $\frac{1}{e}$ response time is related to the critical frequency by $F_c \sim \frac{1}{2 \pi \tau_e}$, we also find
\begin{equation}
    \tau_e \sim \frac{\sqrt{\pi} L h C^\ast}{4 \sigma_s} .
    \label{E:scaleTauE}
\end{equation}
Thus, to achieve a two-order-of-magnitude improvement in response times relative to previous pixel prototypes~\cite{Fabiano2017-e1700345, Do2021-106106, Yang2024-2314983}, we decrease the lateral size $L$ by an order of magnitude (from $\sim 100$~\si{\micro\meter} to $\sim 10$~\si{\micro\meter}) and reduce the polymer thickness by an order of magnitude (from $\sim 100$~\si{\nano\meter} to $\sim 10$~\si{\nano\meter}). (Note that finer patterned PEDOT:PSS features have been demonstrated, but not expressly as optical pixels~\cite{Doshi2024-2271, Lee2025-e12824}.) Finally, by equating the magnitude of the constant phase element's impedance $|Z_{cpe}|=\frac{1}{\left( 2 \pi F \right)^n Y_0}$ (Equation~\ref{E:Zcpe}) to the magnitude of the impedance of an ordinary capacitor $|Z_C|=\frac{1}{2 \pi F C}$, we can estimate the device capacitance $C$ at the critical frequency $F_c$ according to
\begin{equation}
    C \approx Y_0 \! \left(2 \pi F_c\right)^{n-1}.
    \label{E:estimateDeviceCap}
\end{equation}
Note that other methods for estimating capacitance from constant phase elements in Randles circuit models exist~\cite{Brug1984-275, Hirschorn2010-6218}, especially those that rely on the value of $R_2$. However, we find estimating $C$ at $F_c$ to be more consistent for our datasets, because we did not acquire data at low frequencies ($F\ll1$~\si{\hertz}) to resolve $R_2$ precisely. 

\subsection{Additional analysis related to Figure~\ref{F:electrochemChar} in the main article}\label{S:addlFig2}

\begin{figure*}
\centering
\includegraphics[width=160mm]{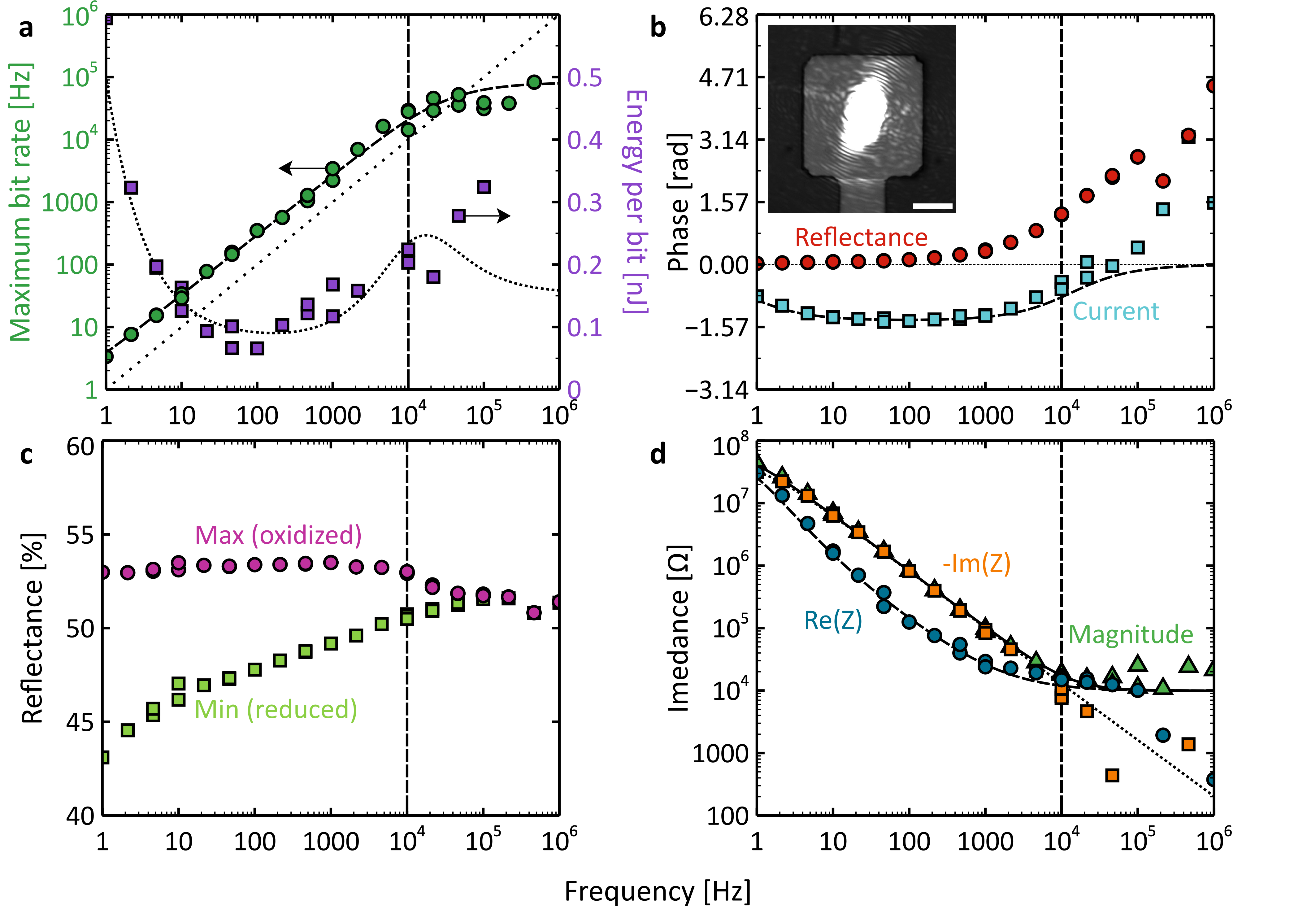}%
\caption{\textbf{Additional frequency-dependent characterization a SPOT.} 
See also Figure~\ref{F:electrochemChar} in the main article. 
Device consists of a PEDOT:PSS~($\app35$~\si{\nano\meter})-\ch{Pt} stack in $1\times$PBS, with lateral dimensions $30\times30$~\si{\micro\meter\squared}. 
Incident laser light at $\lambda_l=635$~\si{\nano\meter} is focused to a spot of roughly $10\times20$~\si{\micro\meter\squared} on the electrode (inset of panel (b)). 
The input potential is a sinusoidal voltage between maximum and minimum values of  $+200$~\si{\milli\volt} and $-800$~\si{\milli\volt}, respectively. 
Circuit model, shown in Figure~\ref{F:simpleCircuit} (see also Equation~\ref{E:modelImpedance}), consists of $R_1=9.9$~\si{\kilo\ohm}, $R_2=80$~\si{\mega\ohm}, and a constant phase element (parameters $Y_0 = 3.7$~\si{\nano\siemens\second\tothe{0.9}} and $n=0.9$).
The vertical dashed line shows critical frequency $F_c\sim10$~\si{\kilo\hertz}.
\textbf{(a)} Maximum bit rate possible (Equation~\ref{E:Fbm}) and energy per bit transferred (Equation~\ref{E:Eb}). The sloped dotted line shows where $F=F_{b,m}$. 
\textbf{(b)} Phase angles of the reflectance $\phi_R$ (Equation~\ref{E:phaseShiftR}) and the current $\phi_I$ (Equation~\ref{E:phaseShiftI}). 
\textbf{(c)} Maximum and minimum reflectance values in the photodiode signal. 
\textbf{(d)} Real and imaginary components of measured impedance (Bode plot; Equations~\ref{E:Zre} and~\ref{E:Zim}, respectively). 
Scale bar (inset of (b)): 10~\si{\micro\meter}. 
 }%
\label{F:electrochemCharExtended}%
\end{figure*}

We provide additional analysis of the frequency-dependent measurement in the main article (Figure~\ref{F:electrochemChar}) here in Figure~\ref{F:electrochemCharExtended}. Panel (a) shows the maximum bit rate (Equation~\ref{E:Fbm}) and energy required to transmit each bit (Equation~\ref{E:Eb}). For this device, above about 50~\si{\kilo\hertz}, the signal-to-noise ratio is so low that the maximum bit rate drops below the sine wave frequency. The energy per bit transmitted has a minimum around 150~\si{\hertz}, which is significant in applications where a tradeoff exists between the power consumed and that data rate. Panel (b) shows the phase angles for the reflectance $\phi_R$ (Equation~\ref{E:phaseShiftR}) and the current $\phi_I$ (Equation~\ref{E:phaseShiftI}); the $\phi_I\sim\frac{-\pi}{2}$ value for $10<F<1000$~\si{\hertz} indicates capacitive behavior. Notice that $F_c$ corresponds to $\phi_I=\frac{-\pi}{4}$. 
Panel (c) shows the maximum and minimum values in the reflectance signal, corresponding to PEDOT:PSS in oxidized and reduced states, respectively. For $F<F_c$, the maximum value stays roughly constant while the minimum value increases with increasing $F$. Under the assumption that the reflectance gives insight into the degree of oxidation of the PEDOT:PSS $\gamma$ (see Figure~\ref{F:electrochemChar}(a) of the main article and Section~\ref{S:expLinSweepReflect}), this suggests that the polymer is being reduced to lesser extents as it is driven faster. Another clue is provided in the reflectance phase angle data in panel (b), which shows an increasing phase lag (increasing $\phi_R$) with increasing frequency. The increasing phase lag may indicate that the speed of the governing physics of the polymer (kinetics, transport, \etc) is limiting its performance. For $F>F_c$, the maximum and minimum values coincide, suggesting that little polymer oxidation-reduction is taking place; likely, much of the charge transfer is then through the device's double layer capacitance~\cite{Lin2020-110435}. Panel (d) shows the complex components of the impedance, indicating the reliability of the simple circuit model (Equations~\ref{E:Zre}, \ref{E:Zim}, and~\ref{E:modelImpedance}). Finally, the model leads to a volumetric capacitance of $C^\ast \sim 38.5$~\si{\farad\per\centi\meter\cubed} for the PEDOT:PSS, similar to other values in the literature~\cite{Kurra2014-17058, Rivnay2015-e1400251, Proctor2016-1433, Volkov2017-1700329, Tybrandt2017-eaao3659, Bianchi2020-11252}. 

The theoretical scaling of the energy cost per bit with SPOT lateral size in the main text is performed as follows. If we change the lateral dimensions of the SPOT but keep the polymer thickness identical in both the updated ($u$) and old ($o$) designs, use the same excitation potentials (between $+200$~\si{\milli\volt} and $-800$~\si{\milli\volt}), and ensure that the entire laser spot remains focused on the SPOT, the peak reflectance amplitude value (the value at low frequencies) will remain unchanged but the SPOT's capacitance and hence the charge transferred per half cycle will be altered. Assuming the PEDOT:PSS scales as an ideal capacitor $C$, we have $Q \sim C V$ where $Q$ is a charge transferred and $V$ is the associated potential change. Under the constraints above, we have $\frac{Q_u}{Q_o} \sim \frac{C_u}{C_o}$. Then, the slope $m$ of the graph relating the relative reflectance amplitude $A_R$ to $Q$ (inset of Figure~\ref{F:electrochemChar}(a) in the main text) will change according to $\frac{m_u}{m_o} \sim \frac{Q_o}{Q_u} \sim \frac{C_o}{C_u}$. We use the new SPOT lateral dimensions to calculate the updated parameters in Equation~\ref{E:modelImpedance} assuming the CPE parameter $n$ is unchanged and that the resistance $R_2$ scales inversely with SPOT area. The updated model parameters and updated slope $m_u$ allow us to estimate the energy cost of the new design. 

\subsection{Assessment of repeatability}\label{S:assessRepeat}

\begin{figure}
\centering
\includegraphics[width=\figWidthCol]{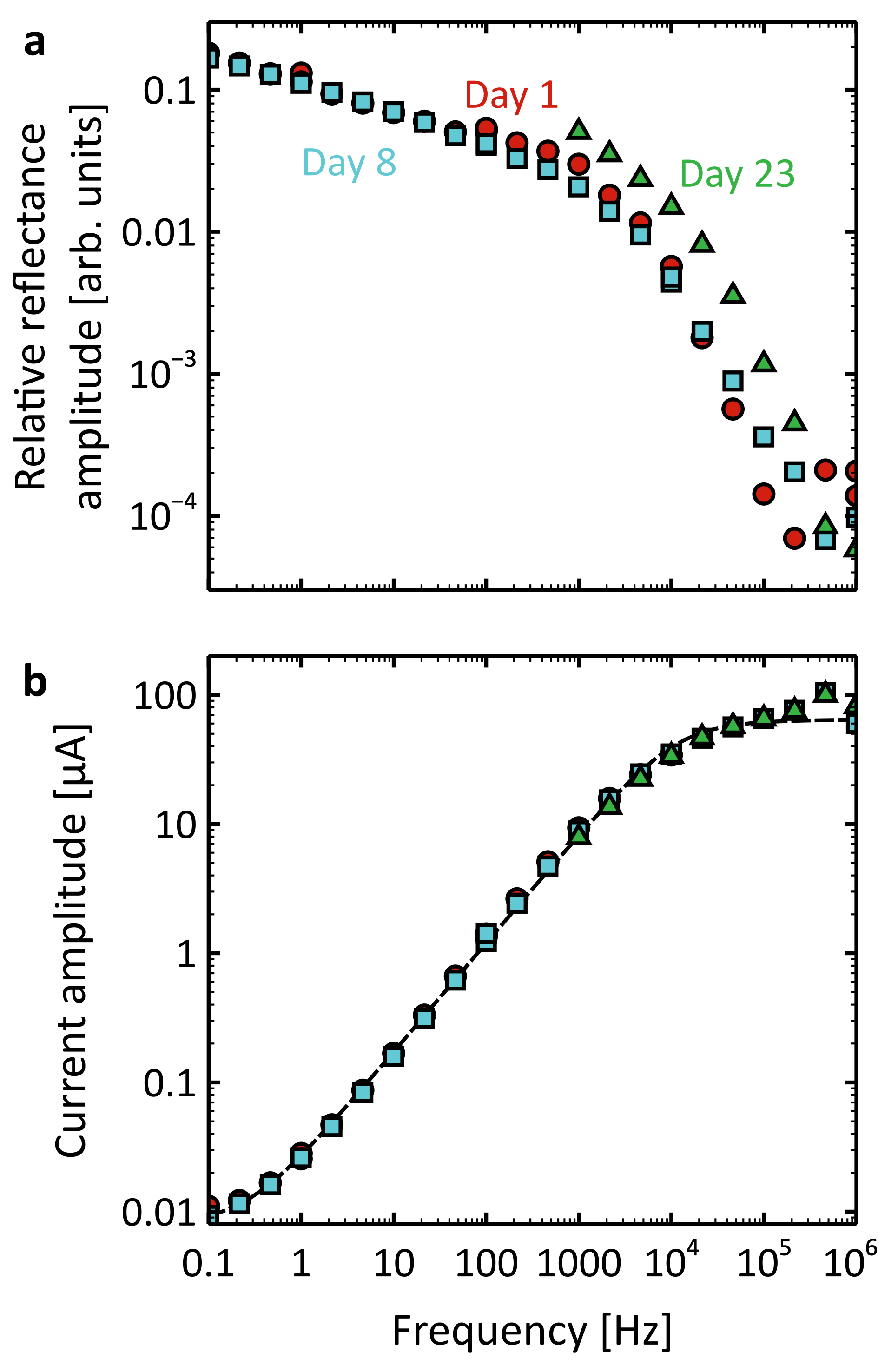}%
\caption{\textbf{Repeatability of frequency-dependent measurements.} 
Device consists of a PEDOT:PSS~($\app73$~\si{\nano\meter})-\ch{Pt} stack in $1\times$PBS, with lateral dimensions $40\times40$~\si{\micro\meter\squared}; laser light at $\lambda_l=635$~\si{\nano\meter} is focused to a $10\times20$~\si{\micro\meter\squared} area on the SPOT; and input potential is a sinusoidal voltage between maximum and minimum values of  $+200$~\si{\milli\volt} and $-800$~\si{\milli\volt}, respectively. 
Experiments were conducted several days apart (days~1, 8, and~23) with the test chip stored dried in the ambient laboratory environment between trials. 
\textbf{(a)} Relative reflectance amplitude. 
\textbf{(b)} Current amplitude. Circuit model, shown in Figure~\ref{F:simpleCircuit} (see also Equation~\ref{E:modelImpedance}), consists of $R_1=7.8$~\si{\kilo\ohm}, $R_2=63.6$~\si{\mega\ohm}, and a constant phase element (parameters $Y_0 = 10.3$~\si{\nano\siemens\second\tothe{0.85}} and $n=0.85$). In this and subsequent figures, the current data at $F\ge 200$~\si{\kilo\hertz} are subject to errors related to the performance of our transimpedance amplifier. 
}%
\label{F:duplicateDay}%
\end{figure}

To assess the repeatability of our frequency-dependent measurements, we test a chip on different days, storing it dry in ambient laboratory air between the trials.  The results are provided in Figure~\ref{F:duplicateDay}.  Some hysteresis in the relative reflectance amplitude is observable, but the current amplitude values measured are very similar (variation of $\le 8$\% from the mean for $F<200$~\si{\kilo\hertz}). See also Figure~\ref{F:areaScaling}.

\subsection{Impact of electrolyte concentration}\label{S:assessElectConc}

\begin{figure}
\centering
\includegraphics[width=\figWidthCol]{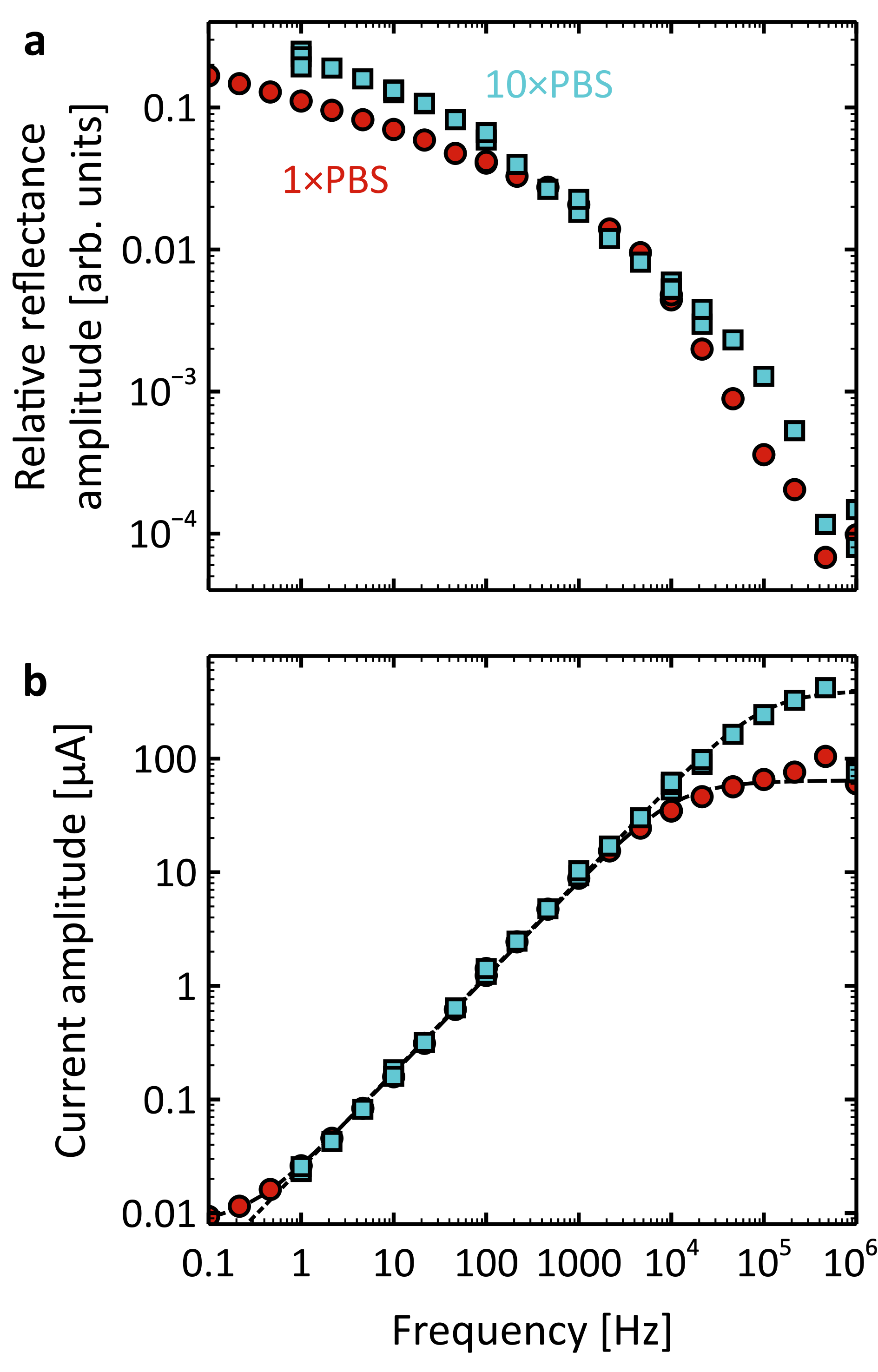}%
\caption{\textbf{Impact of electrolyte concentration.} 
Device consists of a PEDOT:PSS~($\app73$~\si{\nano\meter})-\ch{Pt} stack with lateral dimensions $40\times40$~\si{\micro\meter\squared}; laser light at $\lambda_l=635$~\si{\nano\meter} is focused to a $10\times20$~\si{\micro\meter\squared} area on the SPOT; and input potential is a sinusoidal voltage between maximum and minimum values of  $+200$~\si{\milli\volt} and $-800$~\si{\milli\volt}, respectively. 
\textbf{(a)} Relative reflectance amplitude. 
\textbf{(b)} Current amplitude. 
Circuit model parameters for $1\times$PBS are $R_1=7.8$~\si{\kilo\ohm}, $R_2=63.6$~\si{\mega\ohm}, $Y_0 = 10.3$~\si{\nano\siemens\second\tothe{0.85}}, and $n=0.85$. 
Parameters for $10\times$PBS are $R_1=1.2$~\si{\kilo\ohm}, $R_2=800$~\si{\mega\ohm}, $Y_0 = 10.4$~\si{\nano\siemens\second\tothe{0.85}}, and $n=0.85$. The large $R_2$ value comes because we did not acquire low-frequency data for $10\times$PBS to resolve the roll-over there.
See Figure~\ref{F:simpleCircuit} and Equation~\ref{E:modelImpedance}.
}%
\label{F:pbsConcentration}%
\end{figure}

We conduct frequency-dependent experiments to determine the impact of the electrolyte concentration, using $1\times$PBS (\#SH30256.02, Cytiva Life Sciences) and $10\times$PBS (\#PB5011, Alkali Scientific). The results are provided in Figure~\ref{F:pbsConcentration}, which shows that the current drawn in the two experiments is similar for $F \lesssim 5$~\si{\kilo\hertz}, but is higher with the $10\times$PBS at higher frequencies. From the simple Randles circuit model fits (Figure~\ref{F:simpleCircuit} and Equation~\ref{E:modelImpedance}), we derive the device capacitance $C$ (Equation~\ref{E:estimateDeviceCap}) and approximate the solution resistance as $R_s \sim R_1$. The capacitance values are $C_{1\times}=2.0$~\si{\nano\farad} and $C_{10\times}=1.4$~\si{\nano\farad} for $1\times$PBS and $10\times$PBS, respectively; likewise, the resistance values are $R_{s,1\times}=7.8$~\si{\kilo\ohm} and $R_{s,10\times}=1.2$~\si{\kilo\ohm}. It is apparent that, as expected, the electrolyte conductivity (roughly $\sigma_{s,1\times}=15$~\si{\milli\siemens\per\centi\meter} for $1\times$PBS~\cite{HassanpourTamrin2025-61568} and $\sigma_{s,10\times}=150$~\si{\milli\siemens\per\centi\meter} for $10\times$PBS at 25\si{\celsius}) has little impact on the device capacitance but more significantly alters the solution resistance. 

\subsection{Impact of counter electrode configuration}\label{S:impactCounterElec}

\begin{figure}
\centering
\includegraphics[width=\figWidthCol]{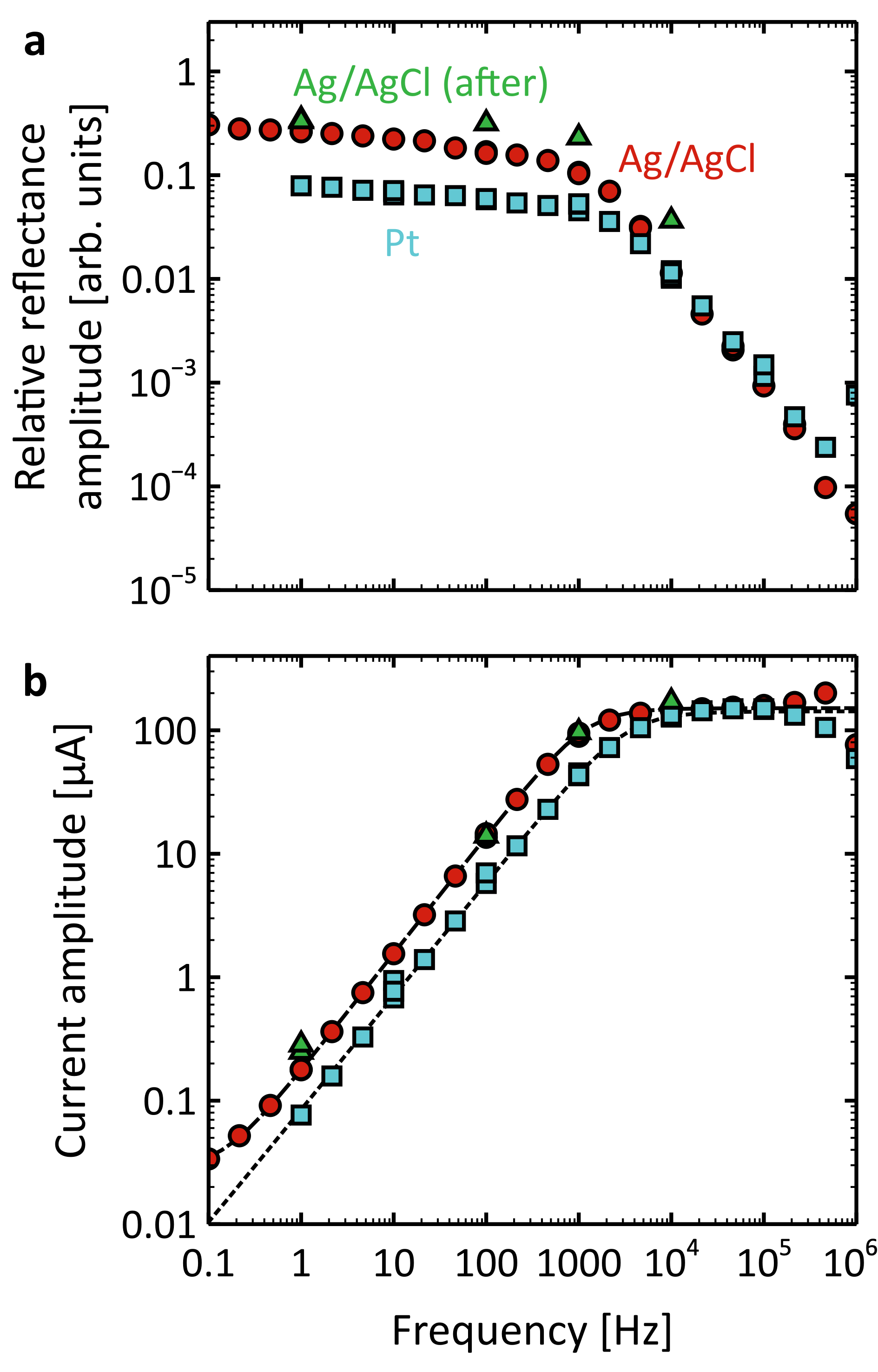}%
\caption{\textbf{Impact of counter electrode configuration.} 
Device consists of a PEDOT:PSS~($\app73$~\si{\nano\meter})-\ch{Pt} stack in $1\times$PBS, with lateral dimensions $100\times100$~\si{\micro\meter\squared}; laser light at $\lambda_l=635$~\si{\nano\meter} is focused to a $10\times20$~\si{\micro\meter\squared} area on the SPOT; and input potential is a sinusoidal voltage between maximum and minimum values of  $+200$~\si{\milli\volt} and $-800$~\si{\milli\volt}, respectively. 
The counter electrode is either a 1~\si{\milli\meter} \ch{Ag}/\ch{AgCl} pellet or a $300\times300$~\si{\micro\meter\squared} square of \ch{Pt} positioned on the same chip as the SPOT (center-to-center distance about 250~\si{\micro\meter}). 
The green triangle points show measurements obtained after reverting the configuration back to the \ch{Ag}/\ch{AgCl} pellet counter electrode. 
\textbf{(a)} Relative reflectance amplitude. 
\textbf{(b)} Current amplitude. 
Circuit model parameters for the \ch{Ag}/\ch{AgCl} configuration are $R_1=3.4$~\si{\kilo\ohm}, $R_2=18.2$~\si{\mega\ohm}, $Y_0 = 59.1$~\si{\nano\siemens\second\tothe{0.95}}, and $n=0.95$. 
Parameters for the \ch{Pt} configuration are $R_1=3.5$~\si{\kilo\ohm}, $R_2=400$~\si{\mega\ohm}, $Y_0 = 31.2$~\si{\nano\siemens\second\tothe{0.92}}, and $n=0.92$. 
The large $R_2$ value comes because we did not acquire low-frequency data for the \ch{Pt} configuration to resolve the roll-over there.
See Figure~\ref{F:simpleCircuit} and Equation~\ref{E:modelImpedance}.
}%
\label{F:refElectrode}%
\end{figure}

To assess the impact of the counter electrode, we test a $100\times100$~\si{\micro\meter\squared} SPOT in two counter electrode configurations: first, with the usual 1~\si{\milli\meter} \ch{Ag}/\ch{AgCl} pellet (Warner Instruments E-205), and second, with a $300\times300$~\si{\micro\meter\squared} square of \ch{Pt} positioned on the same chip as the SPOT (center-to-center distance about 250~\si{\micro\meter}) instead of the \ch{Ag}/\ch{AgCl} pellet. This test is significant, because self-contained microcircuits equipped SPOTs will require on-chip electrodes similar to the \ch{Pt} one we use in this test. The results are shown in Figure~\ref{F:refElectrode}. At frequencies $F \lesssim 10$~\si{\kilo\hertz}, the current in the \ch{Pt} counter electrode configuration is less than that with the \ch{Ag}/\ch{AgCl} pellet, and the reflectance amplitude with the \ch{Pt} electrode is lower. At higher frequencies, the currents and reflectance amplitudes are very similar. The lower reflectance amplitude at $F \lesssim 10$~\si{\kilo\hertz} for the \ch{Pt} electrode is consistent with the lower current there, because it suggests that oxidation and reduction of the PEDOT:PSS cannot occur to the same extent if less charge is available (see Figure~\ref{F:electrochemChar}(a) in the main article). 

Following our analysis in Section~\ref{S:assessElectConc}, from the simple Randles circuit model (Figure~\ref{F:simpleCircuit} and Equation~\ref{E:modelImpedance}), we derive $R_{s,pellet}=3.4$~\si{\kilo\ohm} and $R_{s,on-chip}=3.5$~\si{\kilo\ohm} for the \ch{Ag}/\ch{AgCl} pellet and \ch{Pt} configurations, respectively, and $C_{pellet}=36.7$~\si{\nano\farad} and $C_{on-chip}=14.1$~\si{\nano\farad}. As expected from Equation~\ref{E:RsScaling}, the counter electrode configuration does not influence the solution resistance, which depends primarily on the SPOT (working electrode) geometry~\cite{Newman1966-501}. However, the capacitance of the system is clearly impacted. Treating the capacitance of the counter electrode as being in series with that of the SPOT (at high frequencies where little current travels through $R_2$, see Figure~\ref{F:simpleCircuit} and Equation~\ref{E:modelImpedance}), the total capacitance $C_{total}$ is 
\begin{equation}
    C_{total} = \frac{C_{counter}C_{SPOT}}{C_{counter}+C_{SPOT}}
    \label{E:totalCap}
\end{equation}
where $C_{counter}$ and $C_{SPOT}$ are the capacitance values for the counter electrode and the SPOT, respectively. For the \ch{Ag}/\ch{AgCl} pellet configuration, $C_{counter}=C_{\ch{Ag}/\ch{AgCl}} \gg C_{SPOT}$ such that $C_{total} = C_{SPOT}$.  However, in the \ch{Pt} arrangement, estimating a double layer capacitance for the \ch{Pt} of 64.2~\si{\micro\farad\per\centi\meter\squared}~\cite{Scromeda2008-report}, we have $C_{counter}=C_{\ch{Pt}}\sim 58$~\si{\nano\farad} and $C_{total} \sim 22$~\si{\nano\farad}, which is indeed lower than in the pellet configuration. This implies that circuit designers should account for the capacitance of the counter electrode when integrating SPOTs into their devices. 

\subsection{Impact of SPOT side length}\label{S:impactL}

\begin{figure}
\centering
\includegraphics[width=\figWidthCol]{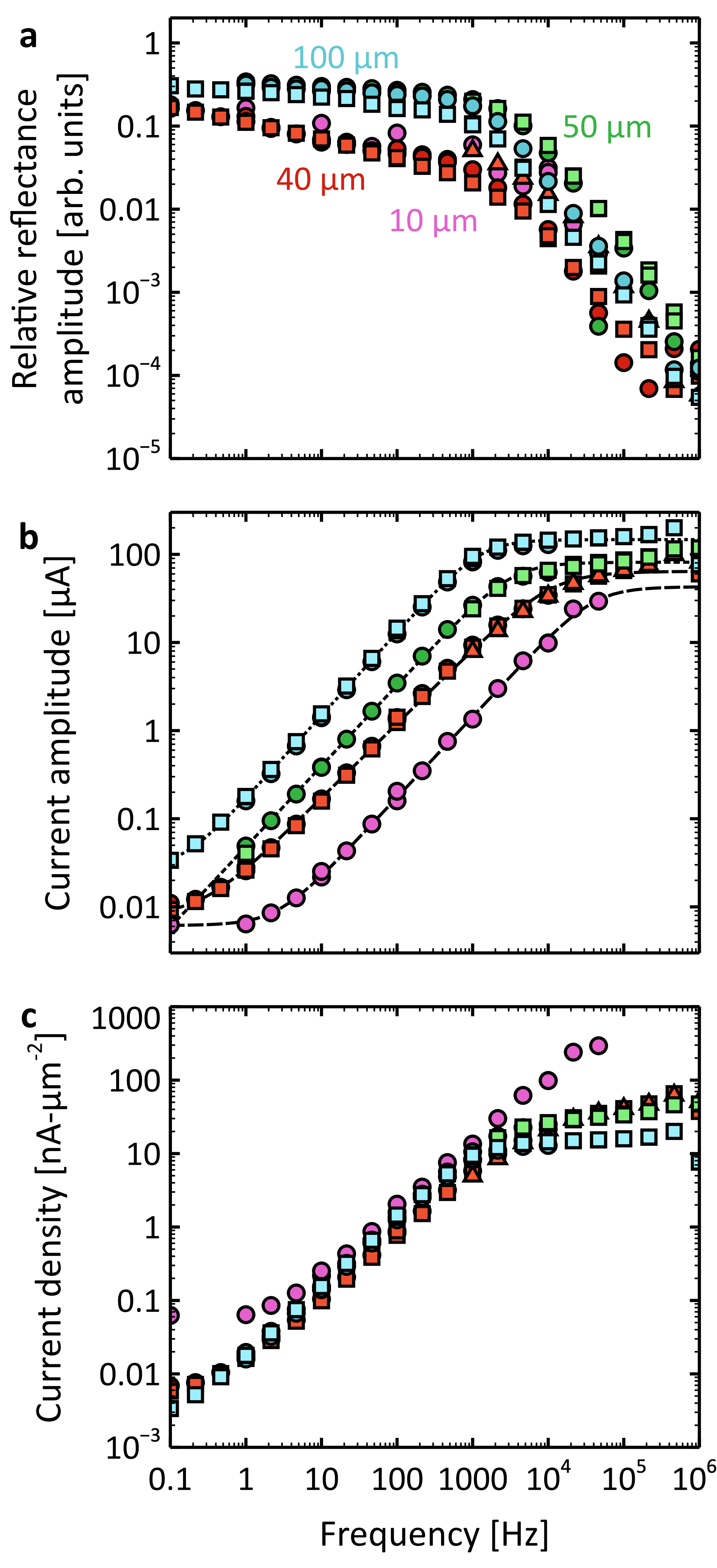}%
\caption{\textbf{Impact of SPOT side length.} 
Devices studied consist of square PEDOT:PSS~($\app73$~\si{\nano\meter})-\ch{Pt} stacks in $1\times$PBS; laser light at $\lambda_l=635$~\si{\nano\meter} is focused to a $10\times20$~\si{\micro\meter\squared} area on the devices; and input potential is a sinusoidal voltage between maximum and minimum values of $+200$~\si{\milli\volt} and $-800$~\si{\milli\volt}, respectively. 
SPOT side lengths are $L=10$~\si{\micro\meter} (obtained on day~1), $L=40$~\si{\micro\meter} (days~1, 8, and~23), $L=50$~\si{\micro\meter} (days~2 and~22), and $100$~\si{\micro\meter} (days~2 and~15); different symbol shapes within similar color  palettes denote different measurement days. 
\textbf{(a)} Relative reflectance amplitude. 
\textbf{(b)} Current amplitude. 
\textbf{(c)} Current density (Equation~\ref{E:currentDensity}). 
Circuit model parameters: $L=10$~\si{\micro\meter}: $R_1=11.6$~\si{\kilo\ohm}, $R_2=81.7$~\si{\mega\ohm}, $Y_0 = 0.94$~\si{\nano\siemens\second\tothe{0.92}}, and $n=0.92$.
$L=40$~\si{\micro\meter}: $R_1=7.8$~\si{\kilo\ohm}, $R_2=63.6$~\si{\mega\ohm}, $Y_0 = 10.3$~\si{\nano\siemens\second\tothe{0.85}}, and $n=0.85$. 
$L=50$~\si{\micro\meter}: $R_1=6.1$~\si{\kilo\ohm}, $R_2=600$~\si{\mega\ohm}, $Y_0 = 17.7$~\si{\nano\siemens\second\tothe{0.92}}, and $n=0.92$.
$L=100$~\si{\micro\meter}: $R_1=3.4$~\si{\kilo\ohm}, $R_2=18.2$~\si{\mega\ohm}, $Y_0 = 59.1$~\si{\nano\siemens\second\tothe{0.95}}, and $n=0.95$. 
See Figure~\ref{F:simpleCircuit} and Equation~\ref{E:modelImpedance}.
See also compiled statistics in Figure~\ref{F:linearRelationships} and in Table~\ref{T:dataFitParams}. 
}%
\label{F:areaScaling}%
\end{figure}

We examine the impact of the device size $L$ by testing the frequency response of four SPOTs with side lengths and thicknesses $L=10$~\si{\micro\meter}, $40$~\si{\micro\meter}, $50$~\si{\micro\meter}, and $100$~\si{\micro\meter}. All devices are on the same chip and have a thickness of $h=73$~\si{\nano\meter}. The results are depicted in Figure~\ref{F:areaScaling}. In panel (c) of this figure, we calculate the current density
\begin{equation}
    I_d = \frac{A_I}{L^2} ,
    \label{E:currentDensity}
\end{equation}
where $A_I$ is the amplitude of the current. The data collapse from a roughly $9\times$ spread to a $3\times$ spread; the aggregation may not be tighter than this because the constant phase element $n$-factors (slopes on the plot) are not identical between the different datasets. The concatenation is suggested by the scalings $C \sim L^2$ (Equation~\ref{E:volCap}), $Z_{cpe} \sim Y_0^{-1}$ (Equation~\ref{E:Zcpe}, in the limit that $n\rightarrow1$ and $Y_0\rightarrow C$), and $A_I=\frac{A_V}{Z}$ (Ohm's law, where $A_V$ is the amplitude of the applied voltage), which together yield $A_I \sim L^2$. The results at the highest frequencies do not collapse together, however, because the resistive scaling is $R_1 \sim R_s \sim L^{-1}$ (Equation~\ref{E:RsScaling}), leading to $A_I \sim L$. 

\subsection{Summary of scaling relationships}\label{S:summaryOfScaling}

\begin{figure*}
\centering
\includegraphics[width=120mm]{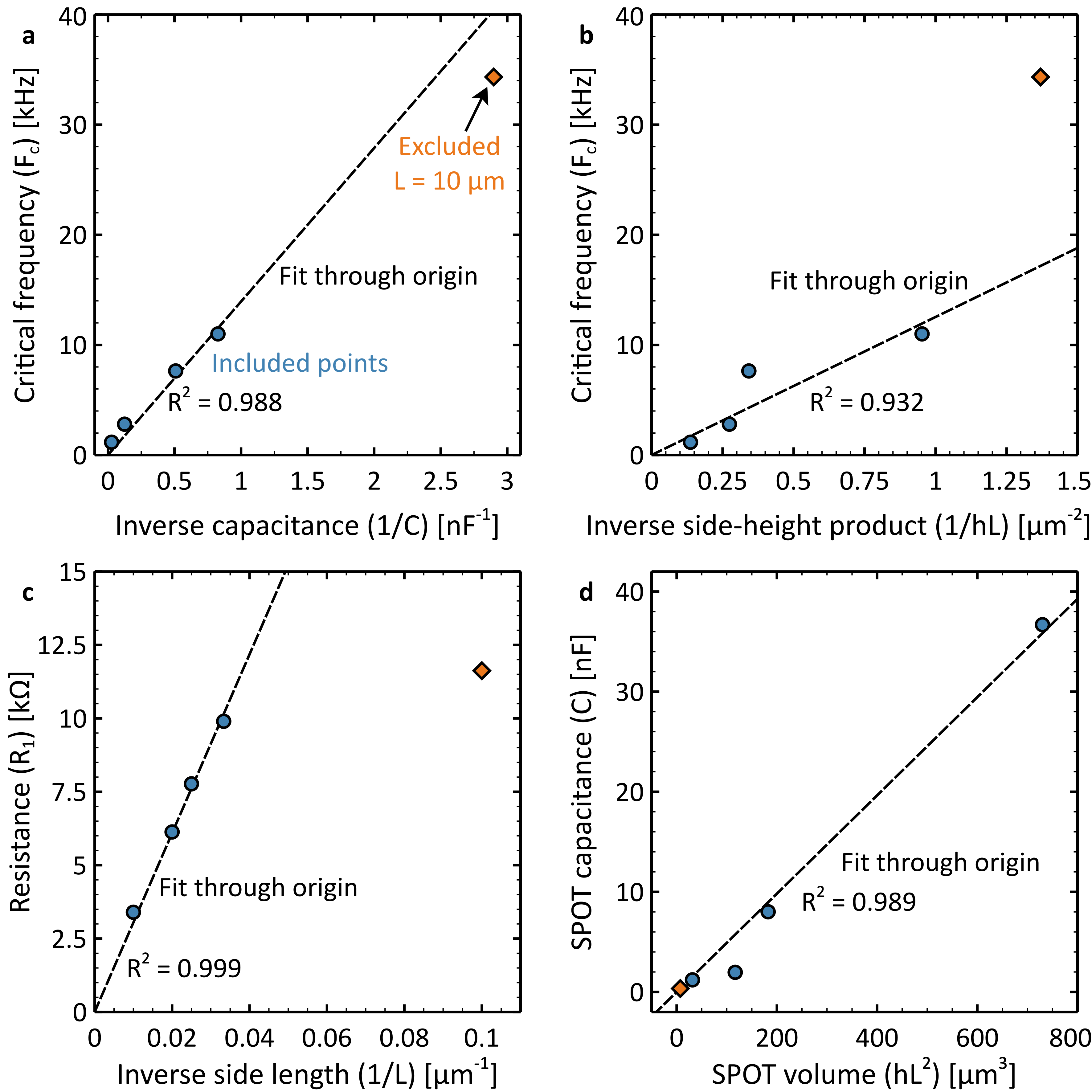}%
\caption{\textbf{Scaling relations for SPOTs.} Devices consist of PEDOT:PSS-\ch{Pt} stacks in $1\times$PBS, having square shapes with lateral side lengths and thicknesses of $L=30$~\si{\micro\meter}/$h\app35$~\si{\nano\meter}, $L=10$~\si{\micro\meter}/$h\app73$~\si{\nano\meter}, $L=40$~\si{\micro\meter}/$h\app73$~\si{\nano\meter}, $L=50$~\si{\micro\meter}/$h\app73$~\si{\nano\meter}, and $L=100$~\si{\micro\meter}/$h\app73$~\si{\nano\meter}. Parameters are derived from fits of Equations~\ref{E:Zmag} and~\ref{E:modelImpedance} to measured current amplitude \vs frequency data. 
Dashed lines show linear fits through the origin. 
The point shown in orange color corresponds to the $L=10$~\si{\micro\meter} SPOT experiment; we have excluded it from the linear fits because its $R_1$ value is an outlier. We speculate that non-ideal effects related to this small electrode size may be complicating our interpretation of this data. 
\textbf{(a)} SPOT critical frequency $F_c$ is inversely proportional to device capacitance $C$  (Equation~\ref{E:criticalFrequencyRC}). 
\textbf{(b)} SPOT critical frequency $F_c$ varies inversely with the product of the SPOT polymer thickness $h$ and side length $L$ (Equation~\ref{E:scaleFc}). 
\textbf{(c)} SPOT resistance $R_1$ is inversely proportional to the device side length $L$ (Equation~\ref{E:RsScaling}). 
\textbf{(d)} SPOT capacitance $C$ is proportional to SPOT volume $hL^2$ (Equation~\ref{E:volCap}). The volumetric capacitance obtained from the linear fit is $C^\ast=49$~\si{\farad\per\centi\meter\cubed}. 
See also raw data in Figure~\ref{F:areaScaling} and parameters listed in Table~\ref{T:dataFitParams}. 
}%
\label{F:linearRelationships}%
\end{figure*}

We used fits of the simple circuit model (Figure~\ref{F:simpleCircuit} and Equation~\ref{E:modelImpedance}) to the frequency-dependent experimental data shown thus far to derive values for $R_1$, $R_2$, $Y_0$, $n$, $F_c$ (Equation~\ref{E:criticalFrequency}), $C$ (Equation~\ref{E:estimateDeviceCap}), and $C^\ast$ (Equation~\ref{E:volCap}).  The results are plotted in Figure~\ref{F:linearRelationships} and are listed in Table~\ref{T:dataFitParams}. The graphs include linear fits through the origin. In these fits, we excluded the data for the $L=10$~\si{\micro\meter} SPOT, because its $R_1$ value is an outlier; we speculate that non-ideal effects related to this small electrode size may be complicating our interpretation of this data. The $C$ \vs $hL^2$ relationship in panel (d) leads to a volumetric capacitance of $C^\ast=49$~\si{\farad\per\centi\meter\cubed} (Equation~\ref{E:volCap}), which is similar to other values in the literature~\cite{Kurra2014-17058, Rivnay2015-e1400251, Proctor2016-1433, Volkov2017-1700329, Tybrandt2017-eaao3659, Bianchi2020-11252}. 

\begin{center}
\begin{table*}
  \caption{\textbf{Tabulated simple circuit model fits and derived parameters.} See Figure~\ref{F:simpleCircuit} and Equation~\ref{E:modelImpedance}. Unless otherwise noted, data were obtained using a 1~\si{\milli\meter} \ch{Ag}/\ch{AgCl} pellet counter electrode in $1\times$PBS electrolyte solution. } 
  \begin{tabular}{ l l l l l l l l l l }
    \hline
    $L$ & $h$ & $R_1$ & $R_2$ & $Y_0$ & $n$ & $F_c$ & $C$ & $C^\ast$ & Notes \\
    \si{\micro\meter} & \si{\nano\meter} & \si{\kilo\ohm} & \si{\mega\ohm} & \si{\nano\siemens\second\tothe{n}} & & \si{\kilo\hertz} & \si{\nano\farad} & \si{\farad\per\centi\meter\cubed} & \\
    \hline
    30  & 35 & 9.9  & 80.0 & 3.70 & 0.90 & 11.0 & 1.21  & 38.5 & \\
    10  & 73 & 11.6 & 81.7 & 0.94 & 0.92 & 34.3 & 0.345 & 47.2 & \\
    40  & 73 & 7.8  & 63.6 & 10.3 & 0.85 & 7.64 & 1.97  & 16.8 & \\
    50  & 73 & 6.1  & 600  & 17.7 & 0.92 & 2.80 & 8.01  & 43.9 & \\
    100 & 73 & 3.4  & 18.2 & 59.1 & 0.95 & 1.17 & 36.7  & 50.3 & \\  
    40  & 73 & 1.2  & 800  & 10.4 & 0.85 & 65.4 & 1.44  & 12.3 & $10\times$PBS \\
    100 & 73 & 3.5  & 400  & 31.2 & 0.92 & 2.78 & 14.1  & 19.4 & \ch{Pt} counter electrode \\
    \hline
  \end{tabular}
  \label{T:dataFitParams}
\end{table*}
\end{center}

\section{Potential step-change experiments}\label{S:ExptVstep}

\subsection{Experimental setup for potential step-change experiments}\label{S:exptSetupVstep}

The experimental setup for the potential step-change experiments (such as Figures~1(d) and 4(c) in the main article) is similar to that of the frequency-dependent experiments (Section~\ref{S:ExptSetupFdep}).  Importantly, however, we remove the transimpedance amplifier from the electrical circuit, because its response time is not fast enough. 

\subsection{Experiments in solutions with varying conductivity}\label{S:expConductivity}

Our procedure for the step time experiments in electrolytes with different conductivities follows that of the standard voltage step-change tests in $1\times$PBS with a few important changes. We test standard conductivity solutions, whose salt content consists of \ch{KCl} ions, all purchased from Environmental Express (formerly Oakton Instruments): 
1~\si{\micro\siemens\per\centi\meter} (00652-20); 
5~\si{\micro\siemens\per\centi\meter} (00652-22); 
10~\si{\micro\siemens\per\centi\meter} (00652-24); 
84~\si{\micro\siemens\per\centi\meter} (00653-16); 
1413~\si{\micro\siemens\per\centi\meter} (00652-30); 
12880~\si{\micro\siemens\per\centi\meter} (00606-10). Since solution conductivity is influenced by temperature, we insert a fine-wire thermocouple into the electrolyte (Dwyer Instruments, LLC, \#5TC-TT-K-36-36, with handheld readout Dwyer Instruments, LLC, HH801A) to account for this factor. We compensate for the temperature $T$ by calculating the actual solution conductivity from that listed on the bottle (each calibration bottle states the conductivity $\sigma_{s,T_{ref}}$ at $T_{ref}=25$\si{\celsius}) according to 
\begin{equation}
    \sigma_s \equiv \sigma_{s,T} = \sigma_{s,T_{ref}} \! \left( 1+\alpha_c\left(T-T_{ref}\right) \right),
    \label{E:correctConcForTemp}
\end{equation}
where $\alpha_c=0.020$\si{\per\celsius} is the temperature coefficient and $\sigma_s \equiv \sigma_{s,T}$ is actual conductivity of the solution in the experiment.  Typical measured experimental temperatures are roughly $20$\si{\celsius}, leading to conductivity corrections of about 10\%. We apply a square wave signal at 1~\si{\hertz}, providing 500~\si{\milli\second} between the steps and thus ample time for the photodiode signal to reach steady state even in the solutions with the lowest conductivities and longest rise/fall times. When changing from one electrolyte to another, we thoroughly rinse all components (objective lens, chip with SPOTs, PDMS fluid barrier, \ch{Ag}/\ch{AgCl} pellet, and thermocouple bead) with DI \ch{H2O} and dry completely with \ch{N2} to ensure there is no salt or water contamination. 

\subsection{Processing potential-step time data}\label{S:processVstepData}

We process the optical step time data as follows. We collect at least ten square wave periods at each voltage pair. To drive the characteristic rise (voltage step-up) and fall (voltage step-down) times, we handle each period separately, deriving times for each and then averaging to find combined statistics. We mark the $\frac{1}{e}$ rise point as the first instant where 
\begin{equation}
    V_p \ge \left(1-\frac{1}{e} \right) \! \left(V_{p,h}-V_{p,l} \right) + V_{p,l} ,
    \label{E:conditionForRiseTime}
\end{equation}
where $V_p$ is the photodetector voltage and $V_{p,h}$ and $V_{p,l}$ are the steady-state (long-time, plateau) photodiode voltage values when the applied voltage is high and low, respectively.  We measure the rise time relative to the instant that the driving voltage signal (applied to the SPOT) rises. Likewise, the fall time is marked by
\begin{equation}
    V_p \le V_{p,h} - \left(1-\frac{1}{e} \right) \! \left(V_{p,h}-V_{p,l} \right) .
    \label{E:conditionForFallTime}
\end{equation}
In order to mitigate the impact of noise, we calculate these values using a four-point moving average of the photodiode voltage signal. Our oscilloscope data rate was typically 100~\si{\kilo\hertz} (or 500~\si{\kilo\hertz} for the fastest rise time tests), giving 10~\si{\micro\second} between points (or 2~\si{\micro\second} at the higher data rate). For the very first points after the driving voltage step change, the algorithm averages as shown in Equation~\ref{E:averaging}. The smoothing that this entails thus introduces an uncertainty of roughly $\pm10$~\si{\micro\second} (100~\si{\kilo\hertz} data rate) or $\pm2$~\si{\micro\second} (500~\si{\kilo\hertz} data rate), which is acceptable given the step times we observe. 
%
{
\renewcommand{\arraystretch}{1.5}
\begin{equation}
    \begin{array}{ccc}
    \text{Point} & \text{Data} & \text{Average} \\
    1 & V_{p,1} & \frac{V_{p,1}+V_{p,2}}{2} \\
    2 & V_{p,2} & \frac{V_{p,1}+V_{p,2}+V_{p,3}}{2} \\
    3 & V_{p,3} & \frac{V_{p,1}+V_{p,2}+V_{p,3}+V_{p,4}}{2} \\
    4 & V_{p,4} & \frac{V_{p,2}+V_{p,3}+V_{p,4}+V_{p,5}}{2} \\
    5 & V_{p,5} & \frac{V_{p,3}+V_{p,4}+V_{p,5}+V_{p,6}}{2} \\
    \ldots & \ldots & \ldots
    \end{array}
    \label{E:averaging}
\end{equation}
}
%

\subsection{Additional potential-step time data}\label{S:additionalVstepData}

PEDOT:PSS films are \textit{p}-type semiconductors~\cite{Rivnay2016-11287} whose hole density changes dramatically with doping state. Specifically, when oxidized (positive potential applied to the electrode), PEDOT:PSS is known as doped and its hole density is high, and when reduced (negative potential applied), it is known as neutral or de-doped and its hole density is very low~\cite{Dingler2022-1600}. In organic electrochemical transistors (OECTs), this materializes as a dramatic change in the source-drain current with gate voltage~\cite{Rivnay2013-7010, Fabiano2017-e1700345, Tybrandt2017-eaao3659, Huang2023-496, Li2025-eadt5186}. 

To explore the impact of doping state on SPOTs, we measure the temporal reflectance response for small voltage steps of only $\Delta V = \pm 100$~\si{\milli\volt} in $1\times$PBS. This voltage step is sufficiently small that the PEDOT:PSS does not experience a large swing in its oxidation (\ie, doping) state, allowing us to compare the rise times at different doping levels by varying the starting (ending) voltage $V_i$ $(V_f)$. This is significantly different from the experiments of Figures~1(e) and~4(c) in the main article, which represent a large swing of the oxidation state of PEDOT:PSS (and thereby a large swing in the doping state) because they were performed between more extreme voltages of $-800$~\si{\milli\volt} and $+200$~\si{\milli\volt}. 

\begin{figure}
\centering
\includegraphics[width=\figWidthCol]{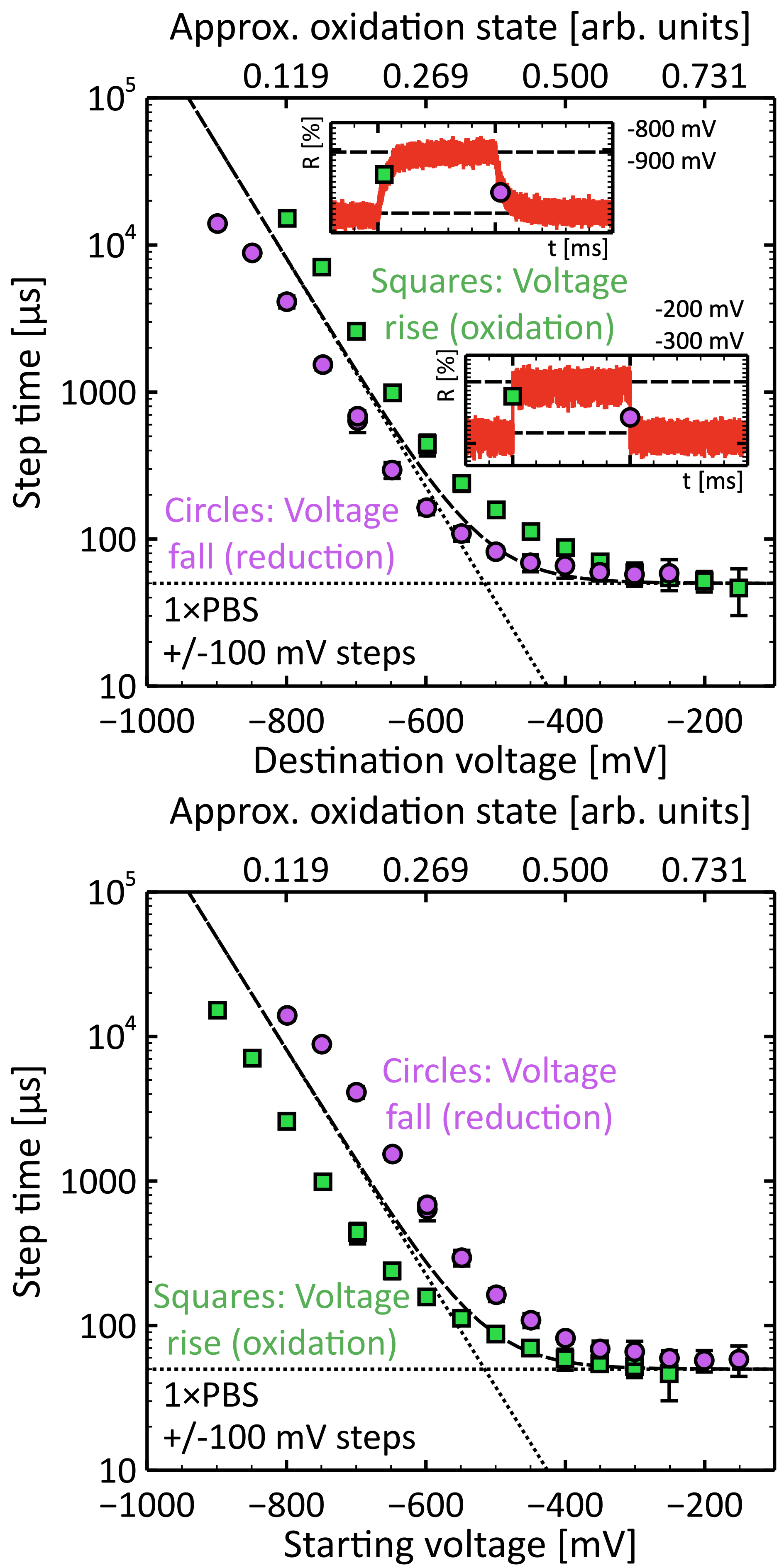}%
\caption{\textbf{Temporal characterization of SPOTs.} 
\textbf{(a)} Time $\left(\frac{1}{e}\right)$ required for reflectance rise and fall for $\Delta V = \pm 100$~\si{\milli\volt} step changes in electrode potential, as a function of the destination (step-to) voltage $V_f$. The error bars represent standard deviations for the measurements. The insets show the reflectance signal changing in response to steps between $-900$~\si{\milli\volt} and $-800$~\si{\milli\volt} (upper) and between $-300$~\si{\milli\volt} and $-200$~\si{\milli\volt} (lower). 
\textbf{(b)} Data from (a), plotted as a function of the starting (step-from) voltage $V_i$.
In both plots, the upper abscissa shows the approximate oxidation state $\gamma$ of the polymer at the voltage indicated on the lower abscissa, calculated according to Equation~\ref{E:oxFrac}. 
Also in both plots, the dashed lines are the sum of an exponentially decaying response (lowest voltages) and a constant response time. 
The device tested consists of a PEDOT:PSS~($\app45$~\si{\nano\meter})-\ch{Pt} stack in $1\times$PBS, with lateral dimensions $30\times30$~\si{\micro\meter\squared}. Incident laser light at $\lambda_l=635$~\si{\nano\meter} is focused to a spot of roughly $10\times20$~\si{\micro\meter\squared} on the electrode. The polymer studied here is cross-linked by thermal treatment (see Section~\ref{S:fabCrossGOPS})~\cite{Doshi2025-2415827}.  
}%
\label{F:stepTimePolyConduct}%
\end{figure}

The results are depicted in Figure~\ref{F:stepTimePolyConduct}, which shows the times as a function of (a) the destination (step-to) voltage $V_f$ and (b) the starting (step-from) voltage $V_i$. On the left side of the graphs (at the most negative voltages), the step times increase exponentially with decreasing $V_f$ and $V_i$. On the right sides of the graphs (at less negative voltages), the times converge to a plateau. On the upper plot, for the same $V_f$, we find the time for oxidation is longer than the time for reduction $(\tau_{e,o} > \tau_{e,r})$. Conversely, on the lower plot, for the same $V_i$, we find $\tau_{e,r} > \tau_{e,o}$.

To analyze these results, as in the main text (Figure~\ref{F:electrochemChar}(a)), we assume equivalence between the optical and electrical rise times. In addition, we include a second abscissa that uses our laser reflectometry data (see Section~\ref{S:expLinSweepReflect} and Equation~\ref{E:oxFrac}) to estimate the oxidation fraction $\gamma$ of the polymer, which we allow to vary from zero (fully reduced) to unity (fully oxidized)~\cite{Cucchi2022-4514}. We include this to give insight to the state of the polymer, in the absence of an independent measure of the polymer's equilibrium doping state $\phi$ as a function of the applied voltage. 

We speculate that the increase in $\tau_e$ with decreasing $V_f$ and $V_i$ is attributable to the doping state of the polymer in its reduced redox state. This slows the rearrangement of holes in the film when the potential is stepped and hence lengthens the time for the reflectance signal to change~\cite{Dingler2022-1600, Huang2023-496, Li2025-eadt5186}. Furthermore, we hypothesize that the convergence and plateauing of step times on the right side of the plot (at less negative voltages) occurs because, in these more oxidized states, the polymer is sufficiently doped that the hole concentration no longer limits the step times. Rather, we suggest that the step times at those potentials are limited by the solution resistance and device capacitance, which in turn depend on the device geometry (PEDOT:PSS thickness $h$ and SPOT side length $L$) and electrolyte concentration (see Sections~\ref{S:FreqScaling} and~\ref{S:stepLowSolnCond}).

Our conjecture for why $\tau_{e,o} > \tau_{e,r}$ for the same $V_f$ at low voltages is as follows. The $\tau_{e,o}$ represents a step from a less-doped state $\phi - \Delta\phi$, and $\tau_{e,r}$ represents a step from a more-doped state $\phi + \Delta\phi$. The oxidation step to the same $V_f$ thus on average covers a lower-hole-mobility state of the polymer relative to the reduction step. Therefore, when ending on the same $V_f$, the time required for the oxidation (voltage rise) step is longer than the time required for the reduction (voltage drop) step, because the lower hole mobility slows rearrangement of holes to achieve a new equilibrium oxidation state. 

Similarly, we explain the fact that $\tau_{e,r} > \tau_{e,o}$ for the same $V_i$ as follows. Beginning at the same doping state of the polymer $\phi$, stepping to a higher voltage implies increasing the doping state to $\phi + \Delta\phi$, such that the hole mobility rises. Conversely, stepping to a lower voltage implies reducing the doping state to $\phi - \Delta\phi$, such that the hole mobility decreases. Thus, for the same $V_i$, the step for reduction, which on average covers a lower-hole-mobility state of the polymer, takes longer than the step for oxidation. 

The difference between the oxidation times (voltage rise, $\tau_{e,o}$) and reduction times (voltage fall, $\tau_{e,r}$) in Figures~1(e) and~4(c) in the main article is also noteworthy. Specifically, these graphics show $\tau_{e,o} < \tau_{e,r}$, as has been observed by others~\cite{Paulsen2020-2003404, Rebetez2021-2105821, Keene2023-eadi3536}. This can be explained by the different forward and reverse PEDOT:PSS reaction rates~\cite{Rebetez2021-2105821} as well as, we speculate, the varying density of holes. Electron transfer to/from the polymer occurs in the thin layer of PEDOT:PSS immediately adjacent to the metal electrode, and holes then redistribute throughout the polymer~\cite{Volkov2017-1700329, Tybrandt2017-eaao3659, Rebetez2021-2105821}. In oxidation, the density of holes (doping state $\phi$) in the metal-adjacent PEDOT:PSS layer quickly rises, increasing the hole mobility and allowing holes to rapidly propagate throughout the thickness of the film. In contrast, during reduction, the hole density in the metal-adjacent polymer layer plummets as holes there recombine with electrons, thereby creating a bottleneck that impedes further reduction progress. This framework explains how, at a given solution conductivity value, the oxidation step time is shorter than the reduction step time. Further work is necessary to validate our hypothesis. 


\subsection{Interpreting step times at low solution conductivity values}\label{S:stepLowSolnCond}

Here we discuss the step time results in Figure~\ref{F:demonstrationFigure}(c) of the main article. Using the Randles model (Figure~\ref{F:simpleCircuit}) under the simplification that the constant phase element can be approximated by a capacitor $C$, we apply a step-change in potential across the circuit $V_s$ and apply Kirchhoff's current law at the junction node, obtaining 
\begin{equation}
    \frac{V_s-V_C}{R_1} = \frac{V_C}{R_2} + C \frac{dV_C}{dt} 
    \label{E:randlesSumCurrents}
\end{equation}
where $V_C$ is the voltage across the capacitor and $t$ is time. Rearranging, we obtain a first-order ordinary differential equation:
\begin{equation}
    \frac{dV_C}{dt} + \frac{1}{C}\frac{R_1+R_2}{R_1 R_2} V_C = \frac{V_s}{C R_1} . 
    \label{E:randlesStepODE}
\end{equation}
Accordingly, the time constant of this circuit is 
\begin{equation}
    \tau_e = C \frac{R_1 R_2}{R_1 + R_2} .
    \label{E:randlesStepTimeConst}
\end{equation}
Assuming PEDOT:PSS acts as a capacitor, it can be charged and discharged through $R_1$ (typically regarded as the solution resistance $R_s$, Equation~\ref{E:RsScaling}) or through $R_2$, which may represent Faradaic reactions on the \ch{Pt} beneath the polymer. From a charging/discharging perspective, these resistors are functionally in parallel, having a combined resistance given by
\begin{equation}
    R_{combined} = \left(\frac{1}{R_1}+\frac{1}{R_2}\right)^{-1} = \frac{R_1 R_2}{R_1 + R_2} , 
    \label{E:Rcombined}
\end{equation}
and we recognize the combined resistance as that dictating the step time constant $\tau_e$ in Equation~\ref{E:randlesStepTimeConst}. 

At solution conductivity values typical of physiological environments (the conductivity of $1\times$PBS is approximately $15$~\si{\milli\siemens\per\centi\meter}~\cite{HassanpourTamrin2025-61568}), the dominant path is $R_{combined} \rightarrow R_1$ since $R_1 \ll R_2$ (see values in Table~\ref{T:dataFitParams}). However, at very low concentrations, $R_2$ becomes important since there $R_1 \simeq R_2$. 

Substituting Equation~\ref{E:RsScaling} for $R_1=R_s$ and $R_2=\frac{\tau_2}{C}$ (where $\tau_2$ is a time constant for the alternate charging path) in Equations~\ref{E:randlesStepTimeConst} and~\ref{E:Rcombined} yields the harmonic mean equation plotted with the dotted line in Figure~\ref{F:demonstrationFigure}(c) in the main article
\begin{equation}
    \tau_e \sim \left(\frac{\sigma_s}{b}+\frac{1}{\tau_2}\right)^{-1}
    \label{E:harmonicMeanCond}
\end{equation}
where $b=\frac{C\sqrt{\pi}}{4L}=210\sci{3}$~\si{\micro\second\micro\siemens\per\centi\meter} and $\tau_2=62\sci{3}$~\si{\micro\second}. Combining $b$ with Equation~\ref{E:volCap} reveals that $C^\ast \sim 39$~\si{\farad\per\centi\meter\cubed}, similar to other literature values~\cite{Kurra2014-17058, Rivnay2015-e1400251, Proctor2016-1433, Volkov2017-1700329, Tybrandt2017-eaao3659, Bianchi2020-11252}. Using this volumetric capacitance with $\tau_2$ gives an estimate of $R_2 \sim 131$~\si{\mega\ohm}, which is in reasonable agreement with the values in Table~\ref{T:dataFitParams}. From Equation~\ref{E:RsScaling}, the solution conductivity for $R_1 \simeq R_2$ occurs at $\sigma_s \sim 3.38$~\si{\micro\siemens\per\centi\meter}. 

Thus, although the exact physical interpretation of $R_2$ in this context is uncertain, it appears that there may be a justification for a charging path by which PEDOT:PSS could have faster step times at low solution conductivity values than predicted by the traditional solution resistance scaling model (Equation~\ref{E:RsScaling}). Finally, we note that solution conductivity values in the \si{\micro\siemens\per\centi\meter} range are less important for applications in biological solutions or implanted devices.



\section{Details about linear sweep reflectometry}\label{S:expLinSweepReflect}

We measure the reflectance as a function of voltage by using the standard electrochemistry-optical setup described in Section~\ref{S:ExptSetupFdep} (including laser light at $\lambda_l=635$~\si{\nano\meter}), but apply a $F=20$~\si{\milli\hertz} (\ie, 64~\si{\milli\volt\per\second}) triangle wave voltage with a minimum of $V=-1100$~\si{\milli\volt} and a maximum of $V=+500$~\si{\milli\volt}. We determine these extreme values by observing the voltage at which the derivative of the reflectance signal is approximately zero during the sweep, and we allow the sweep to run for a few cycles to season the SPOT before acquiring the final data. We calibrate the reflectance using the \ch{Au} reference described previously (Equation~\ref{E:correctReflect}). 

\begin{figure}
\centering
\includegraphics[width=\figWidthCol]{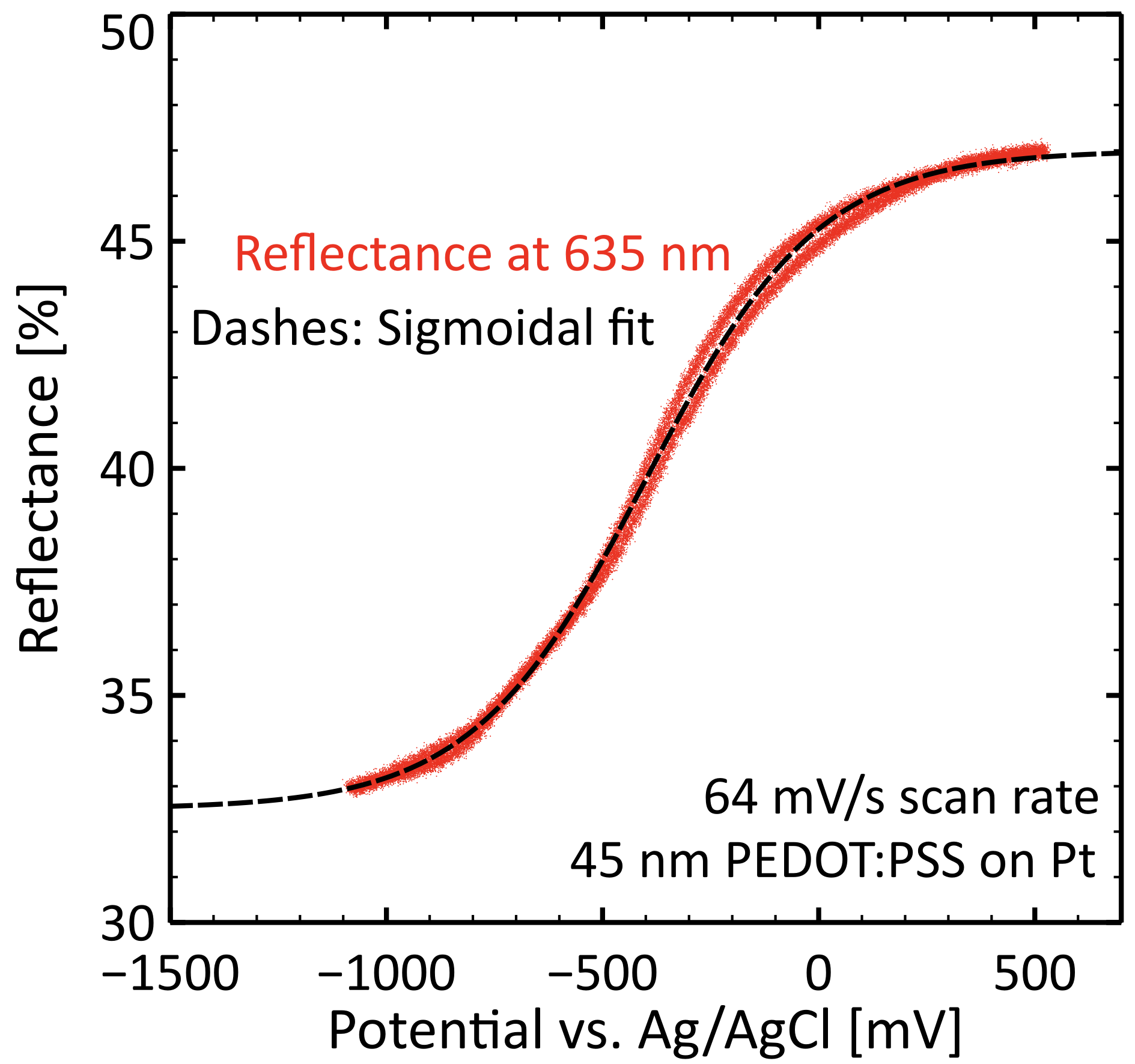}%
\caption{\textbf{SPOT reflectance as a function of voltage.}
Device consists of a PEDOT:PSS~($\app45$~\si{\nano\meter})-\ch{Pt} stack in $1\times$PBS, with lateral dimensions $50\times50$~\si{\micro\meter\squared}; laser light at $\lambda_l=635$~\si{\nano\meter} is focused to a $10\times20$~\si{\micro\meter\squared} area on the SPOT; and input potential is a $F=20$~\si{\milli\hertz} (\ie, 64~\si{\milli\volt\per\second}) triangle wave voltage with a minimum of $V=-1100$~\si{\milli\volt} and a maximum of $V=+500$~\si{\milli\volt}.
The polymer is cross-linked using heat treatment~\cite{Doshi2025-2415827}.
The sigmoidal fit is derived using Equation~\ref{E:linearSweepReflect} with parameters $\chi_1=47.0$\%, $\chi_2=32.5$\%, $\chi_3=0.005$~\si{\milli\volt\tothe{-1}}, and $\chi_4=-400$~\si{\milli\volt}. 
}
\label{F:rampRef}%
\end{figure}

Results for a SPOT consisting of a PEDOT:PSS~($\app45$~\si{\nano\meter})-\ch{Pt} stack in $1\times$PBS, with lateral dimensions $50\times50$~\si{\micro\meter\squared}, are presented in Figure~\ref{F:rampRef}. To fit the reflectance data $\varrho$, we use a sigmoidal function of the form
\begin{equation}
    \varrho = \chi_1 + \frac{\chi_2-\chi_1}{1+\exp\!{\left(\chi_3\left(V - \chi_4\right)\right)}} ,
    \label{E:linearSweepReflect}
\end{equation}
where $\chi_1=47.0$\% and $\chi_2=32.5$\% are the maximum and minimum reflectance values, respectively; $\chi_3=0.005$~\si{\milli\volt\tothe{-1}} is the voltage scaling factor; and $\chi_4=-400$~\si{\milli\volt} is the voltage offset. 

We use this curve to estimate the oxidation fraction $\gamma$ of the PEDOT:PSS as a function of the equilibrium applied voltage, which, similarly to Cucchi~\etal~\cite{Cucchi2022-4514}, we allow to vary between zero (fully reduced at the lowest potentials) and unity (fully oxidized at the highest potentials). This is accomplished in Equation~\ref{E:oxFrac}, in which we have replaced $\chi_1$ and $\chi_2$ with $1$ and $0$, respectively, and the values of $\chi_3$ and $\chi_4$ are unchanged from Equation~\ref{E:linearSweepReflect}. 
\begin{equation}
    \gamma = 1 + \frac{0-1}{1+\exp\!{\left(\chi_3\left(V - \chi_4\right)\right)}} 
    \label{E:oxFrac}
\end{equation}

\section{Details about hyperspectral imaging}\label{S:expHyper}

We acquire hyperspectral images using a Nireos HERA VNIR camera under the illumination of an LED light source (Thorlabs, Inc., SOLIS-1D with DC2200 driver; note this is different from the light source used in the electrochemical characterization experiments).  
Each image consists of a $1024\times1280$ pixel grid, and each pixel contains a 120-point spectrum in the wavelength range 400-1000~\si{\nano\meter}. 
We image chips through an immersion lens (Olympus Co., LUMPLFLN40XW \#34-557) and in $1\times$ PBS (\#SH30256.02, Cytiva Life Sciences) solution.
Focusing the hyperspectral camera is difficult due to the low depth of field of the immersion lens and the slow response of the hyperspectral camera; to accomplish this, we first find the rough position using an adjacent camera (Scientifica Ltd., SciCam+) and then fine-tune the focus with the hyperspectral camera. 
To make electrical contact to PEDOT:PSS-coated chips, we pass an uninsulated needle probe tip (Signatone, Co., SE-T) through the liquid and use a 1~\si{\milli\meter} \ch{Ag}/\ch{AgCl} pellet (Warner Instruments E-205) as a counter electrode.  
Our use of an uninsulated probe tip may introduce a small voltage offset, as discussed later. 
We apply potentials using a BK Precision 4053B function generator and verify these voltage values by monitoring them in a parallel branch to a digital oscilloscope (Picoscope 5442D).
In addition to images of the PEDOT:PSS-coated chips, we acquire several images for calibration.  
These include a spectralon sample (Labsphere Inc., AS-01160-060) to normalize artifacts in the imaging system; an \ch{Al} sample (73~\si{\nano\meter} \ch{Al} on 162~\si{\nano\meter} \ch{SiO2} on 525~\si{\micro\meter} \ch{Si}) to calibrate the reflectance of the spectra; an image with the LED source turned off to measure the background (dark) signal strength; and images of \ch{Au} (120~\si{\nano\meter} \ch{Au} on 20~\si{\nano\meter} \ch{Ti} on 525~\si{\micro\meter} \ch{Si}), \ch{Si} (525~\si{\micro\meter}), and \ch{SiO2} (645~\si{\nano\meter} \ch{SiO2} on 525~\si{\micro\meter} \ch{Si}) to validate the image system calibration. We simulate the spectra using the transfer-matrix method~\cite{Macleod2017-book, Campbell2022-90} using optical constants from the literature~\cite{Segelstein1981-thesis, Olmon2012-235147, Schinke2015-067168, Johnson1974-5056, Rodriguez-deMarcos2016-3622, Cheng2016-9852, Polyanskiy2024-94}, in MATLAB R2024b software.  


We process the hyperspectral image data as follows.  First, we crop all images to a central region that excludes windowing effects from the lens near the perimeter.  The pixel dimensions of the cropped images are $788\times775$. Next, for all images, we subtract the ``dark'' (light source off) images to account for background lighting in the laboratory, and normalize all images by the spectralon image.  
\begin{equation}
    \xi_{image}(x,y,\lambda) = \frac{\xi_{\ch{Al}}(x,y,\lambda) - \xi_{\ch{Al},dark}(x,y,\lambda)}{\xi_{spec}(x,y,\lambda) - \xi_{spec,dark}(x,y,\lambda)}
    \label{E:HSIsubtNorm}
\end{equation}
Here, $\xi(x,y,\lambda)$ denotes the intensity of an image at pixel $(x,y)$ and wavelength $\lambda$; subscripts $image$, $\ch{Al}$, and $spec$ denote the image being calibrated, the aluminum sample, and the spectralon sample, respectively; and the $dark$ subscript denotes an image acquired with the light source turned off. 

\begin{figure}
\centering
\includegraphics[width=\figWidthCol]{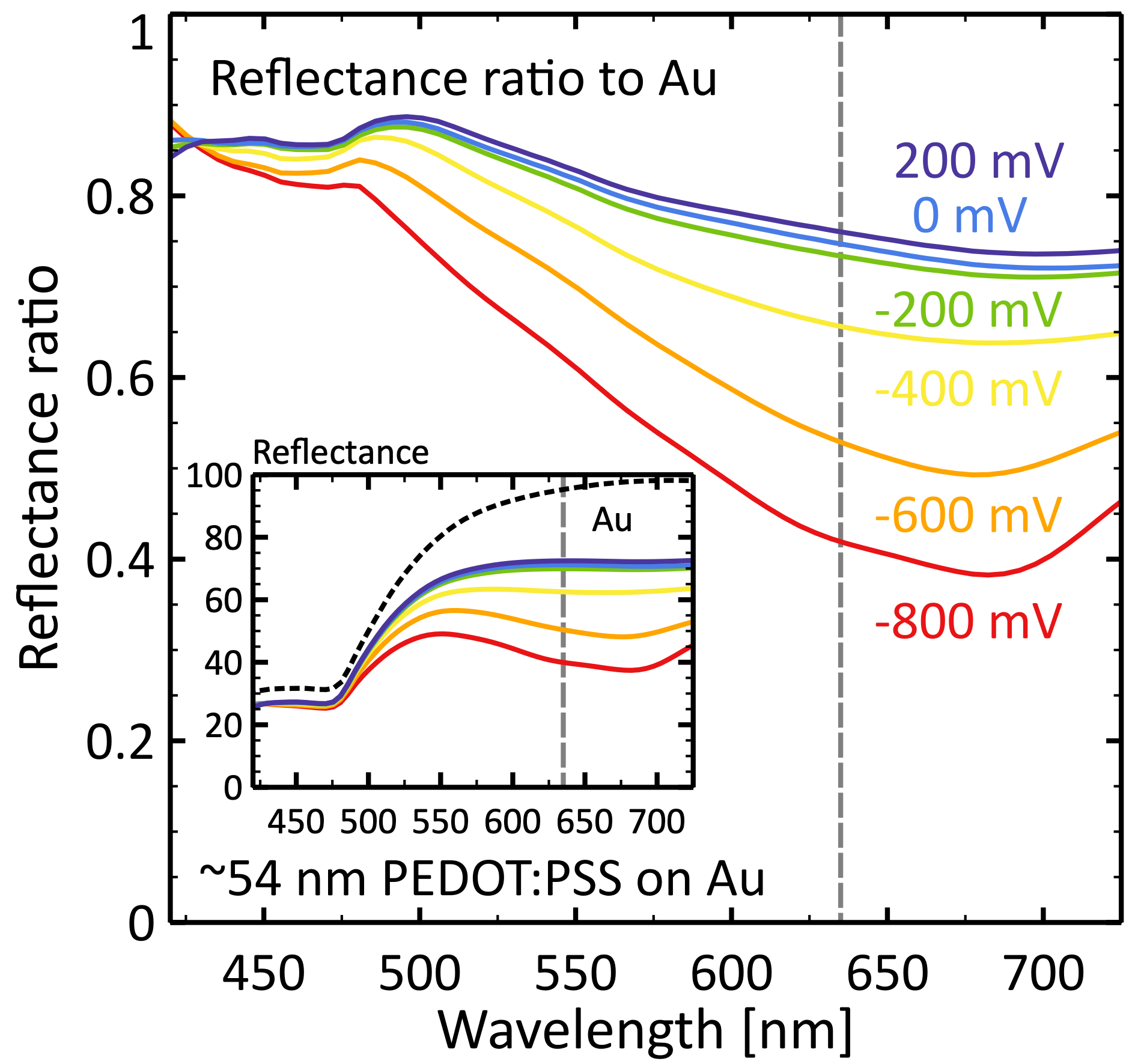}%
\caption{\textbf{Reflectance spectra for SPOT.}
Device consists of a PEDOT:PSS~($\app54$~\si{\nano\meter})-\ch{Au} stack in $1\times$PBS.
The polymer is cross-linked using heat treatment~\cite{Doshi2025-2415827}.
The vertical dashed line corresponds to $\lambda_l=635$~\si{\nano\meter}, the wavelength we use in the temporal and frequency-dependent characterization studies. 
Larger plot shows ratio of reflectance of PEDOT:PSS-\ch{Au} stack to \ch{Au} alone. 
Inset shows reflectance of  PEDOT:PSS-\ch{Au} stack and \ch{Au} individually.
}%
\label{F:plotSpectra}%
\end{figure}

Next, we average all pixels in each image to create a single spectrum that is a function of wavelength only:
\begin{equation}
    \langle\xi_{image}(\lambda)\rangle = \frac{1}{N_x N_y}\sum_{x=1}^{N_x}\sum_{y=1}^{N_y}\xi_{image}(x,y,\lambda) .
    \label{E:averageXYhsi}
\end{equation}
Here, $N_x$ and $N_y$ are the number of $x$ and $y$ pixels, respectively. Thereafter, we calculate the theoretical reflectivity spectrum $\varrho_{\ch{Al},theo}(\lambda)$ of the \ch{Al}-on-\ch{SiO2}-on-\ch{Si} sample that we measured, and use this with our measurement $\langle\xi_{\ch{Al}}(\lambda)\rangle$ to find the reflectance correlation $K(\lambda)$ for the spectra. 
\begin{equation}
    K(\lambda) = \frac{\varrho_{\ch{Al},theo}(\lambda)}{\langle\xi_{\ch{Al}}(\lambda)\rangle}
    \label{E:findCorrHSI}
\end{equation}
Here, the quotient is calculated wavelength-by-wavelength. Finally, we apply this correlation to the other images that we obtained, as in 
\begin{equation}
    \langle \varrho_{image}(\lambda)\rangle = \langle\xi_{image}(\lambda)\rangle  K(\lambda) , 
    \label{E:applyCorrHSI}
\end{equation}
where the product is conducted wavelength-by-wavelength. The results are depicted in Figure~\ref{F:plotSpectra} at several applied potentials. 

Although optical constants for PEDOT:PSS are available~\cite{Lin2020-110435, Dingler2022-1600}, we do not attempt to simulate our spectra with them because we do not have a simultaneous calibration of the swelled thickness of the polymer.

In Figure~\ref{F:plotSpectra}, the vertical dashed line denotes $\lambda_l=635$~\si{\nano\meter}, the wavelength we use in the temporal and frequency-dependent characterization studies. We extract the reflectance values and fit a sigmoid using Equation~\ref{E:linearSweepReflect} with parameters $\chi_1=72.5$\%, $\chi_2=34.0$\%, $\chi_3=0.007$~\si{\milli\volt\tothe{-1}}, and $\chi_4=-550$~\si{\milli\volt}. We can compare the fits from Figures~\ref{F:rampRef} and~\ref{F:plotSpectra}. The difference in the $\chi_1$ and $\chi_2$ parameters is related to the different polymer thickness values and metal substrate composition between the two experiments (45~\si{\nano\meter} PEDOT:PSS on \ch{Pt} \vs 54~\si{\nano\meter} PEDOT:PSS on \ch{Au}). The differences in the $\chi_3$ and $\chi_4$ parameters are more noteworthy, and may indicate current leakage through the uninsulated probe used for the hyperspectral measurements. Regardless, we emphasize that our hyperspectral data show the correct spectral behavior for PEDOT:PSS relative to other studies~\cite{Sonmez2004-1905, Lin2020-110435, Dingler2022-1600}, including the monotonic increase of reflectance with voltage, even if the applied potential values have some uncertainty. 

\section{Details about durability studies}\label{S:expDurability}

We conduct our durability studies using the same custom experimental setup as for the frequency-dependent electrochemical experiments (Section~\ref{S:ExptSetupFdep}). 
We keep the LED off but the laser turned on for the entire trial, except at brief intervals where we photograph the SPOT. 
Since the liquid solution is open to the laboratory environment, it evaporates over time; we add DI \ch{H2O} throughout the experiment (every few hours) to keep the liquid droplet size (roughly 700~\si{\micro\liter}) approximately constant. 
The reflectance data reported in Figure~\ref{F:duration} in the main article should be regarded as only approximate, because we were only able to collect the $V_{p,\ch{Au}}$ and $V_{p,\ch{Au},dark}$ values (see Equation~\ref{E:correctReflect}) at the beginning and end of the experiment, and the room lighting was not held constant. 
However, we estimate that these two quantities were minimally dependent on the ambient conditions, thus making  $V_{p,chip}$ (recorded at each interval) the dominant factor determining the reflectance. 

\begin{figure*}
\centering
\includegraphics[width=120mm]{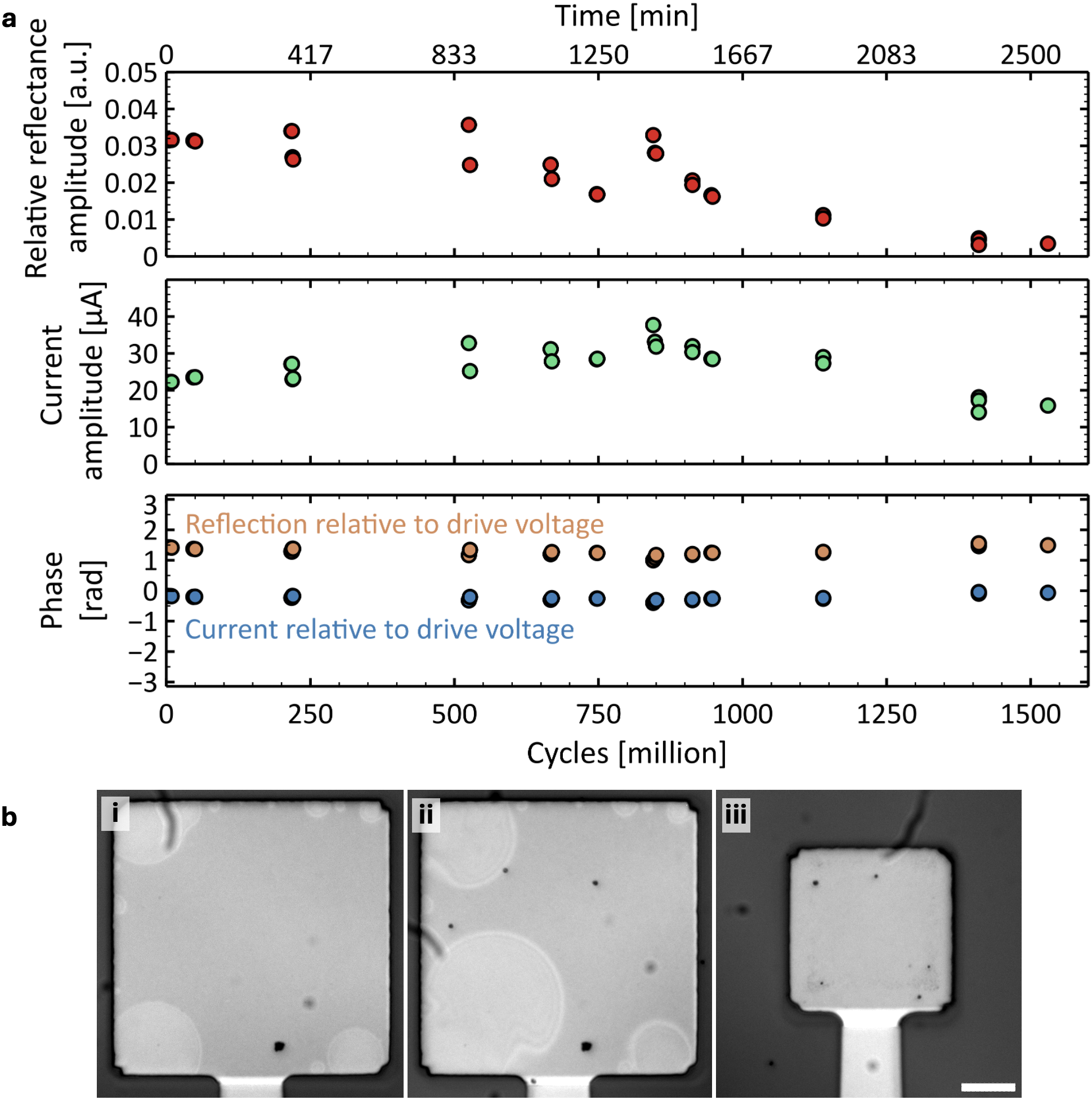}%
\caption{\textbf{Additional results of durability study.} 
Device consists of a PEDOT:PSS~($\app84$~\si{\nano\meter})-\ch{Pt} stack in $1\times$PBS, with lateral dimensions $40\times40$~\si{\micro\meter\squared}. The polymer  is cross-linked using GOPS (see Section~\ref{S:fabCrossGOPS})~\cite{Tseng2024-13384}, and the driving signal applied to the working electrode is a 10~\si{\kilo\hertz} sine wave between maximum and minimum potentials of +100~\si{\volt} and -300~\si{\volt}, respectively.
\textbf{(a)} Charts showing relative reflectance amplitude (top), current amplitude (middle), and phase angle of reflectance and current signals (bottom) as a function of the number of cycles. 
\textbf{(b)} (i,ii) Micrographs of adjacent $50\times50$~\si{\micro\meter\squared} device on same chip that was not actuated in the durability experiment, (i) before and (ii) after the trial. (iii) Micrograph of another adjacent $30\times30$~\si{\micro\meter\squared} device on same chip that was not actuated in the durability experiment, shown after the trial. 
Scale bar (b): 10~\si{\micro\meter}.
}%
\label{F:additionalDurationFigs}%
\end{figure*}

Figure~\ref{F:duration} in the main article presents the durability study results, and Figure~\ref{F:additionalDurationFigs} shows additional data from the experiment. Panel (a) of Figure~\ref{F:additionalDurationFigs} shows summary statistics throughout the experiment, and (b) provides images of adjacent SPOTs on the same chip. From panel (a), the relative reflectance amplitude $A_R$ remains roughly constant to about 500~million cycles, and thereafter decreases, while the current amplitude initially increases and then decreases after about 800~million cycles. The phase angles are consistent throughout the experiment. 


Panel (b) parts (i) and (ii) show a micrograph of a $50\times50$~\si{\micro\meter\squared} adjacent device on same chip that was not actuated before (i) and after (ii) the durability experiment. Some evidence of what may be delamination of the polymer is evident before the test (lighter-gray areas near the edges of the electrode); the associated areas are larger after the durability experiment, but the polymer otherwise appears to be in its original condition. Panel (b) part (iii) shows another adjacent device, this one with a $30\times30$~\si{\micro\meter\squared} area, after the experiment. Some dark spots are observable in panel (b) parts (ii) and (iii), which may be fragments of PEDOT:PSS that detached from the tested device during or after the experiment. The comparable cleanliness of these devices relative to the device undergoing the durability cycling points to the cycling itself as the cause of the polymer degradation shown in Figure~\ref{F:duration}(b) in the main article, rather than simply being immersed in $1\times$PBS for an extended time period~\cite{Richardson-Burns2007-1539, Schander2016-6174, Duc2018-14}. 

PEDOT:PSS films are, mechanically, porous sponge-like networks~\cite{Volkov2017-1700329, Sedghamiz2023-5512} whose thickness can change several-fold with oxidation state~\cite{Savva2018-12023, Biessmann2018-9865, Modarresi2020-6267,  Dingler2022-1600, Doshi2025-205}. This is because PEDOT:PSS is a mixed ion-electron conductor~\cite{Proctor2016-1433, Fabiano2017-e1700345, Savva2018-12023}; cations in the electrolyte move in response to its oxidation state to maintain electroneutrality in the polymer. When biased negatively, PEDOT:PSS films adopt de-doped/reduced oxidation states that are swollen because hydrated cations from the electrolyte enter to electrostatically balance the \ch{SO3-} groups in the PSS. Under positive bias, they become re-doped/oxidized and morphologically shrunken/deflated because cations vacate the film once holes (PEDOT\textsuperscript{+}) balance the \ch{SO3-} charges~\cite{Volkov2017-1700329, Rebetez2021-2105821, Lyu2023-746}. The repeated redox cycling of the polymer during our durability testing presumably caused it to swell/shrink more than a billion times, likely inducing mechanical fatigue that caused it to break down. 

Finally, we note that a second trial with heat-treated PEDOT:PSS~\cite{Doshi2025-2415827} achieved similar results to the GOPS-treated PEDOT:PSS discussed here. 

\section{Details about microcircuit temperature studies}\label{S:expClockbot}

\begin{figure}
\centering
\includegraphics[width=\figWidthCol]{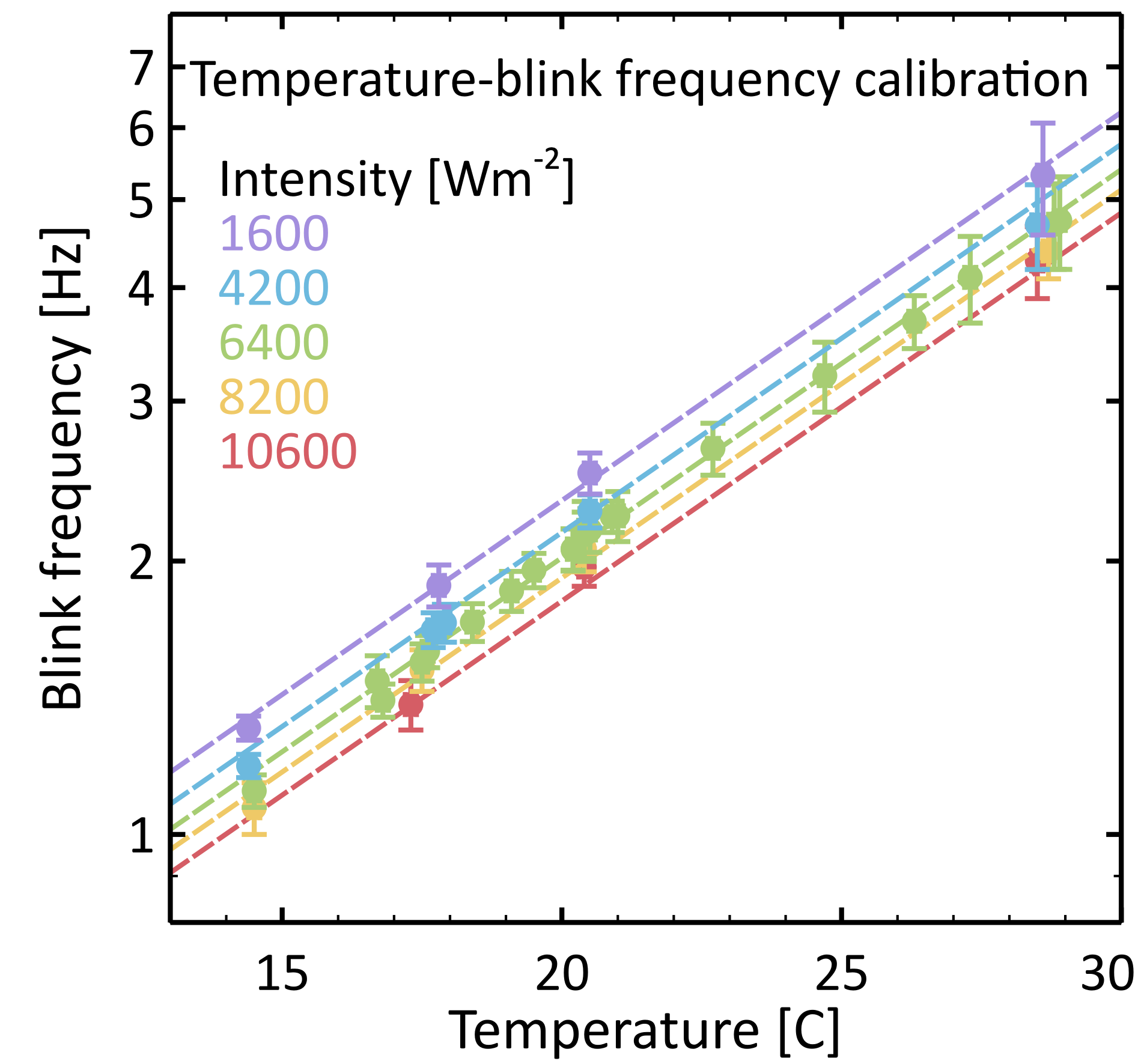}%
\caption{\textbf{Sample measurement data for microcircuit temperature communication.} 
We characterize the blink frequency as a function of the light intensity and the electrolyte ($1\times$PBS) temperature (Equation~\ref{E:tempIntensCalib}). 
}%
\label{F:plotsOfClockbotTdataIntensity}%
\end{figure}

We test the temperature-sensitive microcircuits in a setup similar to that of the electrochemical experiments (Section~\ref{S:ExptSetupFdep}). 
In particular, we submerge the chip containing the microcircuits in a shallow PYREX (Corning, Inc.) Petri dish containing $1\times$PBS, which we position directly atop a Peltier unit (Tark Thermal Solutions, \#430126-503) for heating and cooling.
We drive the unit with a Tektronix 2450 SMU; typical voltages (currents) fall in the range -2000~\si{\milli\volt} (-370~\si{\milli\ampere}) at 27~\dC to 1000~\si{\milli\volt} (215~\si{\milli\ampere}) at 14~\dC. 
A thermocouple probe (Dwyer Instruments, LLC, \#5TC-TT-K-36-36) sealed under a piece of Kapton tape on the bottom of the dish inside the electrolyte and immediately adjacent to the chip allows us to monitor the temperature of the chip and fluid, as indicated on a digital handheld meter (Dwyer Instruments, LLC, HH801A). 
We record a video of the digital on/off reflected signal from the microcircuit's SPOT using a non-immersion lens (Olympus Co., PLN Plan Achromat 10X, \#1-U2B223) and the microscope's camera (Thorlabs, Inc., Zelux CS165MU1), export the raw intensity data for a region of interest (ROI) containing the SPOT using ImageJ~\cite{Schneider2012-671}, and thereafter derive the blink frequency using a MATLAB R2024b script. We fit the temperature ($T$)-intensity ($\varphi$)-blink frequency ($F_b$) data to the function 
\begin{equation}
    F_b = \left(\varsigma_1\varphi^2 + \varsigma_2\varphi + \varsigma_3\right)\exp\!{\left(\varsigma_4 T\right)}, 
    \label{E:tempIntensCalib}
\end{equation}
where the four $\varsigma$ variables are fitting parameters (see Figure~\ref{F:plotsOfClockbotTdataIntensity}). The LED source (SOLIS-3C) serves to illuminate the chip (no laser light); we quantify the light intensity at each power setting at wavelength $\lambda=470$~\si{\nano\meter} through a 10~\si{\micro\meter} aperture (Thorlabs, Inc., PD10D) using a Thorlabs, Inc. S120C meter. 
After releasing the microcircuits from the chip in the dicing process, we image them again in an identical setup, this time with the singulated temperature-sensitive chips on the bottom of the Petri dish.

\section{Details about logo demonstration imaging}\label{S:expLogoImaging}

We image the University of Pennsylvania coat of arms (Figure~\ref{F:setupFabOverview}(c) in the main article) in $1\times$PBS, controlling its oxidation state by landing a probe needle directly on an exposed area of \ch{Pt} and applying a potential relative to an \ch{Ag}/\ch{AgCl} counter electrode. We gather the microscopic images using the Thorlabs, Inc. scope mentioned above.  We obtain the macroscopic images using a DSLR camera (Canon EOS 5D Mark IV with a Canon 100~\si{\milli\meter} f/2.8 macro lens) inside a one-cubic-foot white light box (Glendan), imaging through a 50/50 beam splitter (Thorlabs CCM1-BS013) suspended over the chip to eliminate non-perpendicular light rays. 

\section{Details imaging through chicken skin}\label{S:expChixSkin}

As an additional demonstration, we show that SPOTs can be imaged through tissue. To accomplish this, we place a piece of chicken skin (\textit{Gallus gallus domesticus}, roughly 250~\si{\micro\meter} thick as measured using dial calipers) obtained from a local supermarket over a $50\times50$~\si{\micro\meter\squared} SPOT and image it using 780~\si{\nano\meter} light from a CHROLIS-C1 (Thorlabs, Inc.) LED with a SciCam+ camera (Scientifica Ltd.). The PEDOT:PSS thickness is $\app200\pm50$~\si{\nano\meter}, the excitation potential is a $1$~\si{\hertz} sinusoidal potential relative to the adjacent \ch{Pt} counter electrode, and the electrolyte is the natural fluid from the chicken skin together with some added $1\times$PBS. Although scattering blurs the image~\cite{Ash2017-1909}, recent evidence suggests that deep penetration (albeit with high attenuation) may be possible~\cite{Radford2025-025014}. 

\section{Cost estimates}\label{S:costEst}

Here we estimate the cost to add a SPOT to a microcircuit mote as part of a CMOS (complementary metal-oxide-semiconductor) fabrication run~\cite{Moon2021-1430}. The expense is driven by the area of the wafer occupied by the SPOT and by the cost of the polymer. 

Working with 150~\si{\milli\meter} wafers with 90\% area utilization and assuming the microcircuit motes have $500\times500$~\si{\micro\meter\squared} areas, we estimate each wafer can accommodate about 70,000~motes.  For foundry-based 180~\si{\nano\meter} CMOS processing, a reasonable estimate is US\$10/\si{\centi\meter\squared} of \ch{Si} wafer surface area~\cite{Miskin2020-557}. Assuming each SPOT has an area of $30\times30$~\si{\micro\meter\squared} and also an associated \ch{Pt} counter electrode with an area of $100\times100$~\si{\micro\meter\squared}, this translates to an added cost per microcircuit mote of \app US\$0.001. Note that, in some applications, the counter electrode could be located on the backside of each mote (assumed to be otherwise unoccupied), leading to an added cost per device of just \app US\$0.0001 for each $30\times30$~\si{\micro\meter\squared} SPOT on the microcircuit mote's front side. 
The price of the polymer itself is also small; for our standard 5\% v/v ethylene glycol (EG) in PEDOT:PSS formulation, at a cost of US\$100/\si{\liter} for EG and US\$3000/\si{\liter} for PEDOT:PSS solution, the price to spin about 3~\si{\milli\liter} on a wafer is $<$US\$10, leading to a cost per mote of about \app US\$0.0001. Spinning additional layers for a thicker polymer layer would multiply this cost accordingly. Thus, the total cost of adding a SPOT to a mote, accounting for both occupied area and polymer, is in the range US\$0.0002 - US\$0.001. 

For comparison, we provide rough estimates for similarly-sized micro-communications technologies, namely micro-light-emitting diodes (\textmu{}LEDs)~\cite{Lin2023-042502} and vertical-cavity surface-emitting lasers (VCSELs)~\cite{Pan2024-229}. For simplicity, we consider a single substrate: \ch{GaAs}. We assume the cost of \ch{GaAs} processing is roughly five times that of \ch{Si} processing (\ie, US\$50/\si{\centi\meter\squared}) due to the lower production volumes and specialized tooling required. Assuming again 150~\si{\milli\meter} wafers are used with 90\% area utilization, and assuming that the \textmu{}LED/VCSEL die areas are $30\times30$~\si{\micro\meter\squared}, we find a cost per \textmu{}LED or VCSEL of \app US\$0.0001. In addition, the cost of the area occupied by a \textmu{}LED or VCSEL on each \ch{Si} microcircuit mote is unchanged from the SPOT estimate with the backside counter electrode, namely US\$0.0001 (in both cases we consider $30\times30$~\si{\micro\meter\squared} areas). Thus, the price of adding a \ch{GaAs} \textmu{}LED or VCSEL to a microcircuit mote is roughly US\$0.0002. Note that this estimate neglects the cost of any pick-and-place transferring or bonding required to place the \textmu{}LEDs or VCSELs onto the \ch{Si} microcircuits, and also neglects the price of any required encapsulation steps. 

\section{Future work}\label{S:futureWork}

There are several opportunities for future work. In their present form, SPOTs are well-suited for use in saline environments, which enables significant biomedical and healthcare applications today~\cite{Marblestone2013-137, Cortese2020-9173, Miskin2020-557, Singer2021-2100664, Moon2021-1430, Yuan2022-765, Xu2022-1, Lee2024-1110, Lee2025-1259, Gorski2025-354, VanHouten2023-937, Lassiter2025-eadu8009, Hanson2025-e2500526122, Shen2026-590}. Their utility could be extended to broader settings, however, if incorporated into hydrogels or coated with an ionic liquid polymer layer~\cite{Lu2019-1043, Takalloo2019-60, Zhang2020-1904752, Lin2024-14740}. Additionally, SPOTs have modest signal-to-noise ratios (SNR), which are applicable for immediately-important line-of-slight microscope-based measurements~\cite{Miskin2020-557, Lassiter2025-eadu8009, Hanson2025-e2500526122}. However, their SNR could be improved by integrating retro-reflectors to increase the back-scattered light intensity~\cite{Bender1973-229, Chalasani2011-1158, Wheaton2021-8504, Han2022-eabn0602} or by interrogating the devices with two laser colors simultaneously to examine the resulting reflectance ratio~\cite{Zhou2022-23505}. Also, SPOTs have moderate data bandwidths compared to other similarly-sized optical emitters~\cite{Lin2023-042502, Hofmann2011-A1250, Pan2024-229}, which reflects our deliberate design emphases of simplicity, size minimization, and scalability, and places SPOTs in a complementary position to deliver communication at a massive scale with minimal fabrication overhead. However, SPOTs' bandwidth could be improved by increasing the mobility of ions and holes within the polymer~\cite{Rivnay2016-11287}, by incorporating \ch{Ag} or \ch{Au} nanoparticles into the polymer matrix~\cite{Mumtaz2012-5360}, or by implementing multiple independent reflective modules and segregating their signals in the spatial domain (much like a QR code~\cite{Gu2011-733}). Finally, there is significant potential for integrated systems that combine optical communication with sensing using the same PEDOT:PSS electrode~\cite{Liang2021-2100061, Wang2023-100036, Burtscher2024-7, Kim2025-4032, Li2025-87}, or that leverage alternate electrochromic polymers or blends~\cite{Gu2022-14679, Wang2023-100036}. 

\section{Captions for videos}\label{S:captionsVideos}

\begin{enumerate}
    \item \texttt{PennCoatOfArmsMacro.avi}: A macroscopic image of the University of Pennsylvania coat of arms consisting of microscopic SPOTs in $1\times$PBS solution. Polymer is $\app170$~\si{\nano\meter} on a \ch{Pt} electrode (three spincoats). Potential applied to the \ch{Pt} is a $250$~\si{\milli\hertz} square wave ($-800$~\si{\milli\volt} to $+200$~\si{\milli\volt}) relative to an \ch{Ag}/\ch{AgCl} counter electrode. 
    \item \texttt{PennCoatOfArmsMicro.avi}: An enlarged microscopic image of the University of Pennsylvania coat of arms (depicting the eye of the dolphin) consisting of microscopic SPOTs in $1\times$PBS solution. Polymer is $\app170$~\si{\nano\meter} on a \ch{Pt} electrode (three spincoats). Potential applied to the \ch{Pt} is a $250$~\si{\milli\hertz} square wave ($-800$~\si{\milli\volt} to $+200$~\si{\milli\volt}) relative to an \ch{Ag}/\ch{AgCl} counter electrode. 
    \item \texttt{SPOTstepPotenetialLaserOn.avi}: A $50\times50$~\si{\micro\meter\squared} SPOT consisting of $\app73$~\si{\nano\meter} PEDOT:PSS on \ch{Pt} actuated by a $1$~\si{\hertz} square wave potential ($-800$~\si{\milli\volt} to $+200$~\si{\milli\volt}) relative to an \ch{Ag}/\ch{AgCl} counter electrode in $1\times$PBS solution. A $\lambda_l=635$~\si{\nano\meter} laser spot, focused to roughly $10\times20$~\si{\micro\meter\squared}, is visible. The polymer was deposited using two spincoats. Video speed is real-time $(1\times)$.
    \item \texttt{SPOTstepPotenetialLaserOff.avi}: A $50\times50$~\si{\micro\meter\squared} SPOT consisting of $\app73$~\si{\nano\meter} PEDOT:PSS on \ch{Pt} actuated by a $1$~\si{\hertz} square wave potential ($-800$~\si{\milli\volt} to $+200$~\si{\milli\volt}) relative to an \ch{Ag}/\ch{AgCl} counter electrode in $1\times$PBS solution. Some evidence of damage associated with the laser spot is visible. The polymer was deposited using two spincoats. Video speed is real-time $(1\times)$.
    \item \texttt{SPOTsinePotentialOneSpin.avi}: A $30\times30$~\si{\micro\meter\squared} SPOT consisting of $\app35$~\si{\nano\meter} PEDOT:PSS on \ch{Pt} actuated by a $1$~\si{\hertz} sinusoidal potential ($-800$~\si{\milli\volt} to $+200$~\si{\milli\volt}) relative to an \ch{Ag}/\ch{AgCl} counter electrode in $1\times$PBS solution. The polymer was deposited using a single spincoat (see Section~\ref{S:fabSPOTs}). Video speed is real-time $(1\times)$.
    \item \texttt{SPOTsinePotentialFourSpin.avi}: A $30\times30$~\si{\micro\meter\squared} SPOT (PEDOT:PSS on \ch{Pt}) actuated by a $1$~\si{\hertz} sinusoidal potential ($-800$~\si{\milli\volt} to $+200$~\si{\milli\volt}) relative to an \ch{Ag}/\ch{AgCl} counter electrode in $1\times$PBS solution. The polymer was deposited using four spincoats and has a thickness of $\app200\pm50$~\si{\nano\meter} (some non-uniformity in the coating is visible). Video speed is real-time $(1\times)$.
    \item \texttt{durationBefore.avi}: A $40\times40$~\si{\micro\meter\squared} SPOT consisting of $\app84$~\si{\nano\meter} PEDOT:PSS on \ch{Pt} actuated by a $1$~\si{\hertz} sinusoidal potential ($-400$~\si{\milli\volt} to $+100$~\si{\milli\volt}) relative to an \ch{Ag}/\ch{AgCl} counter electrode in $1\times$PBS solution. The polymer was deposited using a single spincoat and was crosslinked using GOPS~\cite{Tseng2024-13384} (see Section~\ref{S:fabCrossGOPS}). This represents the state of the device before the duration study discussed in the main text (see also Figure~\ref{F:additionalDurationFigs}). Video speed is real-time $(1\times)$.
    \item \texttt{durationAfter.avi}: A $40\times40$~\si{\micro\meter\squared} SPOT consisting of $\app84$~\si{\nano\meter} PEDOT:PSS on \ch{Pt} actuated by a $1$~\si{\hertz} sinusoidal potential ($-400$~\si{\milli\volt} to $+100$~\si{\milli\volt}) relative to an \ch{Ag}/\ch{AgCl} counter electrode in $1\times$PBS solution. The polymer was deposited using a single spincoat and was crosslinked using GOPS~\cite{Tseng2024-13384} (see Section~\ref{S:fabCrossGOPS}). This represents the state of the device after the duration study discussed in the main text (see also Figure~\ref{F:additionalDurationFigs}). It appears that the PEDOT:PSS has wrinkled off of the electrode in some areas, but interestingly, these regions still appear to retain their electrochromism. Video speed is real-time $(1\times)$.
    \item \texttt{microcircuitTemperatureDemo.avi}: A temperature-sensitive microcircuit blinking at 2.24~\si{\hertz} in 20.9\dC $1\times$PBS under a microscope illumination of 6400~\si{\watt\per\meter\squared}. SPOT consists of $\app50$~\si{\nano\meter} PEDOT:PSS on \ch{Pt}. Video speed is real-time $(1\times)$. 
    \item \texttt{SPOTthroughChickenSkin.avi}: Viewing a $50\times50$~\si{\micro\meter\squared} SPOT through a piece of chicken skin (\textit{Gallus gallus domesticus}, roughly 250~\si{\micro\meter} thick as measured using dial calipers) obtained from a local supermarket. PEDOT:PSS thickness is $\app200\pm50$~\si{\nano\meter}. Excitation potential is a $1$~\si{\hertz} sinusoidal potential relative to the adjacent \ch{Pt} counter electrode. The electrolyte is the natural fluid from the chicken skin together with some added $1\times$PBS. The illumination is 780~\si{\nano\meter} light from a CHROLIS-C1 (Thorlabs, Inc.) LED source and the camera is a SciCam+ (Scientifica Ltd.). Video speed is real-time $(1\times)$.
\end{enumerate}

\end{document}